\documentclass[11pt,preprint]{aastex}
\shorttitle{A Global 86\,GHz VLBI Survey}
\shortauthors{Lee et al.}
\begin{document}
\title{A Global 86\, GHz VLBI Survey of Compact Radio Sources}
\author{Sang-Sung Lee\altaffilmark{1}, Andrei P. Lobanov\altaffilmark{1}, 
Thomas P. Krichbaum\altaffilmark{1}, Arno Witzel\altaffilmark{1},
Anton Zensus\altaffilmark{1}, Michael Bremer\altaffilmark{2}, 
Albert Greve\altaffilmark{2}, and Michael Grewing\altaffilmark{2}}
\altaffiltext{1}{Max-Planck-Institut f\"ur Radioastronomie,
Auf dem H\"ugel 69, 53121 Bonn, Germany}
\altaffiltext{2}{Institut de Radio Astronomie Millim\'etrique,
300 rue de la Piscine, 38406 Saint Martin d'H\`eres, France}
\begin{abstract}
We present results from a large 86\,GHz global VLBI
survey of compact radio sources.
The main goal of the survey is to increase by factors of 3--5 
the total number of objects accessible for future 3-mm VLBI imaging.
The survey observations reach the baseline sensitivity  of 0.1\,Jy
and image sensitivity of better than 10 mJy/beam.
The total of 127 compact radio sources have been observed.
The observations have yielded images for 109 sources,
extending the database of the sources imaged at 86\,GHz with VLBI
observation by a factor of 5,
and only 6 sources have not been detected.
The remaining 12 objects have been detected
but could not be imaged due to insufficient closure phase information.
Radio galaxies are less compact than quasars and BL Lacs
on sub-milliarcsecond scale.
Flux densities and sizes of core and jet components of all
imaged sources have been estimated using Gaussian model fitting.
From these measurements, brightness temperatures have been calculated,
taking into account resolution limits of the data.
The cores of 70\% of the imaged sources are resolved.
The core brightness temperatures of the sources peak at $\sim 10^{11}$\,K
and only 1\% have brightness temperatures higher than $10^{12}$\,K.
Cores of Intraday Variable (IDV) sources are smaller in angular size
than non-IDV sources, and so yield higher brightness temperatures.
\end{abstract}
\section{Introduction}
\label{sec:intro}
Very long baseline interferometry (VLBI) at millimeter wavelengths 
offers the best tool for imaging compact radio structures on scales 
of several dozens of microarcseconds.
The first detection of single-baseline interference fringes in 
an 89\,GHz (3.4\,mm) VLBI observation was reported by~\citet{Readhead83}, 
demonstrating the feasibility of 3\,mm-VLBI.
After that, many VLBI observations at 86\,GHz have been made,
probing the most compact regions in active galactic nuclei (AGN).
However, the number of objects detected and imaged at 86\,GHz 
remained small, compared with the number of objects imaged with VLBI 
at lower frequencies.

Sensitive VLBI observations at 86\,GHz
have been made for several sources, including 3C 111~\citep{Doeleman96}, 
3C 454.3~\citep{Krich95,Krich99,Pagels}, NRAO 150~\citep{Agudo150},
NRAO 530~\citep{Bower97}, M87~\citep{Krich06}, 3C 273 and 3C 279~\citep{att01}. 
In order to increase the number of objects imaged at 86\,GHz,
four detection and imaging surveys were conducted during the 1990s, 
with a total of 124 extragalactic radio sources 
observed~\citep[see][]{Beasley97, Lonsdale98, Ranta98, Lobanov00}.
In these surveys, fringes were detected of 44 objects, but 
only 24 radio sources have been
successfully imaged. Table~\ref{table:vlbisurveys} gives an overview 
of these surveys. The low detection and imaging rates of 
the previous 86\,GHz surveys were caused by the relatively poor 
baseline sensitivities, small numbers of telescopes and short observing times.

The results of the survey of a larger number of sources can be used to 
investigate the innermost region
of compact jets and to observationally test inner jet models~\citep{mar95}:
accelerating and decelerating jet models. In the accelerating jet model,
the jet accelerates hydrodynamically from the base of the jet, and as 
the internal energy of the jet plasma is converted into the kinetic energy
of bulk flow, the jet Lorentz factor increases along the jet. In this model,
an ultra relativistic neutral beam is generated from the central engine and
then the neutrons decay into protons and electrons which form a relativistically
flowing plasma. In the decelerating jet model, the central engine produces 
a highly collimated beam of ultra relativistic electron-positron pair plasma 
that scatters photons produced outside the jet (particle cascade). 
The scattered photons emit X-rays and $\gamma$~-rays, which decelerates 
the beam and so decreases the Lorentz factor along the jet. 
 
The theoretical prediction from the jet models leads to the fact 
that the intensity profiles along the jet are different from each other and 
have a distinctive shape in each of these model. The resulting brightness
temperature can be used to probe the difference. The intrinsic brightness
temperature of a distribution of observed brightness temperatures can 
be determined from a statistical modeling~\citep{Lobanov00}.   
By estimating the brightness temperature at several frequencies 
(e.g., 15\,GHz, 43\,GHz, and 86\,GHz) and determining the intrinsic brightness
temperatures, we would be able to constrain the physical conditions
(e.g., dynamics and compositions) of the innermost region of the compact sources.
Moreover the dependence of the intrinsic brightness temperatures on 
the observing frequencies will tell about the feasibility of VLBI 
at higher frequencies (e.g., 150\,GHz, 215\,GHz, etc).

A large global 86\,GHz VLBI survey of compact radio sources was carried out
from 2001 October to 2002 October using the Coordinated Millimeter VLBI 
Array (CMVA)~\citep{rog+95}, which is succeeded by the Global Millimeter 
VLBI Array (GMVA)\footnotemark[1].
\footnotetext[1]{See http://www.mpifr-bonn.mpg.de/div/vlbi/globalmm/index.html.}
The main aim of this VLBI survey is to increase the total number of objects 
accessible for future 3\,mm-VLBI imaging by factors of 3--5, and to provide
the database for the subsequent statistical modeling in order to test  
the inner jet models.

\section{Observation}

   \subsection{Source Selection}
   \label{sec:source}

The source selection of this survey is based on the results from the VLBI surveys at 
22\,GHz~\citep{Moellen96} and 15\,GHz~\citep{Kellermann98}, 
and on flux density measurements from the multi-frequency monitoring programs at Mets\"ahovi 
at 22, 37, and 86\,GHz~\citep{Teras98} and at Pico Veleta at 90, 150, and 
230\,GHz (Ungerechts, priv. comm.). Using these databases, we selected 
the sources with an expected flux density above 0.3\,Jy at 86\,GHz. 
We excluded some of the brightest sources already imaged at 86\,GHz,
and focused on those sources which had not been detected or imaged 
in the previous surveys. 
Objects in the southern sky with low declinations ($\delta \leq -40^\circ$) 
were rejected, in order to optimize the {\it uv}-coverage of the survey data.

\begin{figure*}
\begin{center} 
\includegraphics[width=0.5\textwidth, angle = -90]{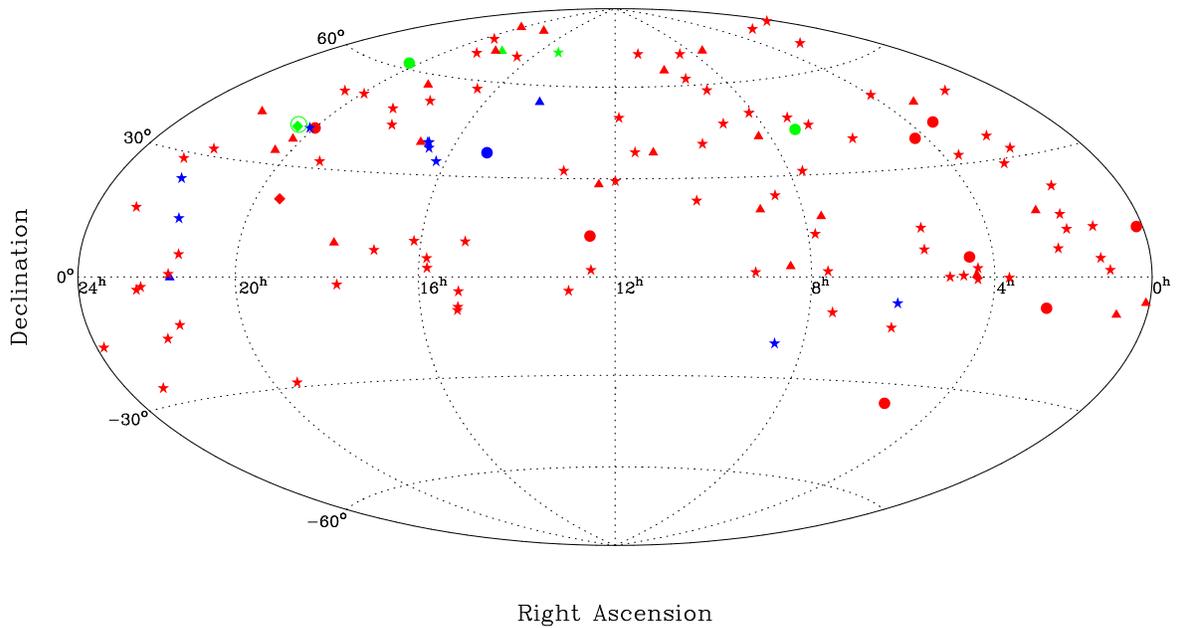}
\caption{The sky-distribution of 86\,GHz VLBI sources: 
109 detected and imaged sources (red), 
12 detected and non-imaged sources (blue) 
and 6 non-detected sources (green).
Symbols: stars are quasars (Q), triangles are BL Lac objects (B), 
circles are galaxies (G), diamonds are unidentified sources (U)
and a single open star represents a star, Cyg X-3 (S).\label{fig:skyplot}} 
\end{center}
\end{figure*}

According to the aforementioned selection criteria, a total of 127 compact 
radio sources was selected and observed, consisting of 88 quasars, 
25 BL Lac objects, 11 radio galaxies, 1 star (Cyg X-3), and 2 unidentified 
sources. Table~\ref{table:sourcelist} lists the general information
of the observed sources, with the columns corresponding to 
(1) source, (2) name, (3) epoch, (4) Right Ascension (J2000),
(5) Declination (J2000), (6) status, (7) redshift,
(8) optical class, (9) optical magnitude, 
and (10) total flux density $S_{\rm 86\,GHz}$. 
In Figure~\ref{fig:skyplot}, 
the sky-distribution of the observed sources is shown.

   \subsection{Observational Strategy}
   \label{sec:obs}

The survey observations were conducted during three sessions of 
the Co-ordinated (global) millimeter VLBI array (CMVA/GMVA) 
on 2001 October, 2002 April and 2002 October, 
as summarized in Table~\ref{table:obs}.
Table~\ref{table:obs} shows the log of the survey observations,
with the columns corresponding to (1) epoch, (2) code of each epoch,
(3) bit rate, (4) frequency channels, (5) sampling mode, 
(6) total observing bandwidth, (7) number of sources, 
and (8) participating telescopes.

Table~\ref{table:antenna} lists the technical information of 
the participating telescopes, with the columns for (1) name, 
(2) abbreviation of the telescope name, (3) diameter, 
(4) typical zenith gain, 
(5) system temperature, (6) aperture efficiency, (7) typical 
zenith SEFD obtained from the formula, $SEFD = T_{\rm sys}/G$, 
(8) baseline sensitivity on baseline to Pico Veleta, assuming 
a recording rate 256 Mbps and a fringe-fit interval of 30\,seconds,
and (9) $7\sigma$ detection threshold.
The participation of the large and sensitive European 
antennas (the 100-m radio telescope at Effelsberg, the 30-m radio 
telescope at Pico Veleta, the 6$\times$15-m interferometer telescopes on 
Plateau de Bure) and the 8 VLBA\footnote{The Very Long Baseline Array (VLBA) 
is an instrument of the National Radio Astronomy Observatory, which is 
a facility of the National Science Foundation operated under cooperative 
agreement by Associated Universities, Inc.}
antennas available at 86\,GHz 
resulted in a typical single baseline sensitivity of $\sim 0.1$\,Jy and 
an image sensitivity of better than 10\,mJy\,${\rm beam}^{-1}$. 

\begin{figure}
\plotone{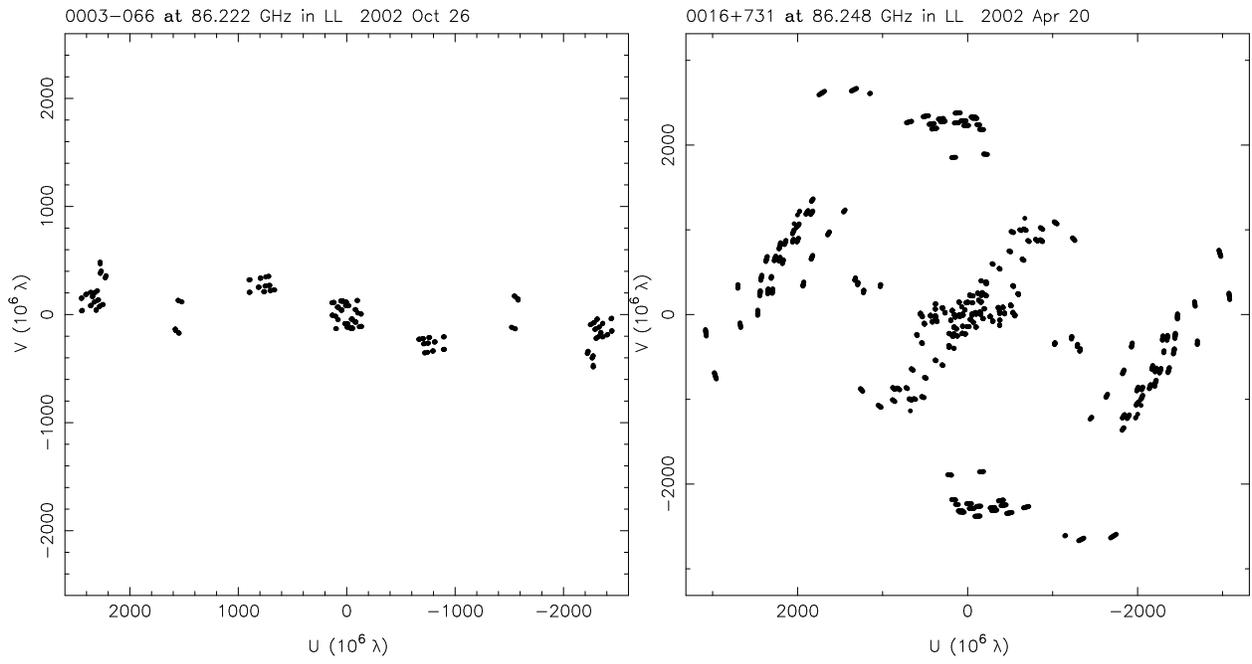}
\caption{{\it uv}-plots of 0003-066 and 0016+731 at low and high 
declinations of -06 and 73 degrees, respectively.\label{fig:uv}} 
\end{figure}

Every source in the sample was observed for 3-4 scans of 7-minute 
duration ({\it snapshot} mode). Although the {\it uv}-coverage of such 
an experiment limits the dynamic range and structural 
sensitivity of images, the large number of the participating antennas 
gives a sufficient {\it uv}-coverage of the sources at low and high 
declinations (Figure~\ref{fig:uv}). The data were recorded either with 
128-MHz or 64-MHz bandwidth using the MkIV VLBI system with 1- and 2- bit 
sampling adopted at different epochs. The observations were made in 
lefthand circular polarization (LCP). Three to four scans per hour were
recorded, using the time between the scans for antenna focusing, pointing 
and calibration. The data were correlated using the MkIV correlator of the
Max-Planck-Institut f\"ur Radioastronomie (MPIfR) in Bonn~\citep{Alef2000}.

\section{Data Processing}

\begin{figure}
\begin{center} 
\includegraphics[width = 0.5\columnwidth,clip]{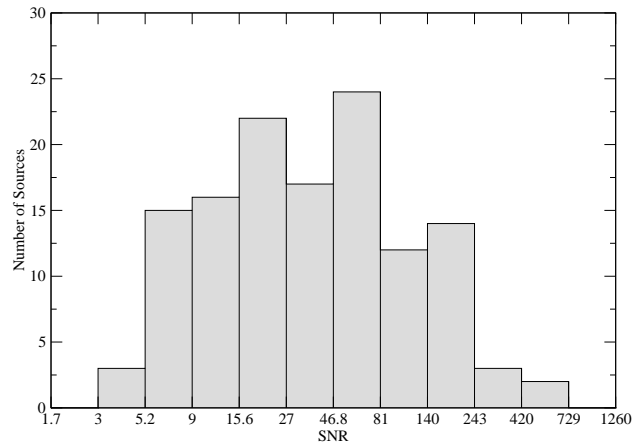}
\caption{ Distribution of fringe detection SNR of the sources.
The highest SNRs are 425 for 1741-038
on the baseline with Pico Veleta -
Plateau de Bure and for 1633+382 with Effelsberg - Pico Veleta.
The X-axis is in logarithmic scale of $\sqrt{3}$. 
The labels on the X-axis are corresponding to
$\sqrt{3}^1$, $\sqrt{3}^2$, ... $\sqrt{3}^{13}$.
\label{fig:DetectionRate} 
}
\end{center}
\end{figure}

In this section, we describe the post-correlation 
processing of the 3\,mm-VLBI survey datasets.
Fringes were searched in two steps using HOPS (Haystack Observatory 
Postprocessing System) and AIPS (The NRAO Astronomical Image Processing 
System). In the first step, the HOPS task {\it fourfit} was used to 
precisely determine phase-residuals. 
The first {\it fourfit} was run with a wide search window (e.g. a width 
of 1\,$\mu$sec for singleband delay, 2\,$\mu$sec for multiband delay 
and 500\,psec\,${\rm sec}^{-1}$ for delay rate) centered at zero in delay.
Since the {\it fourfit} produces baseline-based fringe solutions,
the mean and standard deviation of the detected fringe solutions 
on each baseline were estimated and served as the offset and width 
of the search window for the second {\it fourfit}. The detected fringe 
solutions from the second run were used to interpolate the offset of 
the singleband delay for non-detected scans. In the final run of 
{\it fourfit}, an {\it interpolated} search window was used with a width of
0.02\,$\mu$sec for the singleband delay and centered at
the offset interpolated for each non-detected scan.
After this first step of the fringe search, 
the total number of fringe detections
for the survey data was improved by up to 20\%. 
In the second step of the fringe search, the baseline-based fringe
solutions were imported into AIPS using a modified AIPS task 
MK4IN~\citep{Alef2002}. We then made an antenna-based fringe
fit to the data using the AIPS task FRING. Pico Veleta (PV)
was chosen as reference antenna for most of the data.  
When PV was not available, Fort Davis (Fd) was selected as
an alternative reference antenna. The antenna-based fringe fitting
was done with the solution interval of 7 minutes in order to 
achieve higher signal-to-noise ratio. 
Fringe solutions for strong sources were
used to define coarse search windows for the fringe solutions 
for nearby weaker sources.
With the fringe fit of {\it fourfit} and FRING, 
121 out of 127 observed sources have yielded fringe
detections with SNR$\geq 6$. Figure~\ref{fig:DetectionRate} 
shows the SNR distribution of the fringe detection in the entire survey data.
Only 6 sources (0710+439, 1458+718 (3C 309.1), 1749+701 (4C 09.57), 
2021+614, 2030+407 (Cyg X-3), and 2031+405 (MWC 349)) are not detected.
The highest SNRs of 425 is measured on the ``Pico Veleta--Plateau de Bure'' baseline 
for 1741-038 and the ``Effelsberg-Pico Veleta'' baseline for 1633+382.

The fringe fitted data were amplitude calibrated using 
regular measurements of the system temperatures and antenna gains
and the weather information for each station made during the observations. 
Where possible, time-dependent factors in the antenna power gains 
were accounted for by applying atmospheric opacity corrections.
The AIPS task APCAL was used to calibrate the amplitudes.
As a check of the accuracy and consistency of the amplitude calibration, 
we have investigated (independently for each of the detected sources) 
the calibrated visibility amplitudes using the best fit Gaussian component models 
obtained from the data (the corresponding models are given in Table~\ref{table:modelfit}). 
For each of the sources, the antenna gains were allowed to be scaled 
by a constant factor so as to optimize the fit by the Gaussian model. 

The obtained corrections are within 20\% for most of sources,
which is also reflected in the average correction factors listed 
in Table~\ref{table:gain}.
On average, the gain factors for Pico Veleta, Onsala, North Liberty,
Owens Valley, and Los Alamos remained within 10\% in all three observing sessions.
The average gains for Effelsberg did not change much except for the session C.
Fort Davis and Mauna Kea required average corrections by more than 20\%.
Time-dependent errors may still be present in the calibrated data. 
Therefore we expect overall calibration accuracy of $\sim$20--30\%

From the phase- and amplitude-calibrated data,
the images were made using the Caltech DIFMAP 
software~\citep{Shepherd94}.
After averaging in frequency, the {\it uv}-data were 
averaged in time over 30 seconds and were edited for deviant data points. 

The {\it uv}-data were then fitted with a simple Gaussian model. 
First, a single circular Gaussian component was applied to fit the data.
In case that the single-component model did not represent 
the data satisfactorily, a multiple-component model was applied.
Self-calibration and CLEAN deconvolution were applied 
to produce final images of the detected sources.

\begin{figure}
\begin{center} 
\includegraphics[width = 0.5\columnwidth, clip]{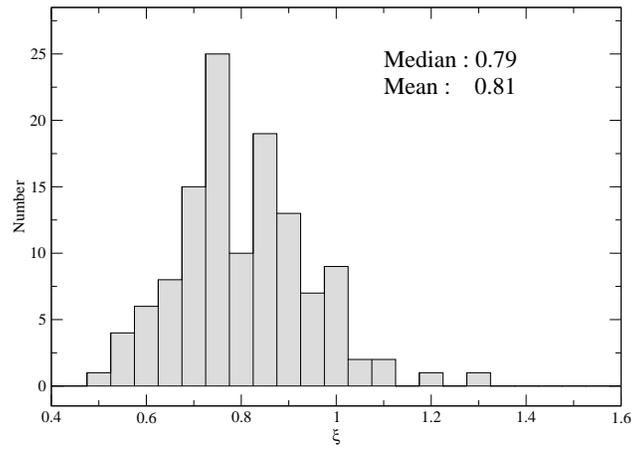}
\caption{ Distribution of the image quality factor $\xi_{\rm r}$.
The median and mean of the distribution are presented.
\label{fig:Qual}
}
\end{center}
\end{figure}

The noise in the final image can be expressed quantitatively 
by the quality $\xi_{\rm r}$ of the residual noise.
Suppose that a residual image has an rms $\sigma_{\rm r}$ and 
the maximum absolute flux density $|s_{\rm r}|$. For Gaussian 
noise with a zero mean, the expectation of $s_{\rm r}$ is
\begin{equation}
|s_{\rm {r,exp}}| = \\
\sigma_{\rm r} { \left[ \sqrt{2} \ln{\left( \frac{N_{\rm pix}}
{\sqrt{2\pi}\sigma_{\rm r}} \right) }\right]}^{1/2} ,
\end{equation}
where $N_{\rm pix}$ is the total number of pixels in the image. 
The quality of the residual noise is given by 
\begin{equation}
\xi_{\rm r} = s_{\rm r} / s_{\rm {r,exp}}.
\end{equation}
When the residual noise approaches Gaussian noise, 
$ \xi_{\rm r} \rightarrow 1 $. If $ \xi_{\rm r} > 1 $,
not all the structure has been adequately recovered; 
if $ \xi_{\rm r} < 1 $, the image model has
an excessively large number of degrees of freedom~\citep{Lobanov06}.
The values of $\xi_{\rm r} $ of the images in the survey 
are presented in Column (14) of Table~\ref{table:imageparameter}
and the distribution of them is shown in Figure~\ref{fig:Qual},
implying that the images adequately represent the structure 
detected in the visibility data.

\begin{figure}    
\begin{center} 
\includegraphics[width=1.00\columnwidth]{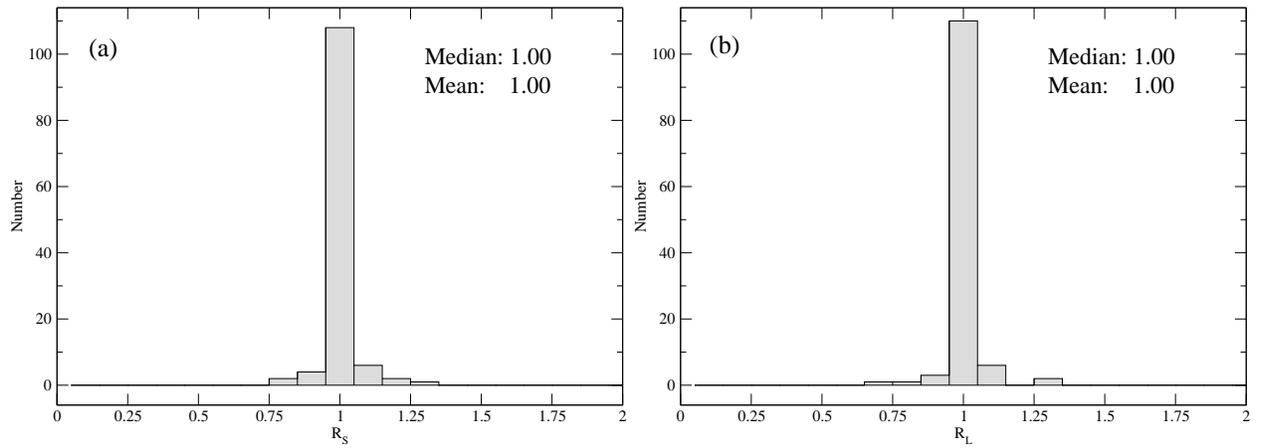}
\caption{Distributions of the correlated flux density ratios 
(a) $R_{\rm S} (= S_{\rm S}/S_{\rm S}^{\prime}) $ and 
(b) $R_{\rm L} (= S_{\rm L}/S_{\rm L}^{\prime})$. 
The means and medians for the distributions are presented in each panel.
The correlated flux densities $S_{\rm S,L}$ are listed in 
Table~\ref{table:imageparameter}.
\label{fig:R}
}
\end{center}
\end{figure}

The initial CLEAN cycles for the region around the core component
are conducted with using natural weighting 
and without using {\it uv}-tapering. Once the CLEAN models 
satisfactorily fit the visibility at the longest baselines, 
{\it uv}-tapering was applied to the data on long baselines
in order to recover faint emission 
further out from the core component.
We did not modify the visibility amplitudes, except for introducing 
an overall, time-constant gain correction factor 
wherever it was required for improving the agreement 
between the CLEAN model and the data.
In addition to the check for the antenna gain corrections 
described above,
we investigated the changes of the visibility amplitudes 
with and without introducing the time-constant gain correction factor, 
by using the correlated flux densities $S_{\rm S}$, $S_{\rm L}$, 
at the shortest and longest baselines
$B_{\rm S}$, $B_{\rm L}$, 
(presented in Columns 4--7 in Table~\ref{table:imageparameter}),
for each of the sources. 
The correlated flux densities $S_{\rm S,L}$ obtained 
after introducing the antenna gain corrections
are compared with the flux densities $S_{\rm S}^{\prime}$ and 
$S_{\rm L}^{\prime}$ before the gain corrections.
As shown in Figure~\ref{fig:R} for the distributions of the ratios 
$R_{\rm S} = S_{\rm S}/S_{\rm S}^{\prime}$ and 
$R_{\rm L} = S_{\rm L}/S_{\rm L}^{\prime}$,
the visibility amplitudes on the shortest and longest baselines for 
each of the sources
were not changed for most of the sources during the hybrid imaging. 
For a small number of peculiar sources,
the ratios of the visibility amplitudes fall within a range of 0.75--1.25.
This analysis shows again that the amplitude calibration error of 
this survey observations is 20\%--30\%.

We introduce a zero-baseline flux to recover 
a faint structure in the extended region 
by adding a fake visibility at 
the point of origin in the Fourier plane ({\it uv}-plane).
Since the shortest baseline of this survey observations
is about 50--100\,M$\lambda$, 
this may result in the faint structures of the extended regions 
appearing to be negative because the flux in every pixel 
of the map is offset by a small negative amount. 
The effect may be countered to some extent by adding the fake
visibility at the origin of the {\it uv}-plane.
The measured total flux density of each
source $S_{86}$ (listed Column 3 in Table~\ref{table:imageparameter}) 
is used as the zero-baseline flux.

\section{Estimating Parameters}

In order to extract quantitative information from the images, 
circular Gaussian-component models were used to fit the self-calibrated 
{\it uv}-data yielding the following parameters:
total and peak flux densities, positions, and sizes of each 
component. Uncertainties of the models were estimated,
based on
the signal-to-noise ratio (SNR) of detection of a given model 
fit component, using an analytical (first order)
approximation~\citep{Fomalont99}. 
The general fit parameters of a component in 
VLBI images of radio sources are {\it$S_{\rm tot}$} - 
total flux density, {\it $S_{\rm peak}$} - peak flux density, 
{\it $\sigma_{\rm rms}$} - post-fit rms, $d$ - size, 
$r$ - radial distance (for jet components), 
$\theta$ - position angle 
(measured for jet components, with respect to the location of 
the core component). The uncertainties of the fit parameters 
can be estimated by adopting approximations given by~\cite{Fomalont99}: 
\begin{eqnarray}
\label{un1}
\sigma_{\rm peak} = \sigma_{\rm rms}{\left({1+\frac{S_{\rm peak}}{\sigma_{\rm rms}}}\right)}^{1/2}, & 
\sigma_{\rm tot} = \sigma_{\rm peak}{\left({1+\frac{S_{\rm tot}^2}{S_{\rm peak}^2}}\right)}^{1/2},
\end{eqnarray}
\begin{eqnarray}
\label{un2}
\sigma_{d} = d\frac{\sigma_{\rm peak}}{S_{\rm peak}}, & 
\sigma_{r} = \frac{1}{2}\sigma_{d}, & 
\sigma_{\theta} ={\rm atan}\left(\frac{\sigma_{r}}{r}\right),
\end{eqnarray}
where $\sigma_{\rm peak}$, $\sigma_{\rm tot}$, $\sigma_{d}$, 
$\sigma_{r}$, and $\sigma_{\theta}$ are the uncertainties 
of total flux density, peak flux density, post-fit rms, size, 
and radial distance of a component, respectively. 
When the size, $d$, of a component was determined,
the resolution limits~\citep{Lob2005} should be taken into 
account. So, the minimum resolvable size of a component in 
an image is given by
\begin{equation}
d_{\rm min}=\frac{2^{1+\beta/2}}{\pi}{\left[{\pi ab\ln2\ln{\frac{SNR}{SNR-1}}}\right]}^{1/2},  \\
\end{equation}
where {\it a} and {\it b} are the axes of the restoring beam, 
{\it SNR} is the signal-to-noise ratio, and $\beta$ is the 
weighting function, which is 0 for natural weighting or 2 
for uniform weighting. When $d<d_{\rm min}$, the uncertainties 
should be estimated with $d=d_{\rm min}$.

We use the results of the model fitting to estimate brightness 
temperatures of the core and jet components.
The rest frame brightness temperature $T_{\rm b}$ of 
the emission region represented by a Gaussian component 
is
\begin{equation}
\label{tb1}
T_{\rm b} = \frac{2\ln{2}}{{\pi}{k}}\frac{S_{\rm tot}\lambda^2}{d^2}(1+z),
\end{equation}
where $\lambda$ is the wavelength of observation, $z$ is the redshift, 
and $k$ is the Boltzmann constant. Practically, the brightness 
temperature can be calculated by simplifying (\ref{tb1}):
\begin{equation}
\label{tb1-1}
T_{\rm b} =1.22\times 10^{12} \frac{S_{\rm tot}}{d^2\nu^2}(1+z) \,\,{\rm K},
\end{equation}	
where the total flux density $S_{\rm tot}$ is measured in Jy, 
the size of the circular Gaussian component $d$ in mas, and 
the observing frequency $\nu$ in GHz. If $d < d_{\rm min}$, 
then the lower limit of $T_{\rm b}$ is obtained with $d=d_{\rm min}$.

\section{Results}

Out of 127 sources, 109 sources have been imaged and  model fitted. 
The first 3\,mm-VLBI maps 
for 90 sources were made in this survey, increasing the number 
of sources ever imaged with 3\,mm-VLBI observations up to 110. 

In Figure~\ref{fig:images}, we present two plots and one contour map 
for each source at each epoch. 
In the left panel, the plot of the visibility amplitudes 
against {\it uv}-radius is shown. The corresponding {\it uv}-sampling
distribution is given in the inset.
The X-axis of the plot of the visibility amplitude represents 
the {\it uv}-radius which is the length of the baseline used to obtain
the corresponding visibility point. The {\it uv}-radius is given
in the units of $10^6\lambda$, where $\lambda$ is the observing wavelength. 
The Y-axis of the plot shows the amplitude 
of each visibility point (i.e., correlated flux density) in units of Jy.
The {\it uv}-sampling distribution in the inset of the left panel
describes the overall distribution of the visibility in the {\it uv}-plane,
whose maximum scale equals that of the {\it uv}-radius. In the right panel, 
the contour map of each source is shown, with the X- and Y-axis in the units 
of milliarcsecond. For each source, the source name and the observation data are given in 
the upper left corner of the map. The lowest contour level is identified
in the lower right corner of the map. The shaded ellipse represents the FWHM 
of the restoring beam in the image. In all of the images, the contours 
have a logarithmic spacing and they are 
drawn at -1, 1, 1.4,...,$1.4^n$ of the lowest flux density level. 
For 12 sources (0133+476, 0149+218, 
0212+735, 0234+285, 0238-084, 0316+413, 0355+508, 0415+379, 0430+052, 
0716+714, 1928+738, and 2255-282), multi-epoch images are presented. 
Most sources are centered on the brightest component (VLBI core),
but for some sources with a larger structure, we have shifted the center
to fit the image in the box. 

In Table~\ref{table:imageparameter}, parameters of the images presented in 
Figure~\ref{fig:images} are summarized. 
For each image, 
Table~\ref{table:imageparameter} lists the source name, the observing epoch, 
the total flux density, $S_{86}$, obtained from pointing and calibration 
measurements made at Pico Veleta during the observation, the correlated flux densities, 
$S_{\rm S,L}$, measured on the shortest and longest baselines, $B_{\rm S,L}$, 
the parameters of the restoring beam (the size of the major axis, $B_{\rm a}$ 
and the minor axis, $B_{\rm b}$, and the position angle of the beam, 
$B_{\rm PA}$), the total flux, $S_{\rm t}$, the peak flux density, 
$S_{\rm p}$, the off-source RMS, $\sigma$, and the quality of the residual 
noise in the image.

Table~\ref{table:modelfit} lists the parameters of each model-fit component:
the total flux, $S_{\rm tot}$, peak flux density, $S_{\rm peak}$, size, $d$,
radius, $r$ (only for jet components), position angle, $\theta$
(the location of the jet component with respect to the core component), 
and measured brightness temperature, $T_{\rm b}$. 
For sources with multiple components, parameters of the core component are 
followed by those of jet components. For sources observed at multiple epochs,
individual epochs are marked. 
The estimated uncertainties are given next to each parameter. The upper limits of size, $d$, 
and the lower limits of brightness temperature, $T_{\rm b}$, are in italic 
with brackets.

\section{Discussion}

\subsection{Source compactness}

For all imaged sources, we discuss the source compactness, showing
the distributions of the total flux density $S_{86}$, the CLEAN flux 
density $S_{\rm CLEAN}$, and the correlated flux densities $S_{\rm S,L}$ 
measured on the shortest and longest baselines, listed in 
Table~\ref{table:imageparameter}. In Figure~\ref{fig:com1}, we present 
the distributions of the flux densities and source 
compactness. The distribution of the total flux density $S_{86}$ 
({\it top left panel}) peaks at 1.3\,Jy, and shows that almost all sources 
are brighter than 0.3\,Jy, which corresponds to the flux limit of 
our source selection. The median value of the CLEAN flux density 
$S_{\rm CLEAN}$ ({\it middle left panel}) is 0.6\,Jy and the peak of 
the distribution is around 0.5\,Jy, indicating that much of the emission 
at 86\,GHz from the compact radio sources is resolved out at milliarcsecond 
scales. The source compactness on milliarcsecond scales 
$S_{\rm CLEAN}/S_{86}$ is also shown in Figure~\ref{fig:com1} 
({\it top right panel}). The median compactness on milliarcsecond 
scales of our sample is 0.51. 

While the median correlated flux density at the longest baseline 
$S_{\rm L}$ is 0.22\,Jy ({\it bottom left panel}), quite a few 
sources have considerable flux at long baselines 
(e.g., Pico Veleta and Kitt Peak). Among 95 sources whose correlated 
flux density can be measured at projected baselines longer than 2000\,M$\lambda$, 
82 sources have a correlated flux density 
greater than 0.1\,Jy. From the distribution of the source compactness on 
sub-milliarcsecond scales $S_{\rm L}/S_{\rm S}$ ({\it middle right panel}) 
we can see that most of the imaged sources are resolved. 
A few sources have a slightly greater flux density on the longest baseline 
than on the shortest baseline, since they are very compact and faint, 
giving a large scatter of visibility points on the long baselines. 
Although most of the imaged sources are resolved,
they are highly core-dominated in flux ({\it bottom right panel}).
Some of the extremely compact sources have the core dominance index 
$S_{\rm core}/S_{\rm CLEAN}$ larger than unity due to the uncertainty of 
the model fit and CLEAN flux.

The overall sample of imaged sources consist of 78 quasars, 22 BL Lac objects, and 
8 radio galaxies. Despite the significant difference in the number 
of sources between the optical classes, the dependence of 
sub-milliarcsecond scale compactness $S_{\rm L}/S_{\rm S}$ 
on the optical class is apparent in the distribution. 
Quasars and BL Lacs have similar distributions 
(the average is 0.54 for quasars and 0.48 for BL Lacs, 
and the median is 0.48 for quasars and 0.42 for BL Lacs),
and radio galaxies have a relatively different distribution
(the average is 0.38 and the median is 0.41).
The dependence is also evident in Figure~\ref{fig:com2}, which shows
the normalized mean visibility function in terms of {\it uv}-radius,
averaged for Quasars, BL Lacs, and radio galaxies.
The normalized mean visibility amplitudes for radio galaxies are, 
on average, lower than those for quasars and BL Lacs. At long 
{\it uv}-radii ranging from 700\,M$\lambda$ to 2500\,M$\lambda$, 
the amplitudes for the radio galaxies are quite distinct from 
those of the quasars and BL Lac objects in the sample.
Overall, the radio galaxies are less compact than the others, 
but BL Lacs and quasars are similar in compactness.
According to the unification paradigm of AGN \citep{UP95},
it is expected that quasars and BL Lacs on sub-milliarcsecond scales 
are still more compact than radio galaxies since the former are seen 
at smaller viewing angle and brightened by Doppler boosting. 
Our results from the 86\,GHz VLBI survey are consistent with this.

\subsection{Brightness temperature $T_{\rm b}$}

Figure~\ref{fig:C4-Tb1} shows the distributions of flux density 
and angular size for the core components. Most of the cores are 
smaller than 0.1\,mas in angular size. The cores of 77 sources 
are resolved and 32 sources have unresolved core components. 
Most of the unresolved 
sources are quasars (23), and a few sources are BL Lacs (7) and 
radio galaxies (2).

Figure~\ref{fig:C4-Tb2} shows the distributions of the measured 
core brightness temperatures in the source frame. The median value 
of these brightness temperatures is $7\times 10^{10}$\,K. The tail 
of the distribution extends up to $5\times 10^{12}$\,K. Only about 
1\,\% of the imaged sources yield brightness temperatures greater 
than $1.0 \times 10^{12}$\,K, which is the maximum value of the 
inverse Compton limit~\citep{Kellermann69}, and about 8\,\% have 
brightness temperatures higher than $3.0\times 10^{11}$\,K, which 
corresponds to the equipartition limit~\citep{Readhead94}. 
This distribution shows brightness temperatures lower by a factor 
of 10 than those derived from the VSOP survey 
at 5\,GHz~\citep[see][]{hor+04} and 
VLBA 2\,cm Survey~\citep[see][]{Kovalev2005}.
Higher brightness temperatures of compact radio sources can be 
explained by Doppler boosting, transient non-equilibrium events, 
coherent emission, emission by relativistic protons, or a combination 
of these effects~\citep[see][]{Kar00, Kellermann03}. 
Such a substantial decrease in the brightness temperatures measured 
at 86 GHz may be caused by two reasons. 
Most of the extragalactic radio sources may be resolved at 86\,GHz, 
as indicated by the compactness index derived in this paper. 
Alternatively, opacity and other physical conditions 
can change along the jet, causing observations at 86\,GHz 
to probe regions of the flow in which the brightness temperature 
is intrinsically lower~\citep[due to gradients in the physical 
conditions in the flows, see e.g.,][]{mar95}. 
Both these possibilities will be investigated in a follow-up paper.

\subsection{Intraday variable sources}

In order to identify intraday variable (IDV) sources in our sample, 
we used the list of IDV sources compiled in~\citet[][see references 
therein]{Kovalev2005}. Clear identifications can be made for most 
of the objects except 6 sources: 1044+719, 1150+497, 1842+681, 1923+210, 
2013+370, and 2023+336. From the references given in~\citet{Kovalev2005}, 
we identify them as ``non-IDV'' sources. In total, 26 sources 
are identified as ``IDV'' sources in our sample.

Figure~\ref{fig:IDV} shows the distributions of the correlated flux 
density at the longest baseline $S_{\rm L}$ ({\it top left panel}), 
and the core flux density $S_{\rm core}$ ({\it top right panel}),
as well as the distributions of the size $d_{\rm core}$ ({\it middle 
right panel}) and brightness temperature $T_{\rm b}$ ({\it bottom 
right panel}) of the cores. The source compactness on 
sub-milliarcsecond scales $S_{\rm L}/S_{\rm S}$ ({\it middle left panel})
and the core dominance $S_{\rm core}/S_{\rm CLEAN}$ ({\it bottom left panel })
are also compared for the IDV and non-IDV sources. 
The statistics of the distributions are summarized in Table~\ref{table:IDV}. 

The IDV and non-IDV sources have different mean values (0.36\,Jy and 0.31\,Jy)
median values (0.26 Jy and 0.21 Jy) of the correlated flux densities at 
the longest baselines, $S_{\rm L}$.
The Kolmogorov-Smirnov (K-S) test shows that there is 
a 17\% chance that the IDV and non-IDV samples are derived from 
a common distribution. This is a somewhat inconclusive result 
due to a few points at higher flux densities in the non-IDV sample,
which affects strongly the statistical results. If we exclude those outliers,
then the mean of the non-IDV sample gets smaller than that of the IDV sample
and the probability decreases to 15\%. However, it is difficult to 
conclude that IDV sources have a higher flux density $S_{\rm L}$ 
than non-IDV sources in our sample. 

The distributions of the compactness index $S_{\rm L}/S_{\rm S}$
for IDV and non-IDV sources have means of 0.53 and 0.51 with medians
of 0.46 and 0.44. The K-S test shows that a common parent distribution
for IDV and non-IDV sources is acceptable at a 100\% level.
In Figure~\ref{fig:radplotIDV}, it is shown that the sub-milliarcsecond
compactness $S_{\rm L}/S_{\rm S}$ for the IDV sources is, on average, 
similar to the non-IDV sources.

For the core dominance $S_{\rm core}/S_{\rm CLEAN}$,
the K-S test implies a single parent population 
for IDV and non-IDV sources
with the K-S probability at a 80\% level. 
We conclude therefore that IDV sources are similar to non-IDV sources
in core-dominance.

We find different results for the core parameters such as 
the core flux density, the core size, and the core brightness temperature
of IDV and non-IDV sources.
The distributions of the core flux density $S_{\rm core}$ for 
IDV and non-IDV sources have means of 0.70\,Jy and 0.62\,Jy, 
with medians of 0.42\,Jy and 0.46\,Jy.
They have a 34\% probability of being derived from a common population.
This is quite distinctive from the results for the other parameters.
The distributions of the core sizes $d_{\rm core}$ for IDV and non-IDV sources
have means of 0.039\,mas and 0.057\,mas with medians 
of 0.035\,mas and 0.043\,mas. The cores of IDV sources are smaller in angular size
than those of non-IDV sources. The K-S test also yields a probability 
of less than 4\% that the core size has the same parent distribution 
for IDV and non-IDV sources.
The mean values of the core brightness temperature for IDV and non-IDV sources 
are $10^{11.1\pm 0.1}$\,K and $10^{10.8\pm0.1}$\,K,
and the respective median values are $10^{11.1}$\,K and $10^{10.8}$\,K, respectively. 
A common parent population of the core brightness temperature 
for IDV and non-IDV sources is rejected at the 92\% level. 
This implies that, although IDV sources have similar core
flux densities to those of non-IDV sources, their brightness temperatures
are higher than those of the non-IDV sources due to the smaller angular core size.

\section{Summary}
\label{summary}

We have conducted the largest global 86\,GHz VLBI survey
of compact radio sources during three GMVA sessions in 
2001 October, 2002 April, and 2002 October.
Participation of sensitive European telescopes augmented by the VLBA antennas 
ensured high baseline and image sensitivities.
The total of 121 out of 127 sources observed have been detected at least on one baseline 
and 109 sources have been imaged with a typical 
dynamic range exceeding 50.
The survey observations have resulted in an increase by a factor of five 
of the total number of sources imaged at 86\,GHz with VLBI.

We have used two-dimensional, circular Gaussian components 
to fit the observed visibilities and 
parameterize the source structure. 
Using the results of these fits, the source compactness 
and brightness temperatures have been derived. 

We find that almost all of the survey objects are resolved 
and the cores of about 70\% of the imaged sources are resolved. 
Radio galaxies are less compact than quasars and BL Lacs.
BL Lacs are similar to quasars in the compactness at sub-milliarcsecond scales.

The distribution of the core brightness temperatures peaks at $\sim 10^{11}$\,K 
and only 1\% of the cores have brightness temperatures higher than $10^{12}$\,K. 
This shows apparently lower brightness temperatures than those derived from
other VLBI surveys at lower frequencies (e.g., 5\,GHz and 15\,GHz).

IDV sources in our sample are similar to non-IDV sources in
compactness at sub-milliarcseconds. The cores of IDV sources are 
smaller in angular size and so yield a higher brightness temperature 
than non-IDV sources, since the core flux densities of both samples 
are similar to each other.

\acknowledgments

We thank David Graham for his constant support for the mm-VLBI observation and 
correlation.
We gratefully thank the staff of the observatories participating in the GMVA;  
the MPIfR Effelsberg 100-m telescope, 
the IRAM Plateau de Bure Interferometer, 
the IRAM 30-m telescope, 
the Mets\"ahovi Radio Observatory, 
the Onsala Space Observatory, 
and the VLBA. 
IRAM is supported by INSU/CNRS (France), MPG (Germany) and IGN (Spain).
The VLBA is an instrument of the National Radio Astronomy Observatory, 
which is a facility of the National Science Foundation 
operated under cooperative agreement by Associated Universities, Inc.
This research has made use of the NASA/IPAC Extragalactic Database, 
which is operated by the Jet Propulsion Laboratory, California Institute of 
Technology, under contract with the National Aeronautics and Space 
Administration.
S.-S.L. would like to acknowledge support from Korea Science and Engineering 
Foundation under grant M06-2004-000-10009.

\clearpage
\clearpage
\clearpage
\clearpage
\clearpage
\clearpage
\begin{figure*}[p]    
\begin{center} \includegraphics[width=0.7\textwidth] {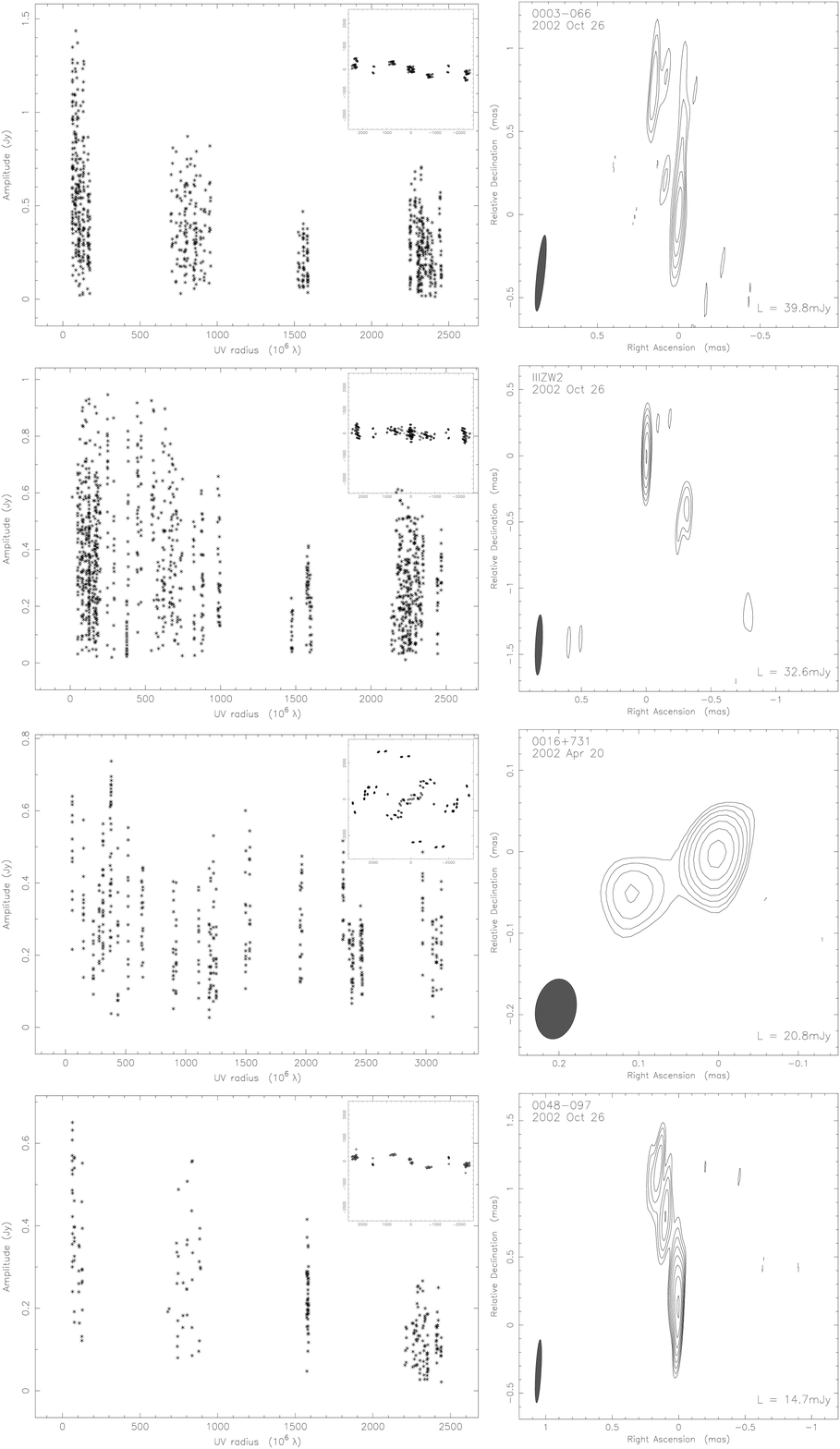}
\caption{123 contour maps of 109 sources with the distributions of
the {\it uv}-sampling and of the visibility amplitude against {\it uv}-radius. 
In the left panel, the X-axis represents the visibility amplitude (correlated flux density)
in Jy, averaged over 30 seconds, and the Y-axis shows the {\it uv}-distance in $10^6 \lambda$. 
The corresponding {\it uv}-sampling distribution is given in the inset.
In the right panel, a contour map of the CLEANed image is shown. 
The axes of the maps show the relative offset from the center of image in milliarcsecond.
Minimum contour level is shown in the lower-right corner of each map.
The contours have a logarithmic spacing and are drawn at 
-1, 1, 1.4,...,$1.4^n$ of the minimum contour level. 
Image parameters of each image are summarized in Table~\ref{table:imageparameter}.}
\label{fig:images}
\end{center}
\end{figure*}    

\clearpage
\setcounter{figure}{5}

\begin{figure*}[p]    
\begin{center} \includegraphics[width=0.7\textwidth] {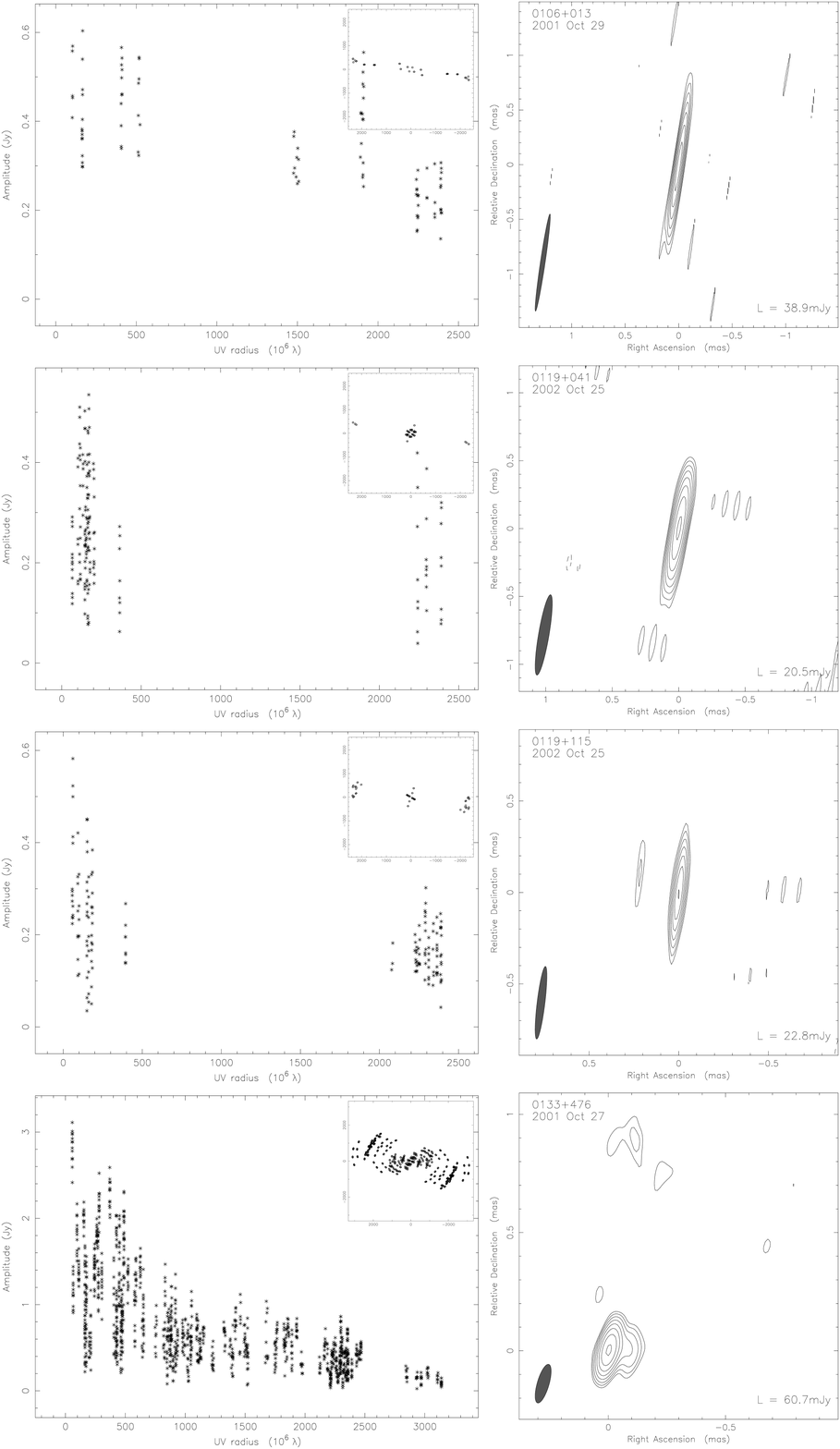}
\caption{{\it continued.}}
\end{center}
\end{figure*}    

\clearpage
\setcounter{figure}{5}

\begin{figure*}[p]    
\begin{center} \includegraphics[width=0.7\textwidth] {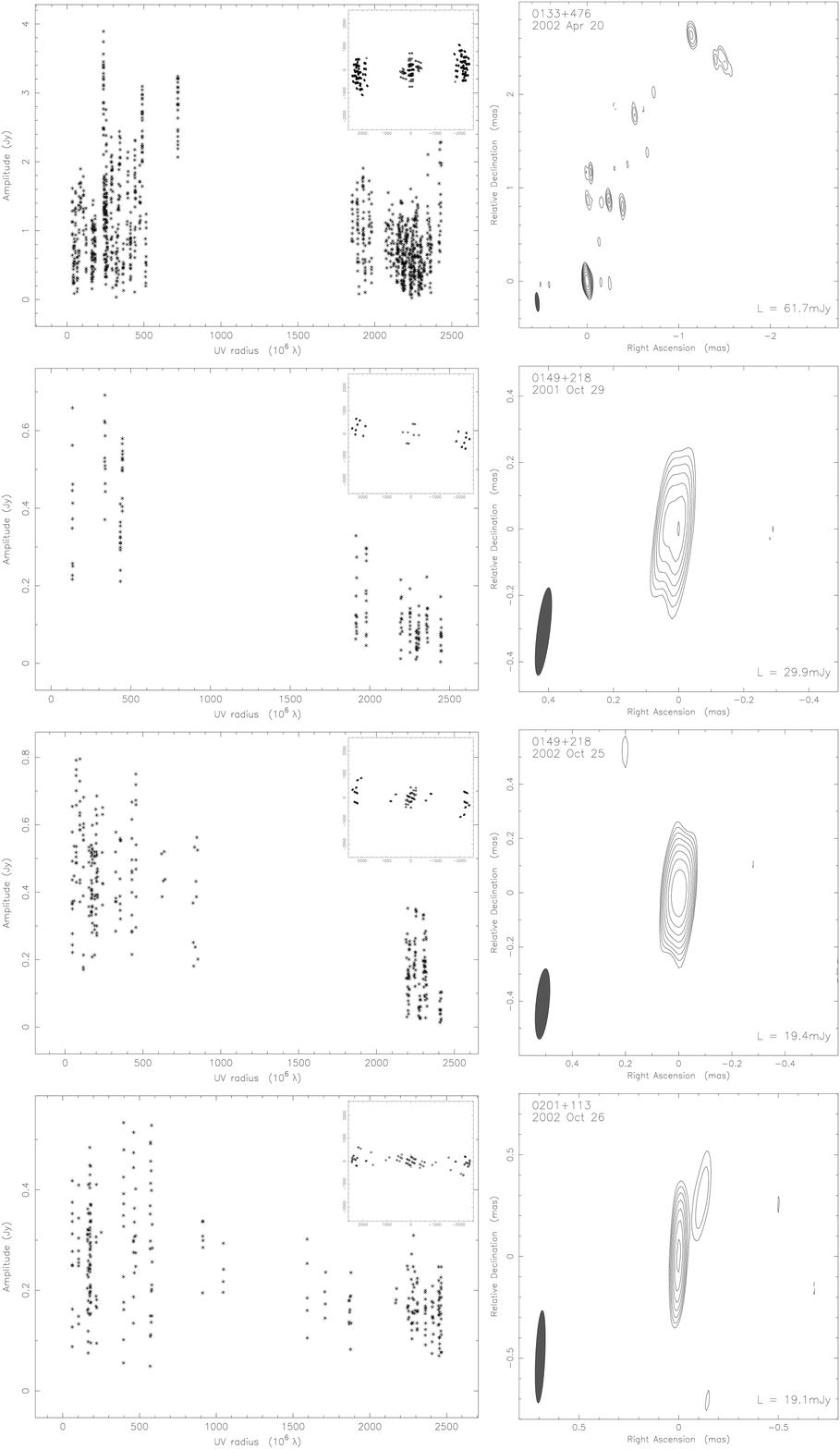}
\caption{{\it continued.}}
\end{center}
\end{figure*}    

\clearpage
\setcounter{figure}{5}

\begin{figure*}[p]    
\begin{center} \includegraphics[width=0.7\textwidth] {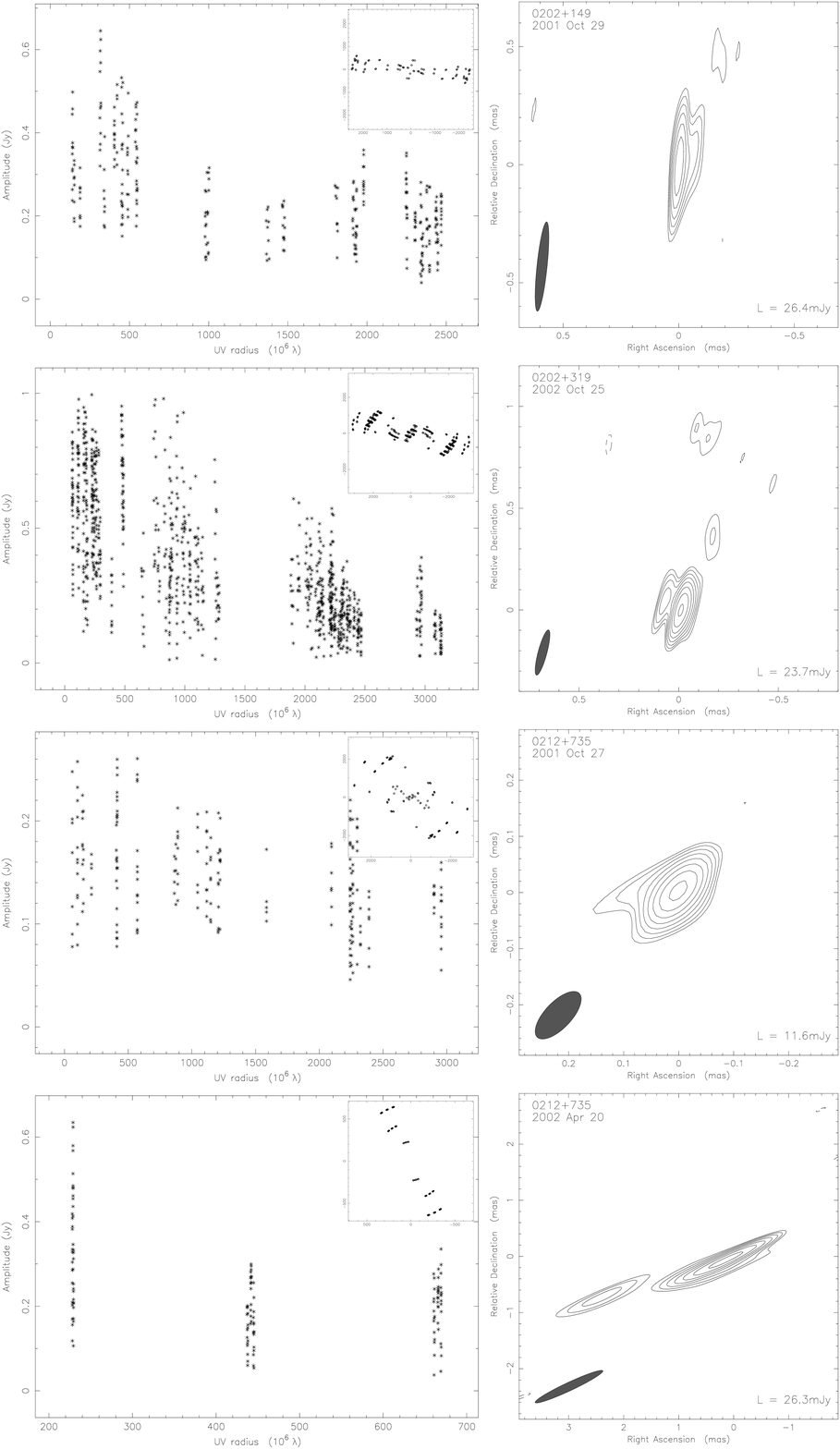}
\caption{{\it continued.}}
\end{center}
\end{figure*}    

\clearpage
\setcounter{figure}{5}

\begin{figure*}[p]    
\begin{center} \includegraphics[width=0.7\textwidth] {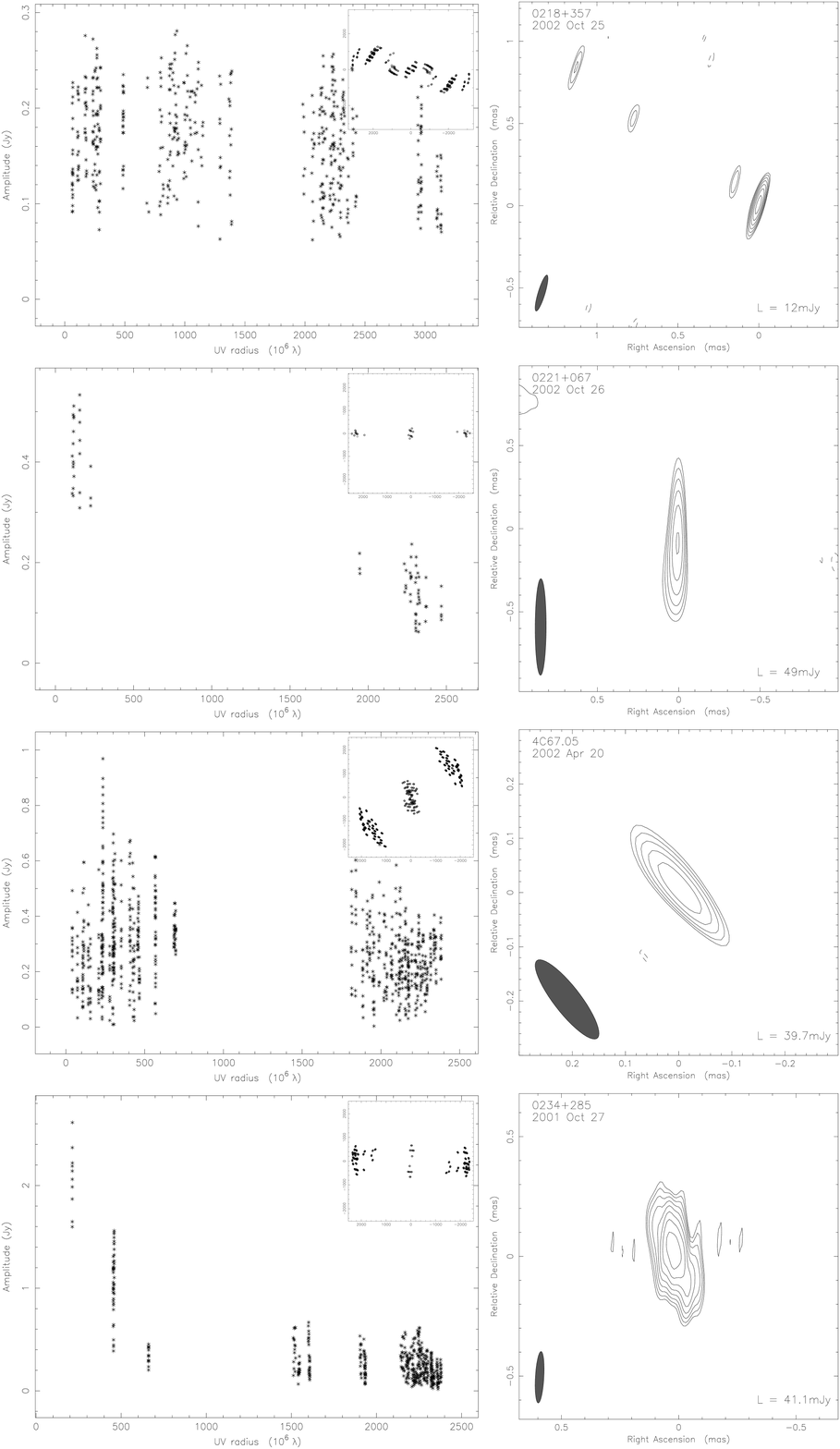}
\caption{{\it continued.}}
\end{center}
\end{figure*}    

\clearpage
\setcounter{figure}{5}

\begin{figure*}[p]    
\begin{center} \includegraphics[width=0.7\textwidth] {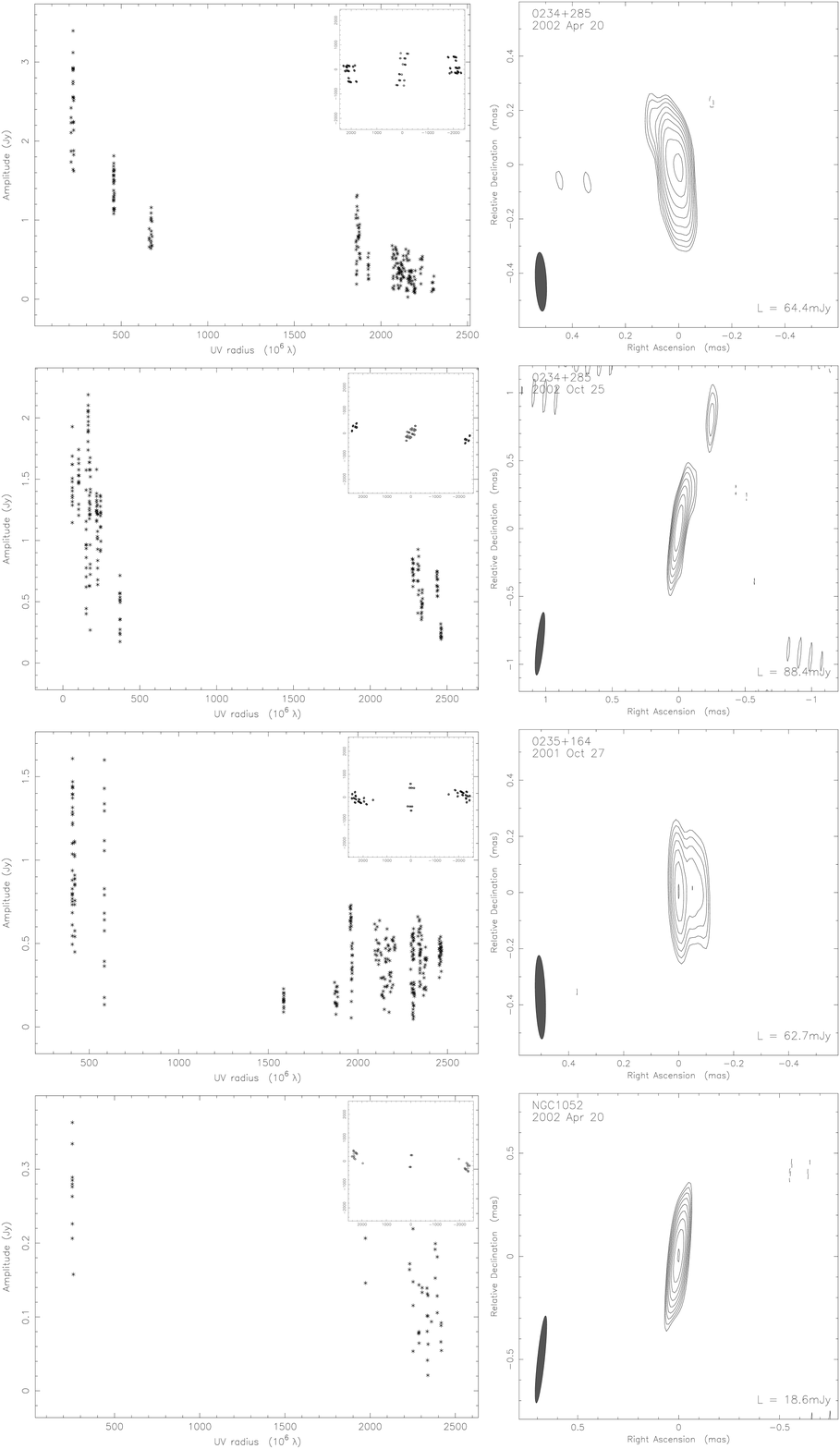}
\caption{{\it continued.}}
\end{center}
\end{figure*}    

\clearpage
\setcounter{figure}{5}

\begin{figure*}[p]    
\begin{center} \includegraphics[width=0.7\textwidth] {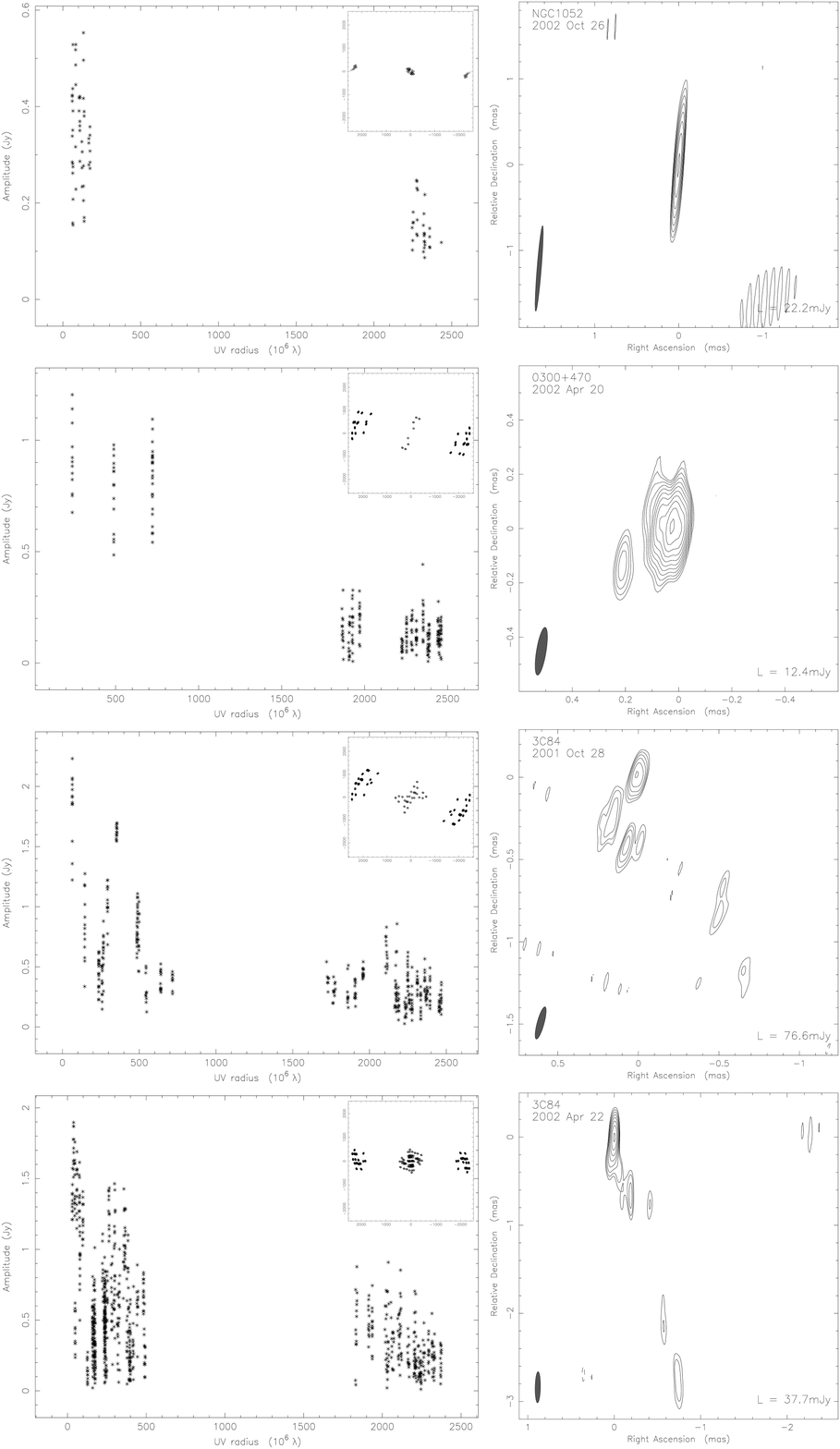}
\caption{{\it continued.}}
\end{center}
\end{figure*}    

\clearpage
\setcounter{figure}{5}

\begin{figure*}[p]    
\begin{center} \includegraphics[width=0.7\textwidth] {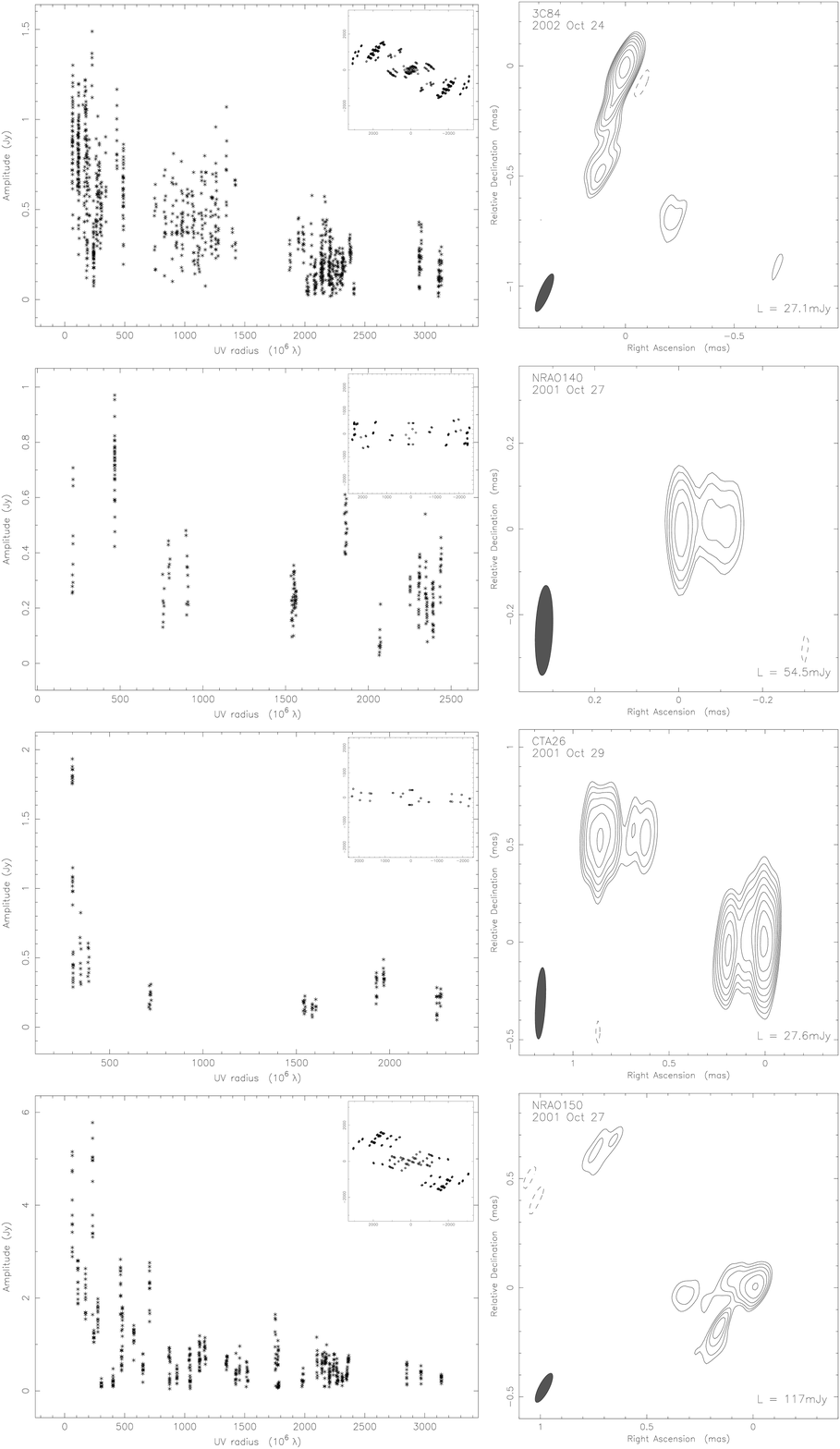}
\caption{{\it continued.}}
\end{center}
\end{figure*}    

\clearpage
\setcounter{figure}{5}

\begin{figure*}[p]    
\begin{center} \includegraphics[width=0.7\textwidth] {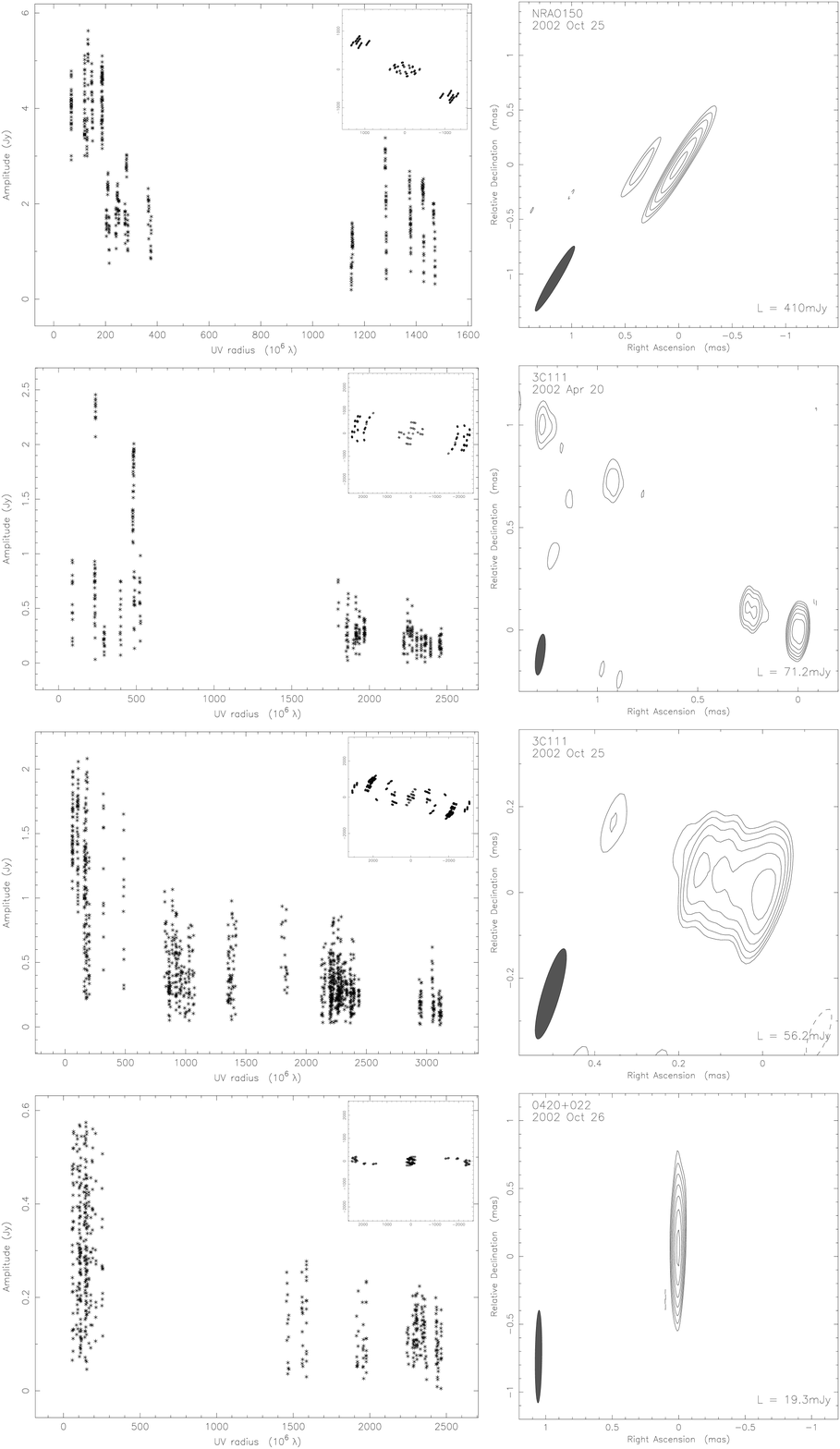}
\caption{{\it continued.}}
\end{center}
\end{figure*}    

\clearpage
\setcounter{figure}{5}

\begin{figure*}[p]    
\begin{center} \includegraphics[width=0.7\textwidth] {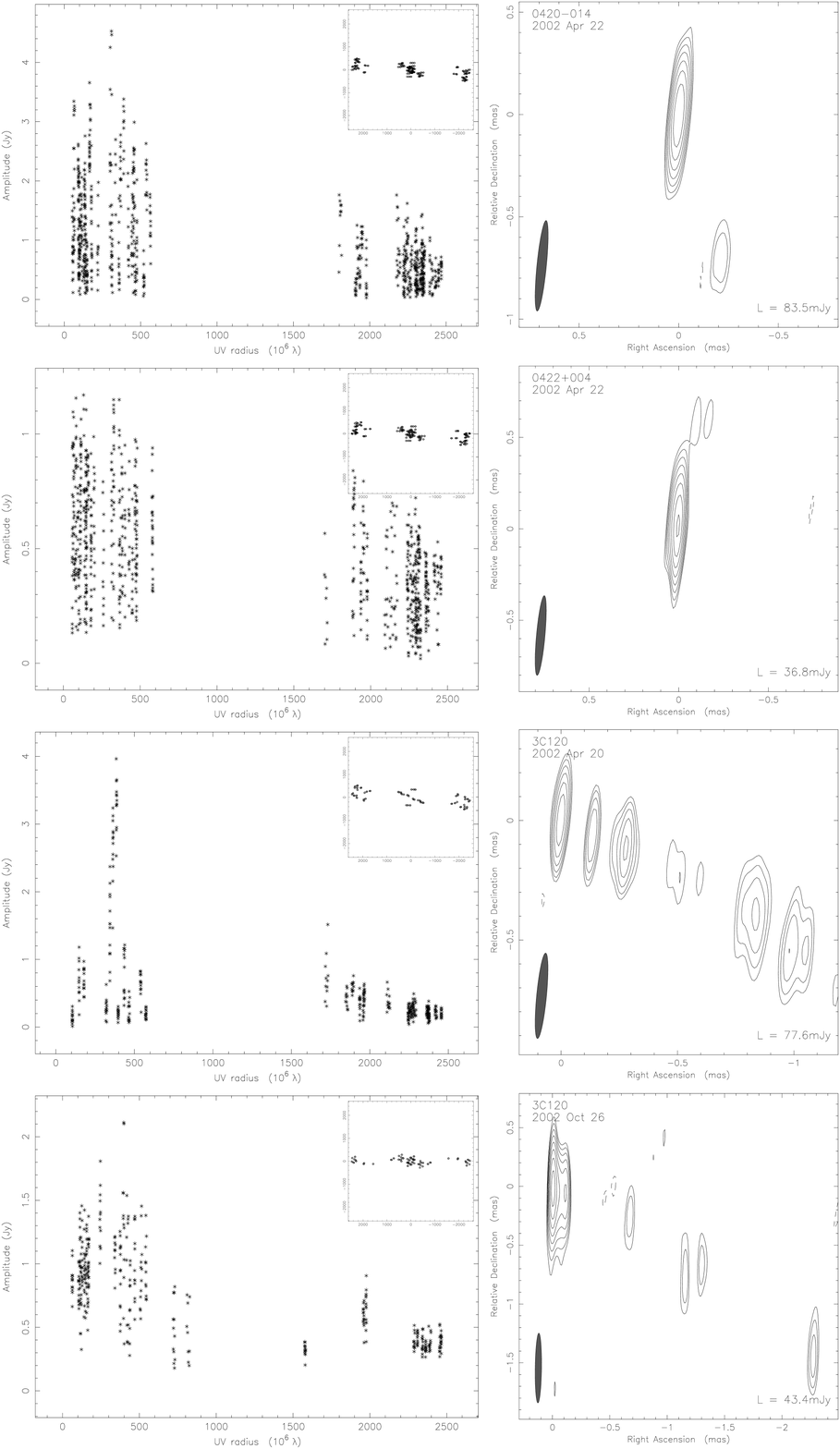}
\caption{{\it continued.}}
\end{center}
\end{figure*}    

\clearpage
\setcounter{figure}{5}

\begin{figure*}[p]    
\begin{center} \includegraphics[width=0.7\textwidth] {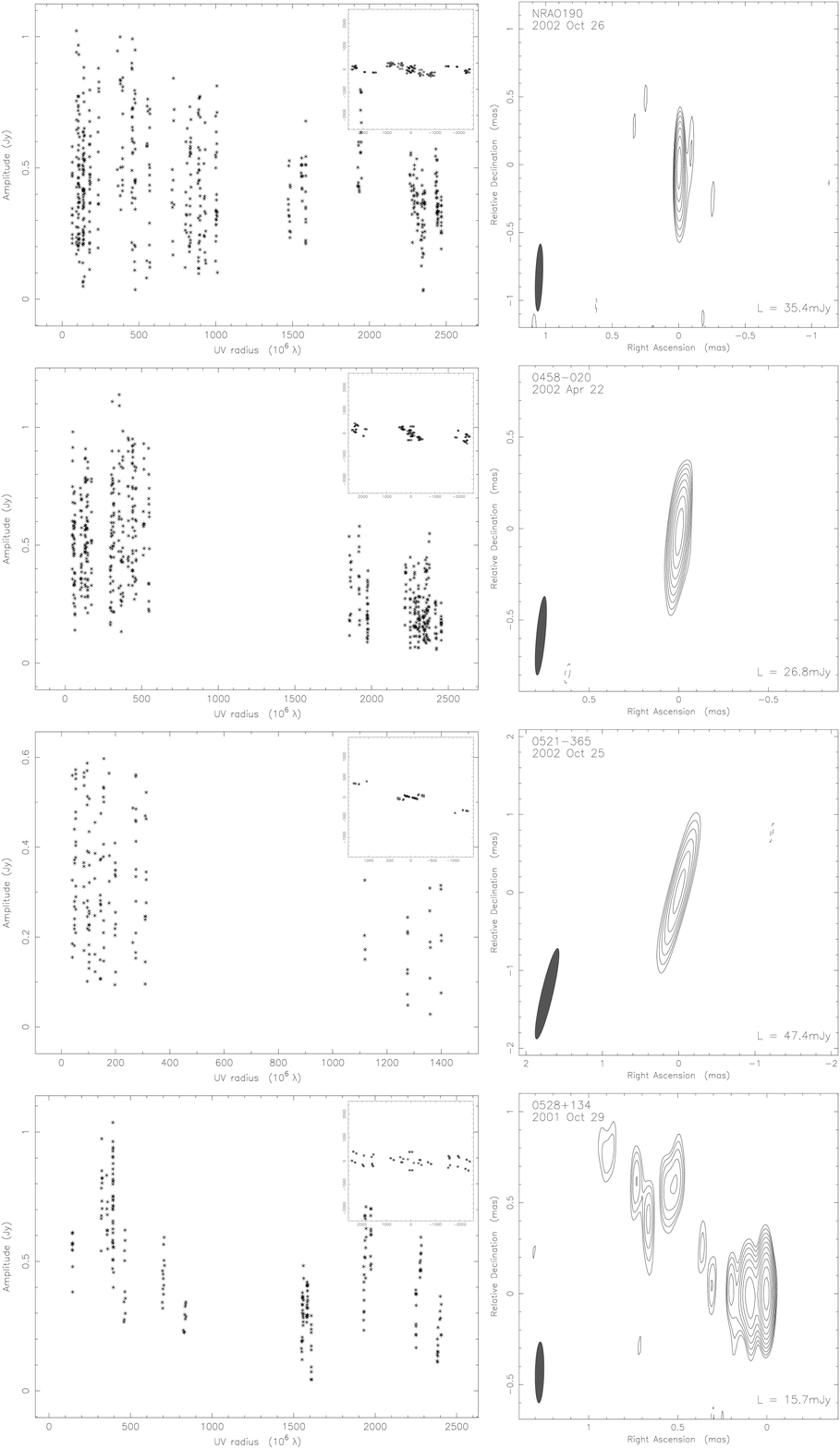}
\caption{{\it continued.}}
\end{center}
\end{figure*}    

\clearpage
\setcounter{figure}{5}

\begin{figure*}[p]    
\begin{center} \includegraphics[width=0.7\textwidth] {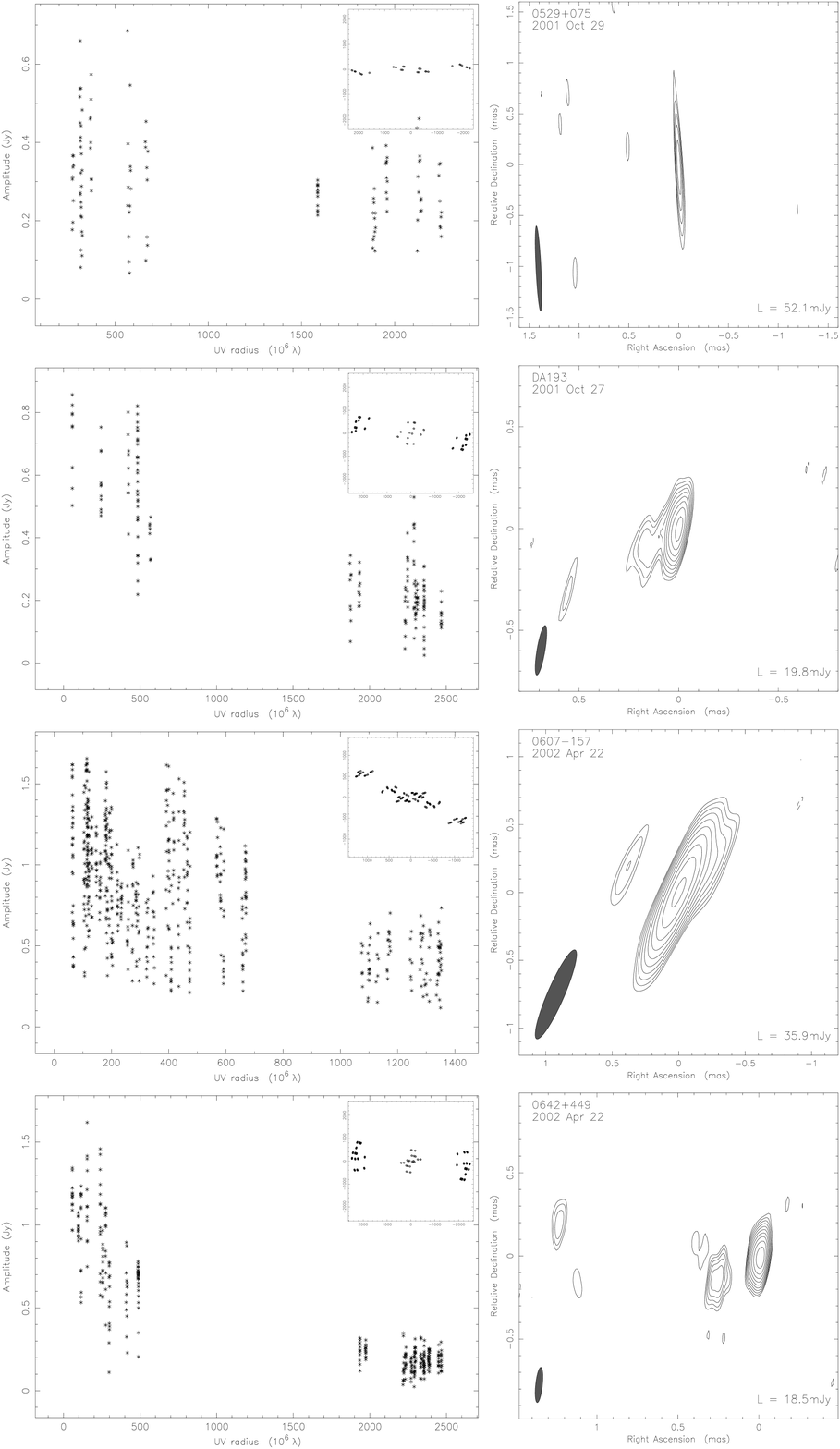}
\caption{{\it continued.}}
\end{center}
\end{figure*}    

\clearpage
\setcounter{figure}{5}

\begin{figure*}[p]    
\begin{center} \includegraphics[width=0.7\textwidth] {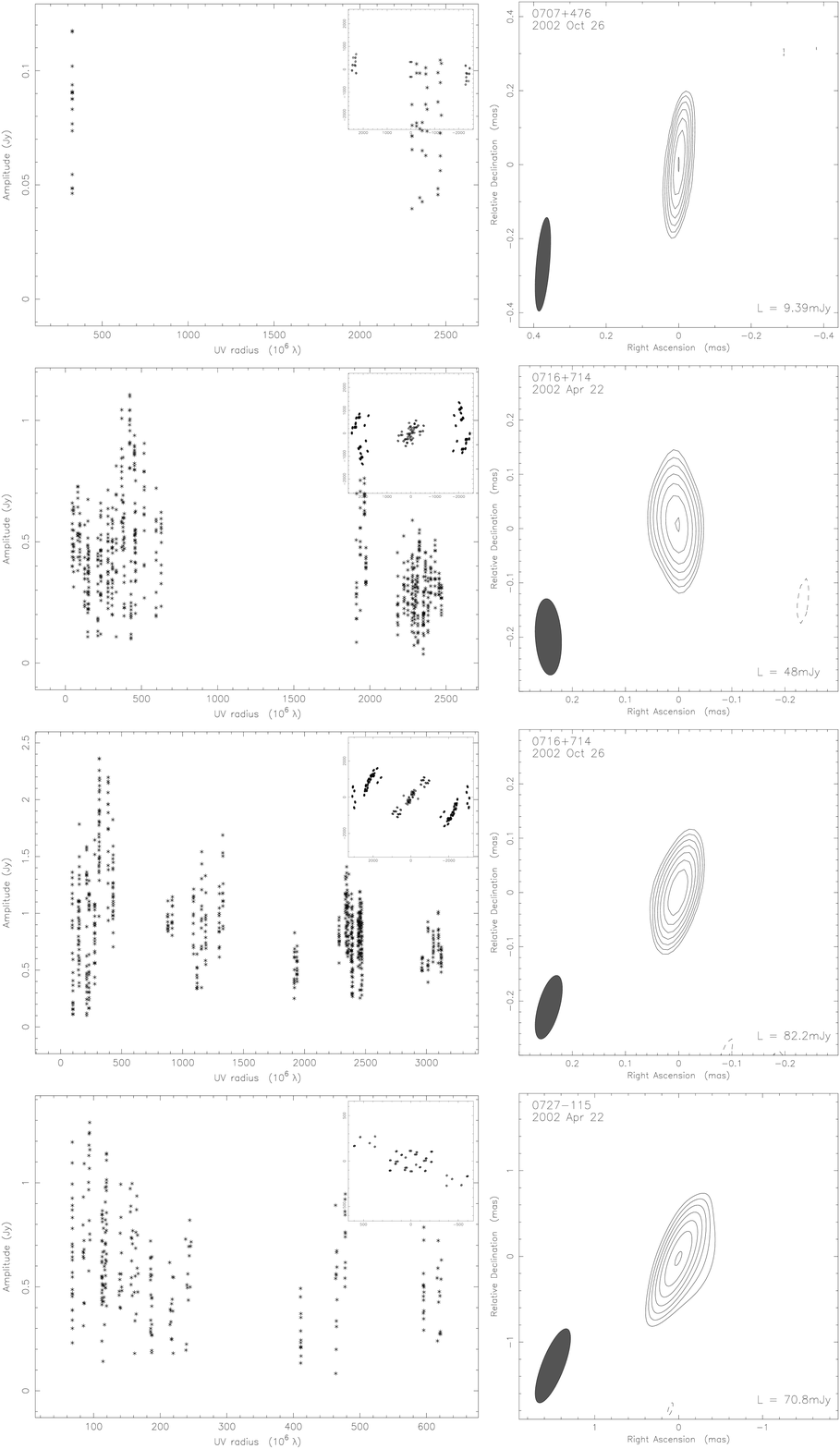}
\caption{{\it continued.}}
\end{center}
\end{figure*}    

\clearpage
\setcounter{figure}{5}

\begin{figure*}[p]    
\begin{center} \includegraphics[width=0.7\textwidth] {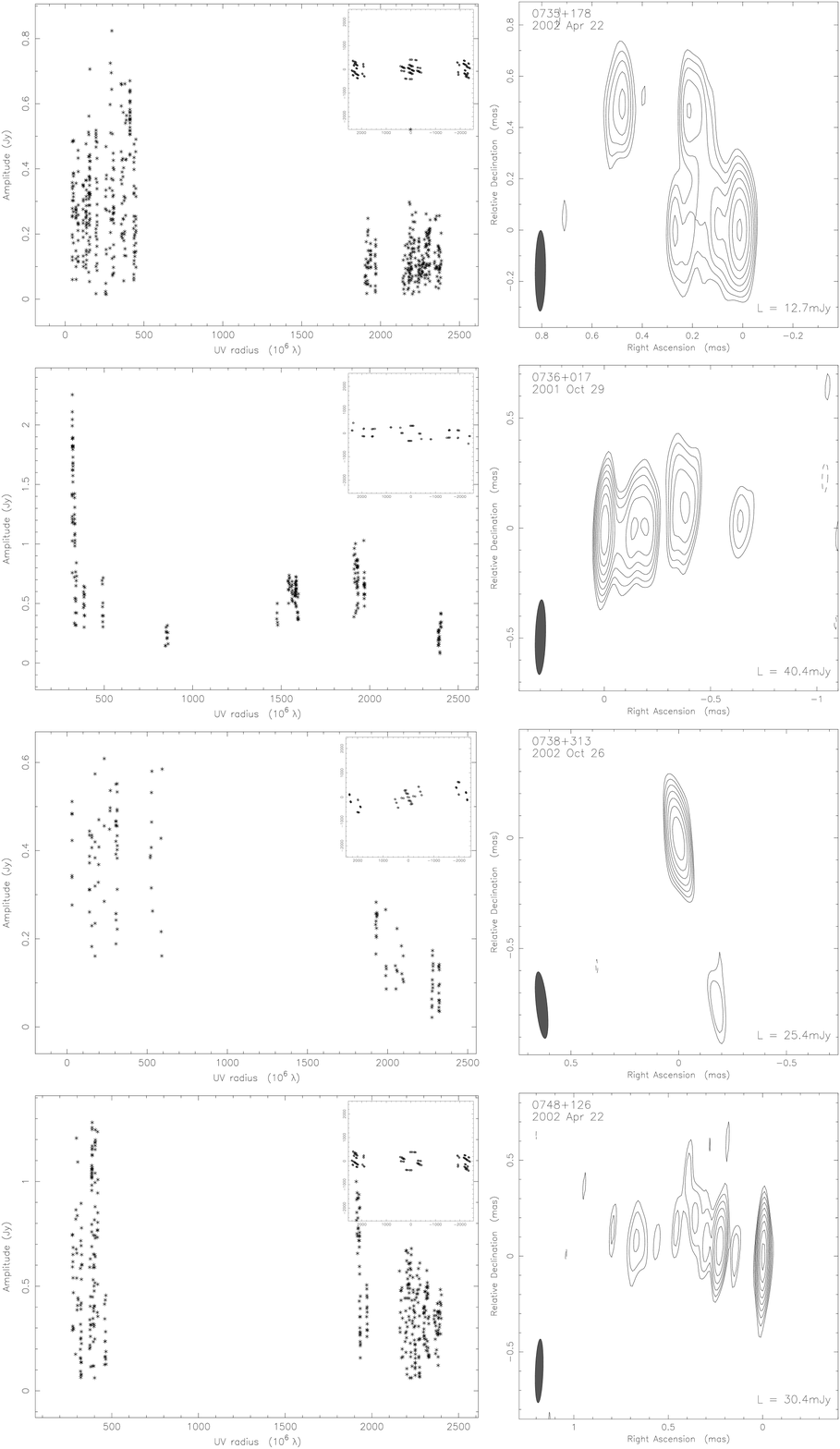}
\caption{{\it continued.}}
\end{center}
\end{figure*}    

\clearpage
\setcounter{figure}{5}

\begin{figure*}[p]    
\begin{center} \includegraphics[width=0.7\textwidth] {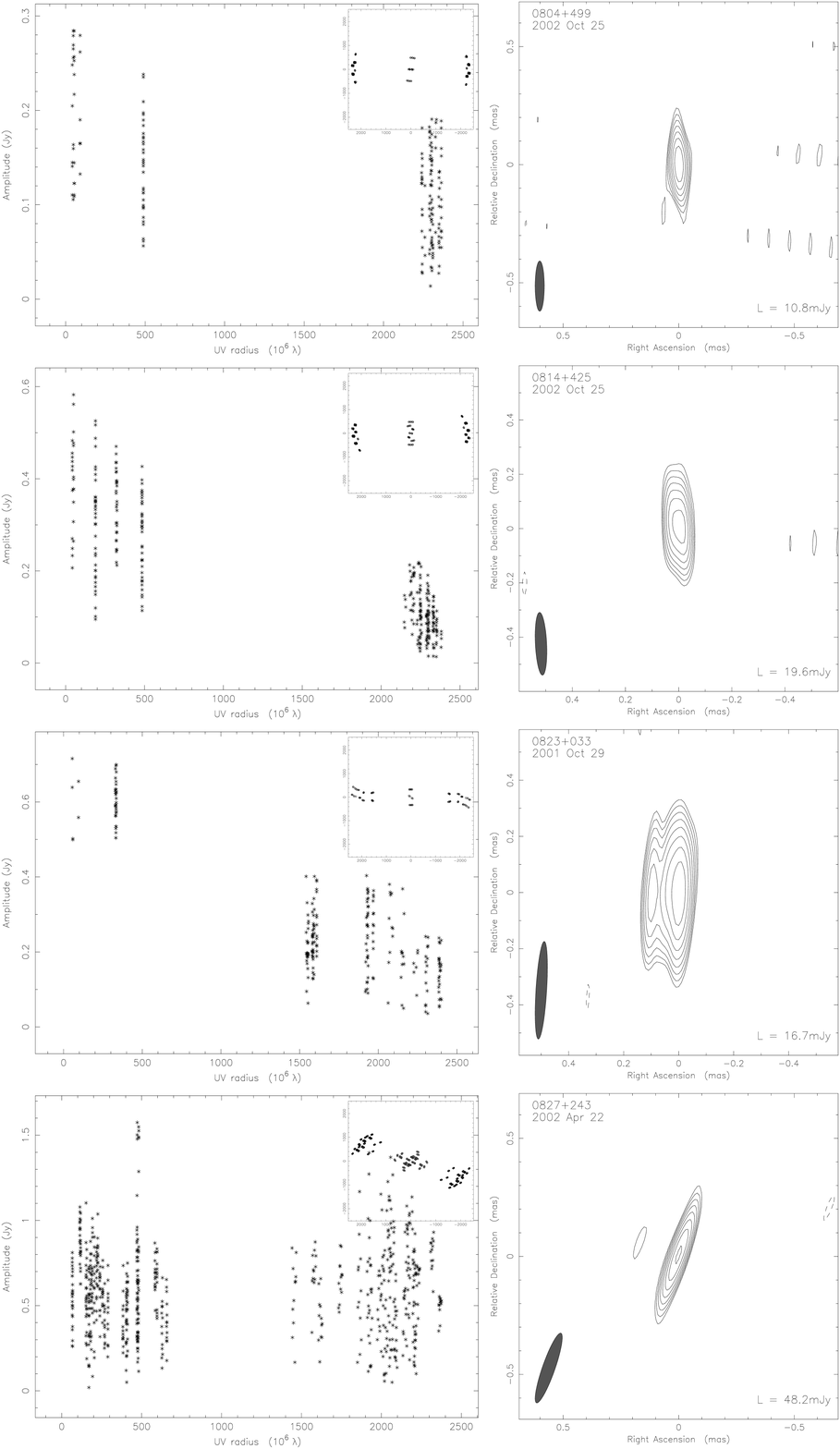}
\caption{{\it continued.}}
\end{center}
\end{figure*}    

\clearpage
\setcounter{figure}{5}

\begin{figure*}[p]    
\begin{center} \includegraphics[width=0.7\textwidth] {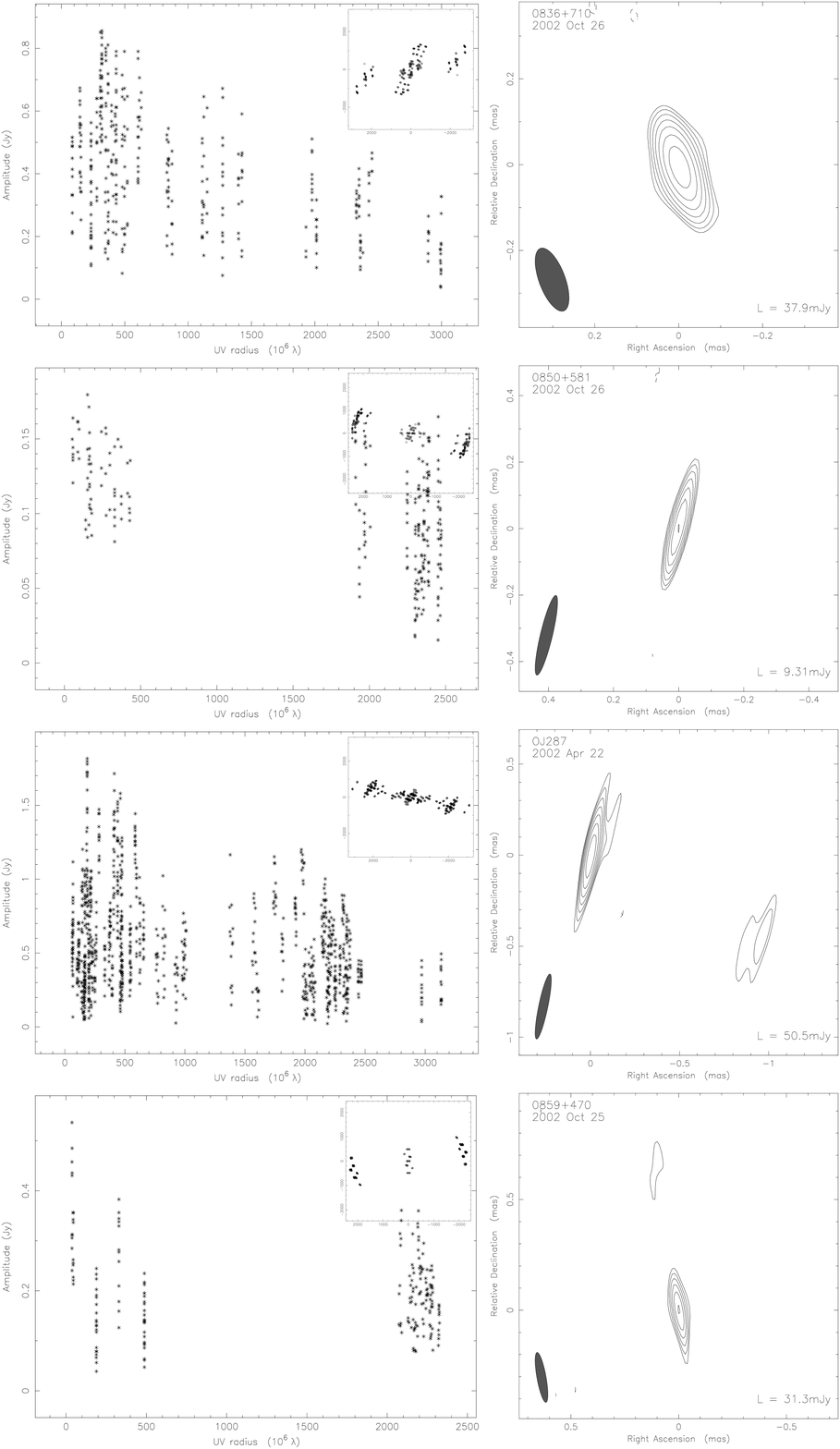}
\caption{{\it continued.}}
\end{center}
\end{figure*}    

\clearpage
\setcounter{figure}{5}

\begin{figure*}[p]    
\begin{center} \includegraphics[width=0.7\textwidth] {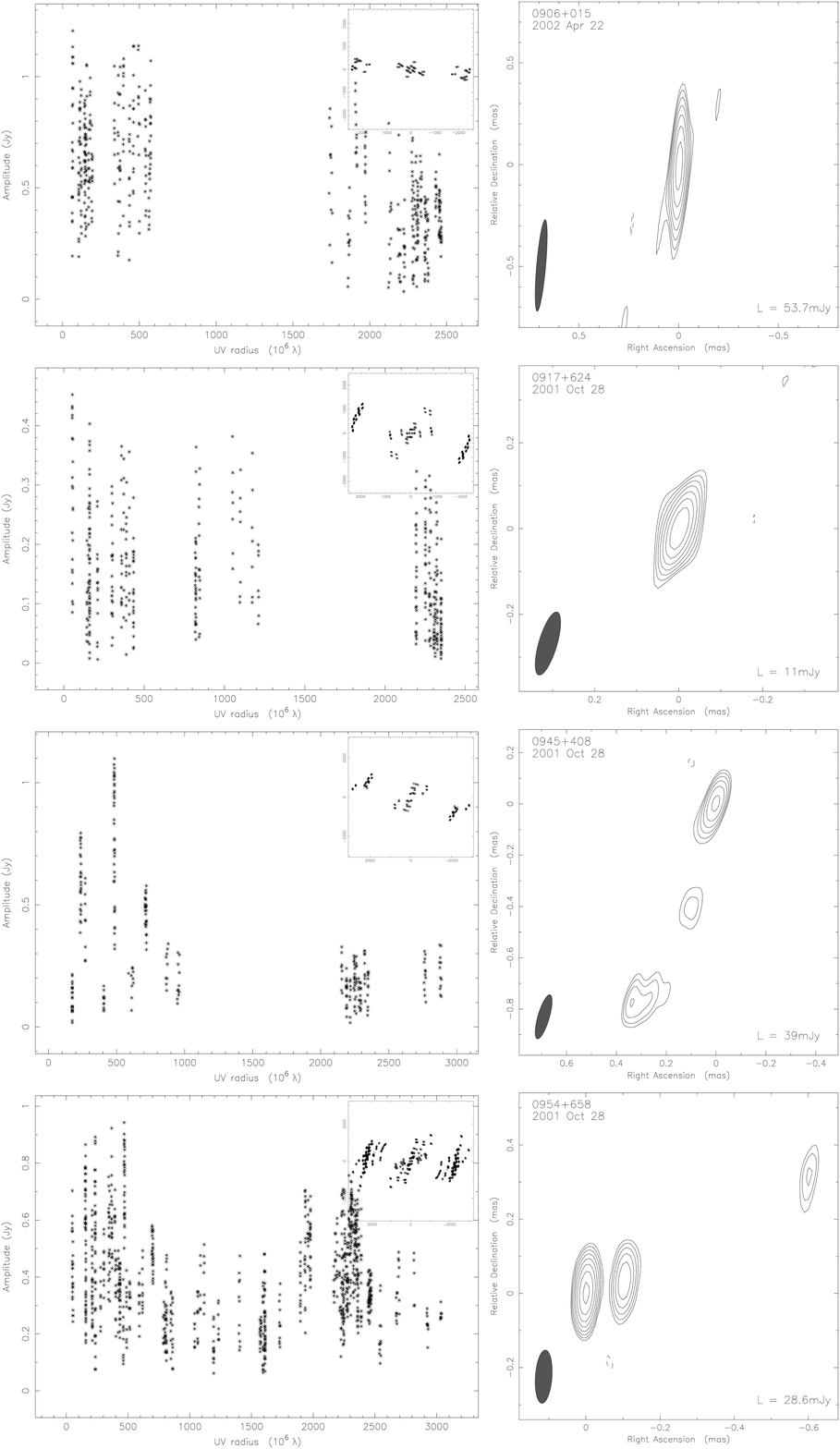}
\caption{{\it continued.}}
\end{center}
\end{figure*}    

\clearpage
\setcounter{figure}{5}

\begin{figure*}[p]    
\begin{center} \includegraphics[width=0.7\textwidth] {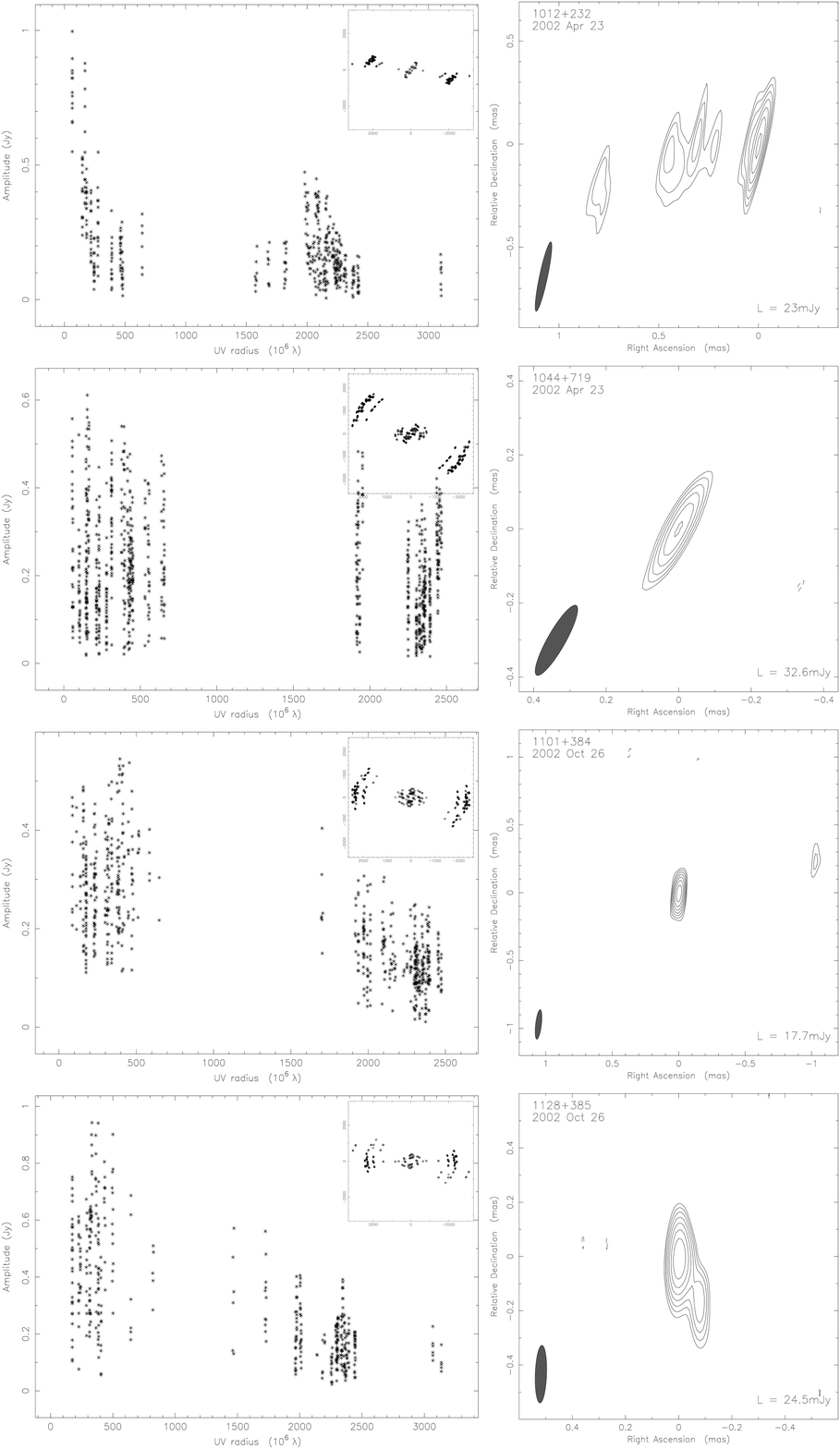}
\caption{{\it continued.}}
\end{center}
\end{figure*}    

\clearpage
\setcounter{figure}{5}

\begin{figure*}[p]    
\begin{center} \includegraphics[width=0.7\textwidth] {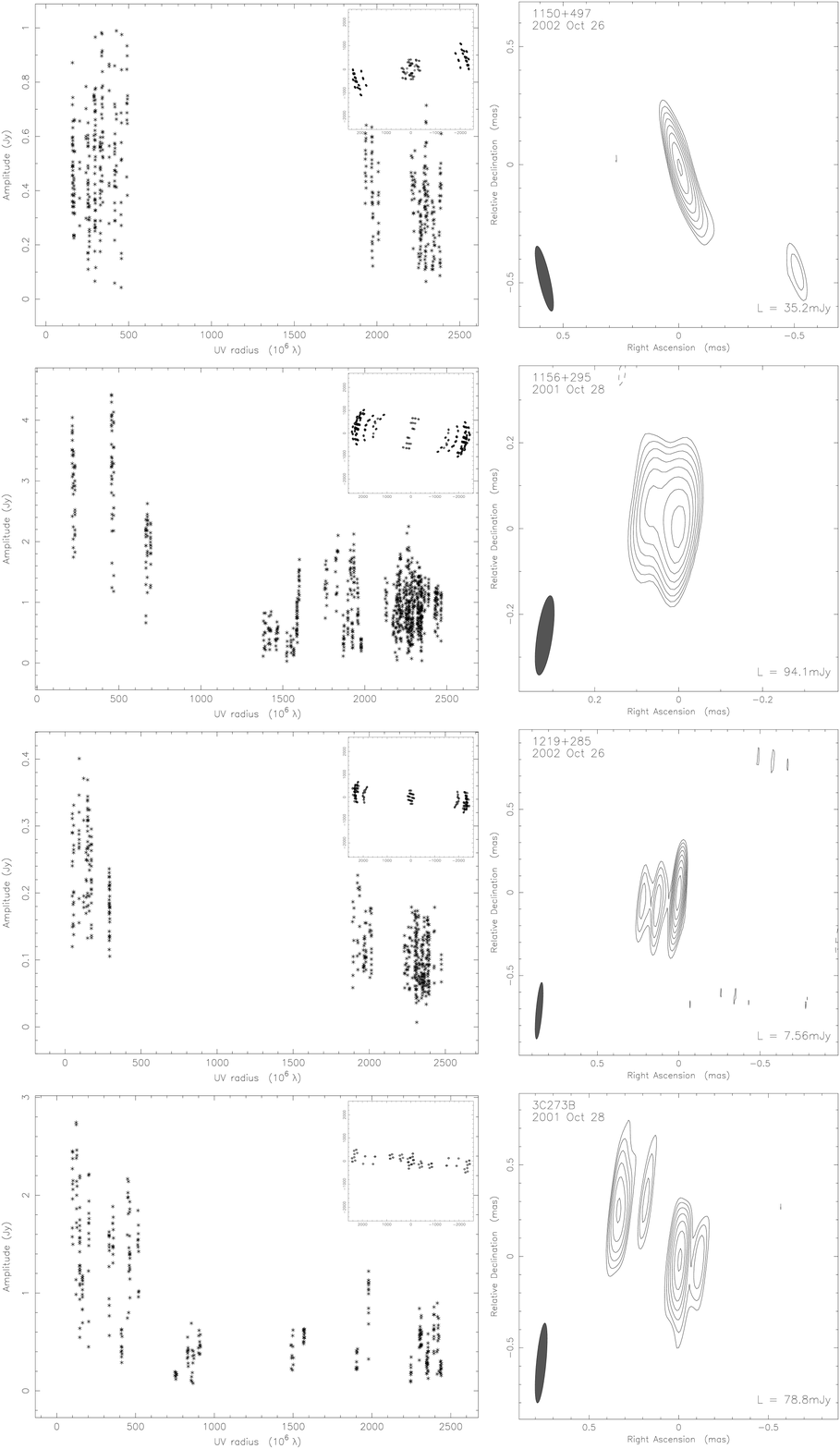}
\caption{{\it continued.}}
\end{center}
\end{figure*}    

\clearpage
\setcounter{figure}{5}

\begin{figure*}[p]    
\begin{center} \includegraphics[width=0.7\textwidth] {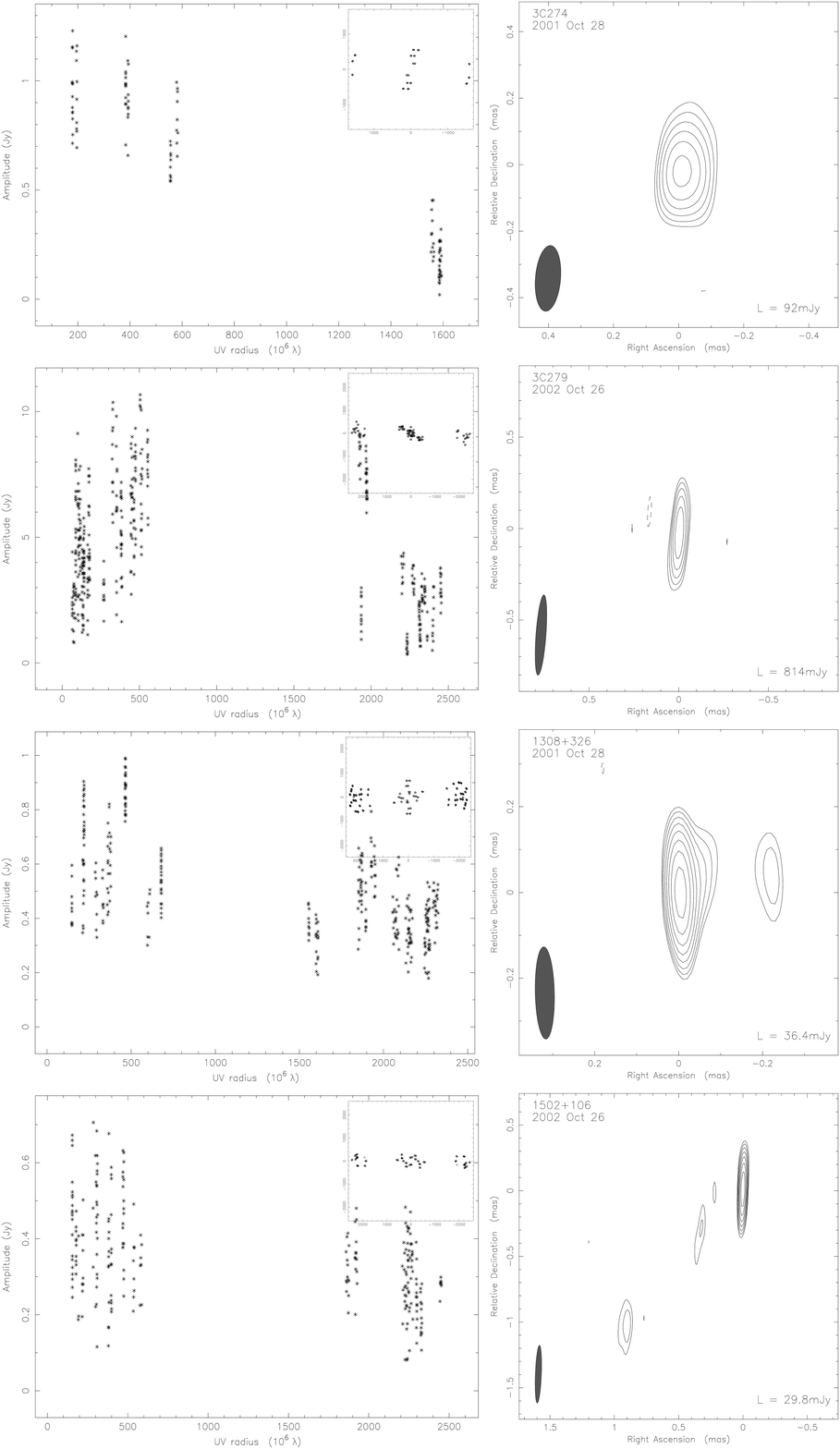}
\caption{{\it continued.}}
\end{center}
\end{figure*}    

\clearpage
\setcounter{figure}{5}

\begin{figure*}[p]    
\begin{center} \includegraphics[width=0.7\textwidth] {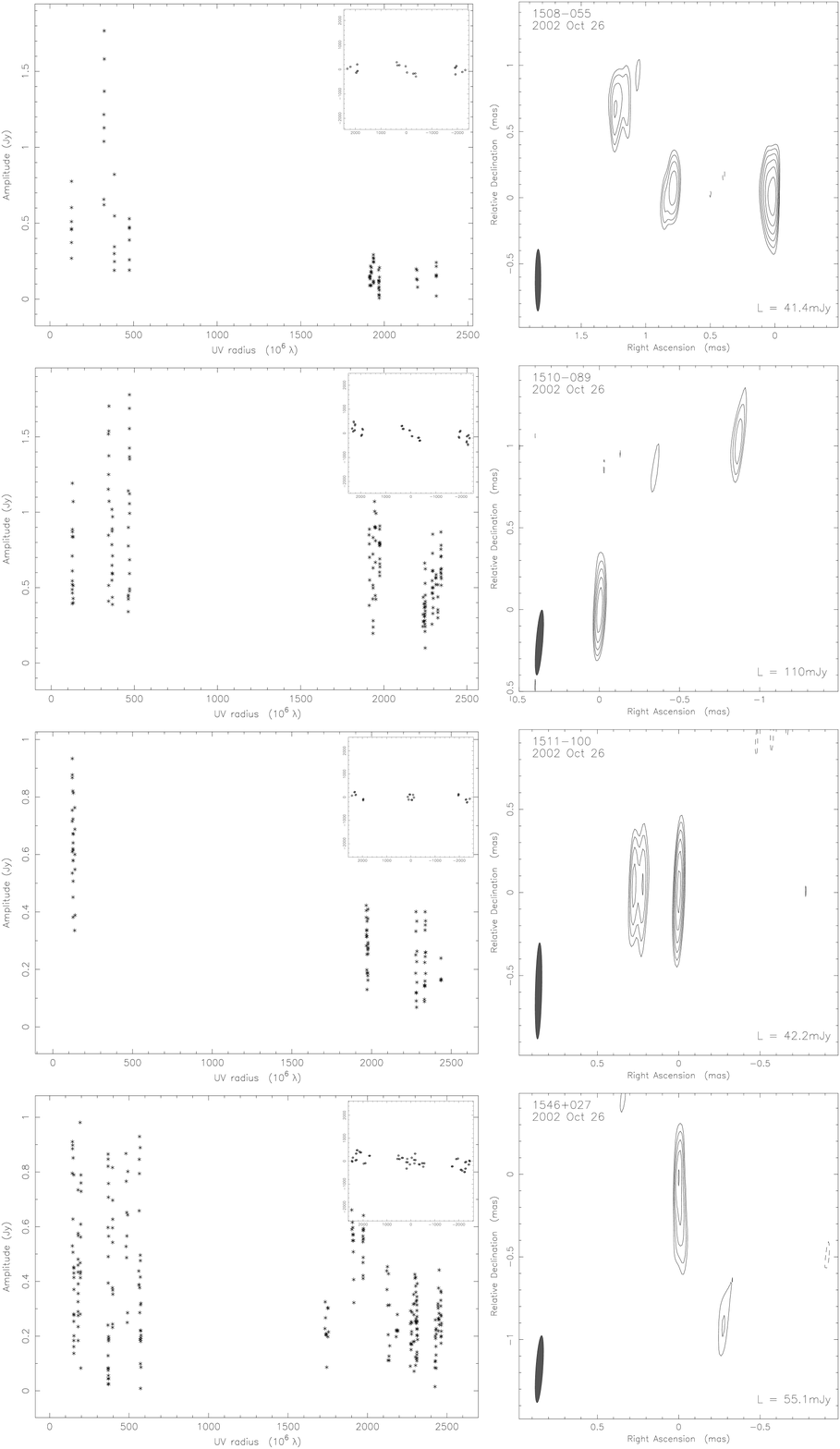}
\caption{{\it continued.}}
\end{center}
\end{figure*}    

\clearpage
\setcounter{figure}{5}

\begin{figure*}[p]    
\begin{center} \includegraphics[width=0.7\textwidth] {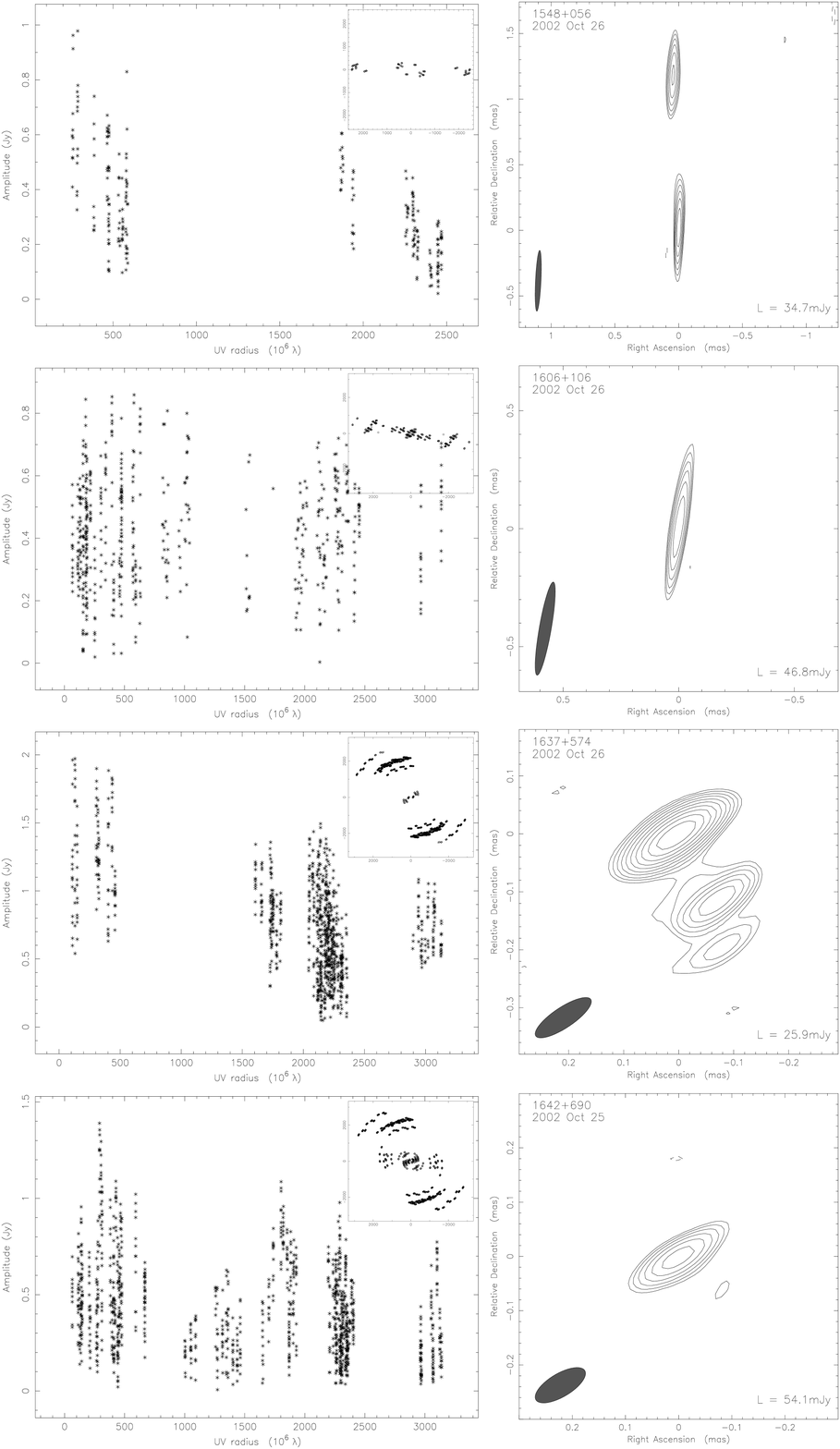}
\caption{{\it continued.}}
\end{center}
\end{figure*}    

\clearpage
\setcounter{figure}{5}

\begin{figure*}[p]    
\begin{center} \includegraphics[width=0.7\textwidth] {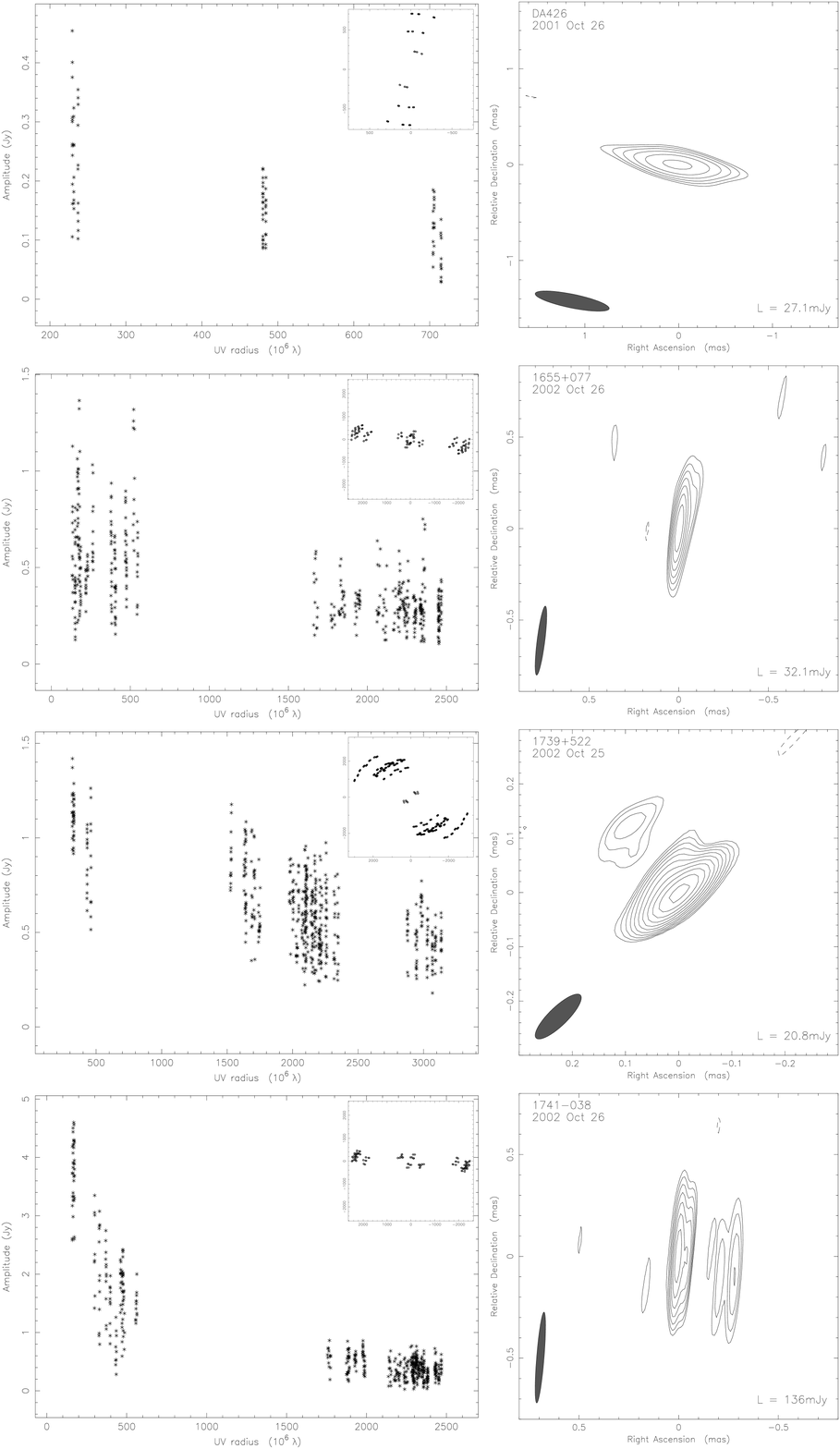}
\caption{{\it continued.}}
\end{center}
\end{figure*}    

\clearpage
\setcounter{figure}{5}

\begin{figure*}[p]    
\begin{center} \includegraphics[width=0.7\textwidth] {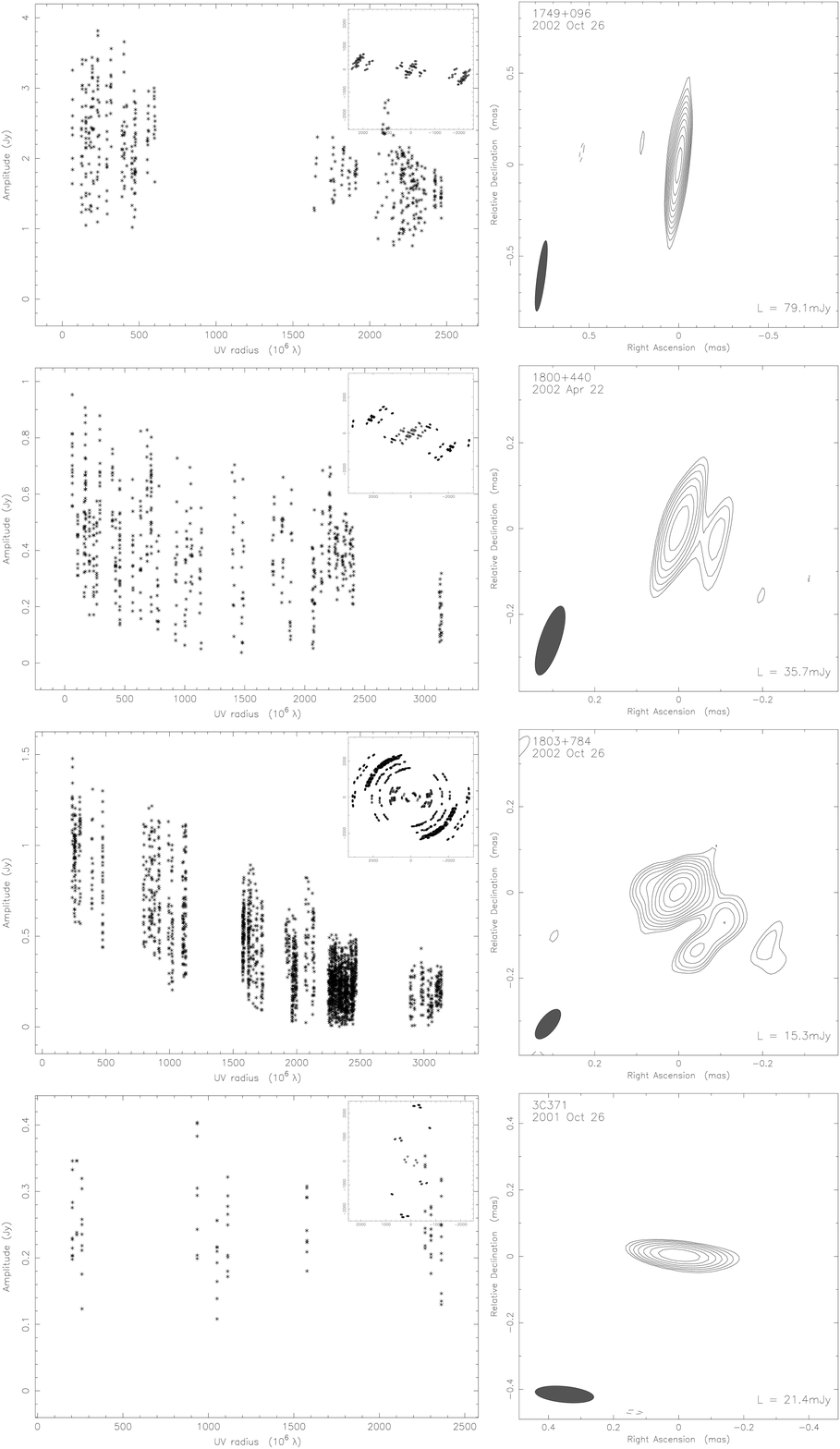}
\caption{{\it continued.}}
\end{center}
\end{figure*}    

\clearpage
\setcounter{figure}{5}

\begin{figure*}[p]    
\begin{center} \includegraphics[width=0.7\textwidth] {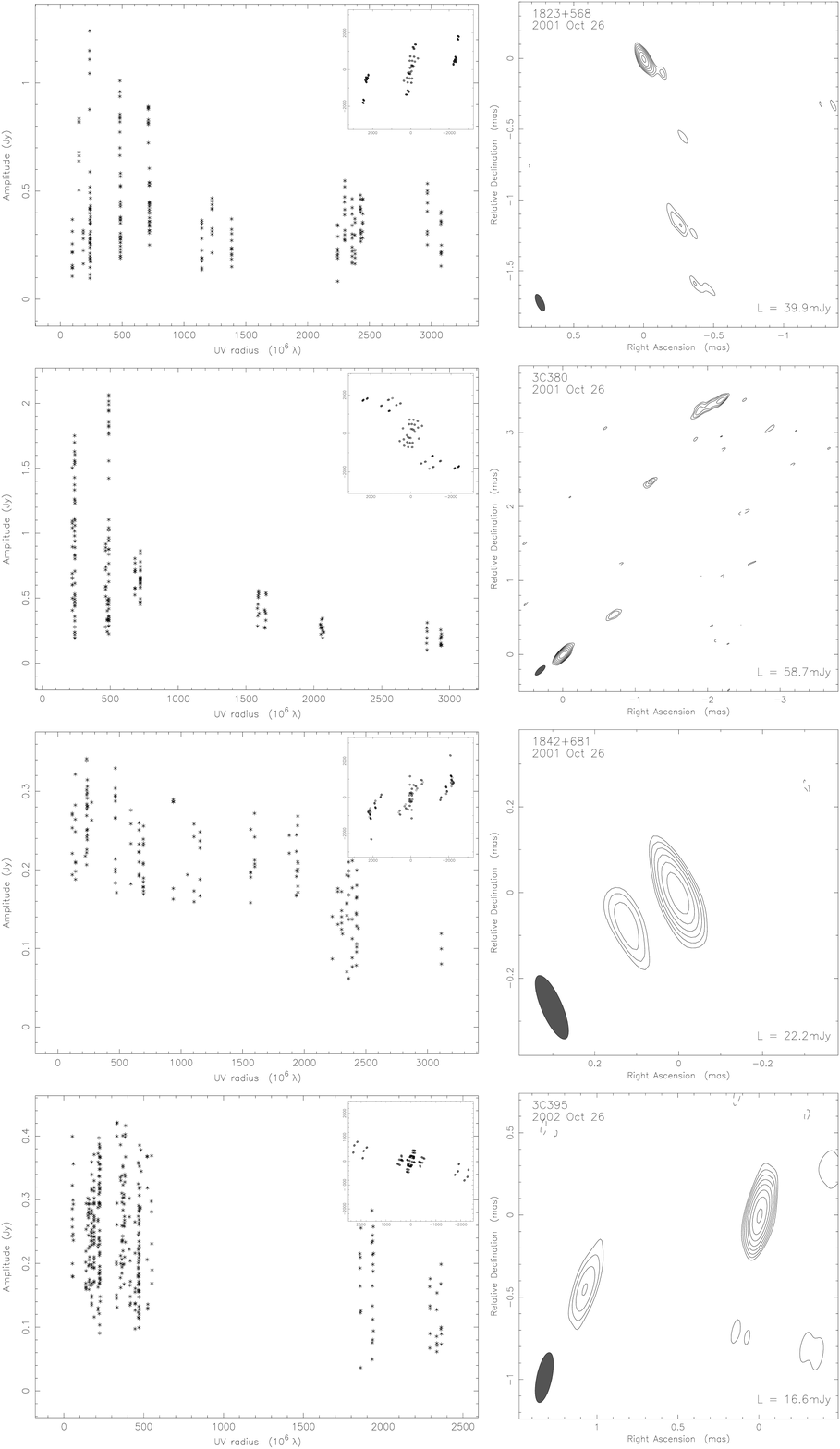}
\caption{{\it continued.}}
\end{center}
\end{figure*}    

\clearpage
\setcounter{figure}{5}

\begin{figure*}[p]    
\begin{center} \includegraphics[width=0.7\textwidth] {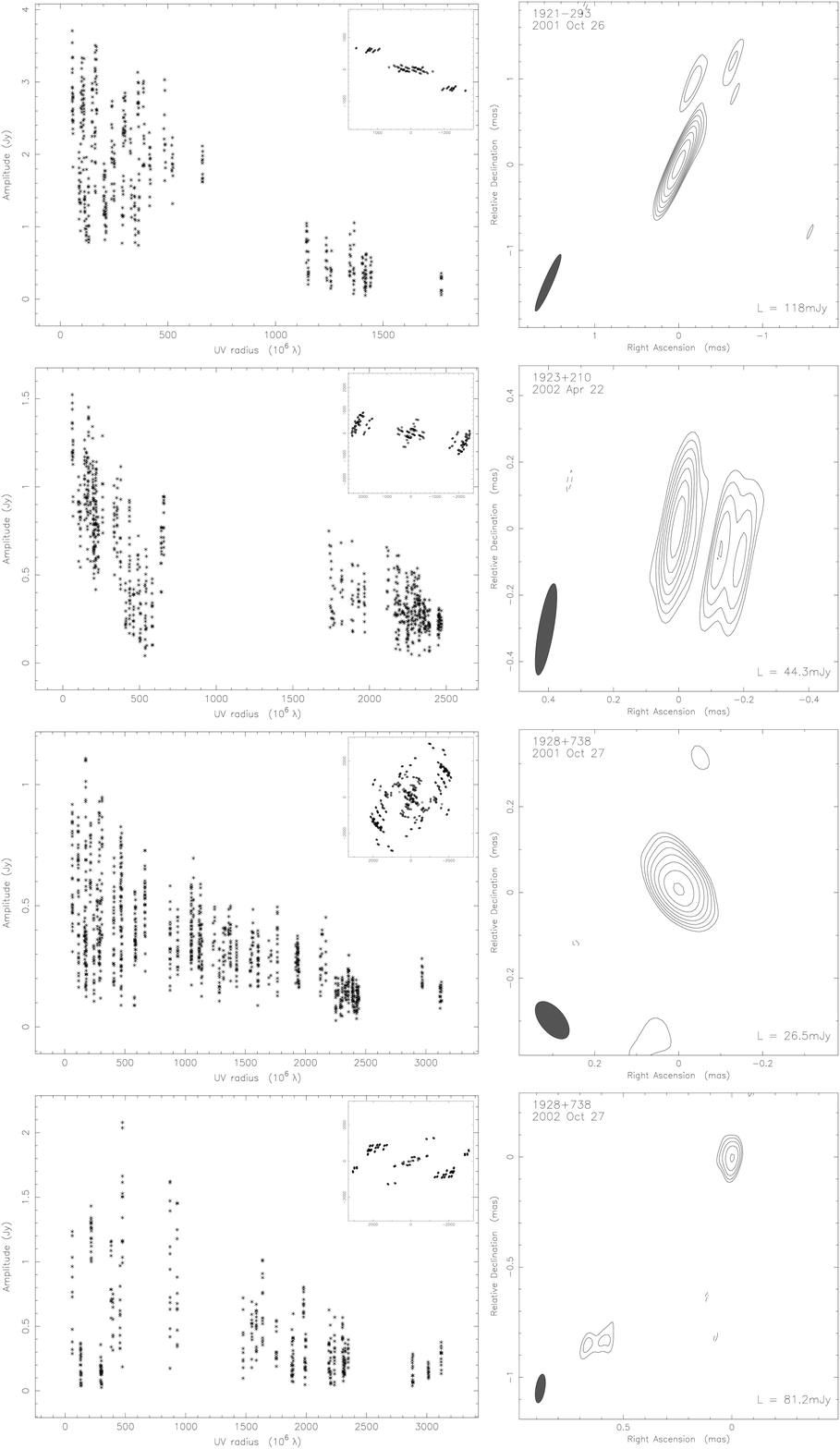}
\caption{{\it continued.}}
\end{center}
\end{figure*}    

\clearpage
\setcounter{figure}{5}

\begin{figure*}[p]    
\begin{center} \includegraphics[width=0.7\textwidth] {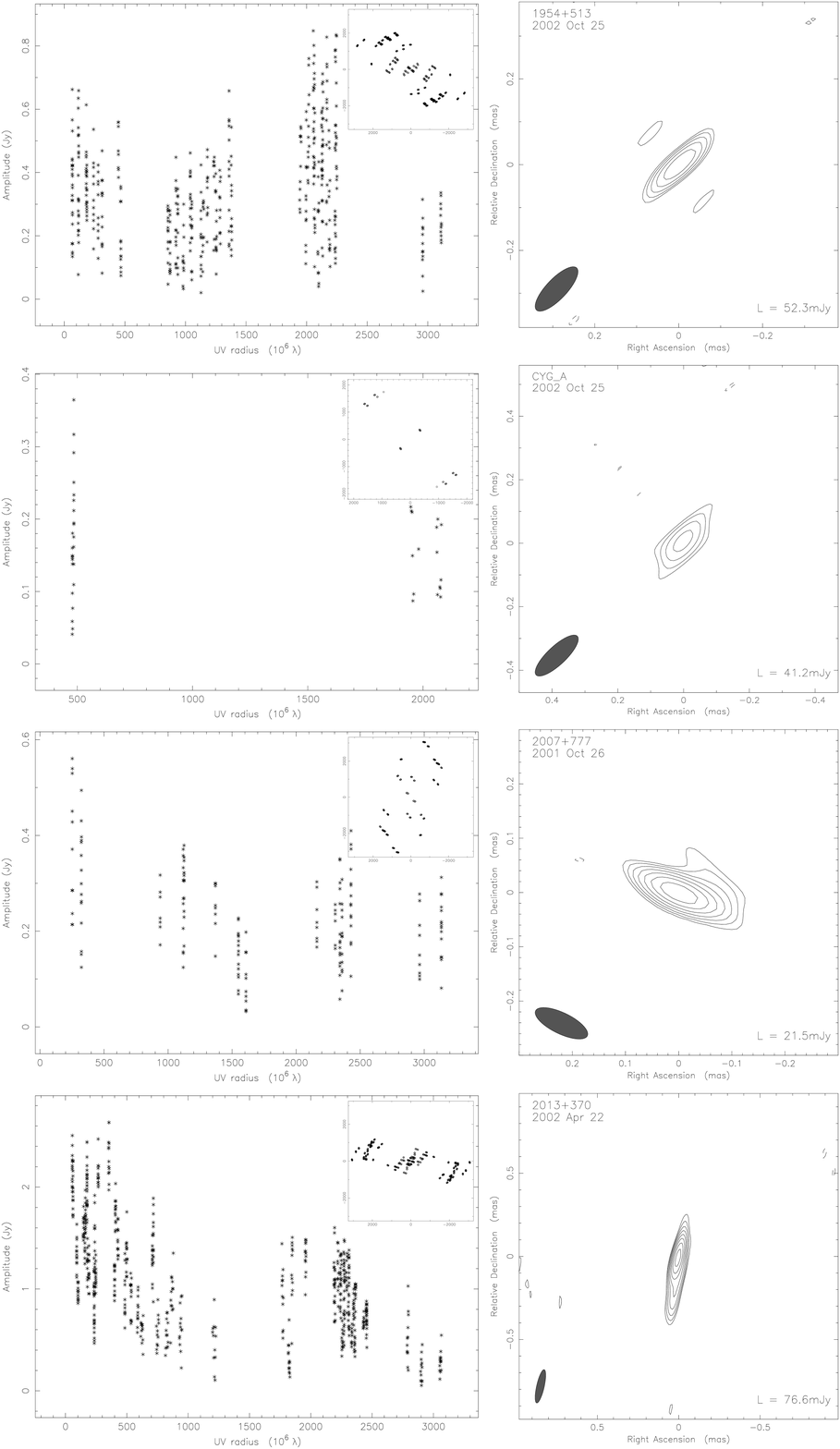}
\caption{{\it continued.}}
\end{center}
\end{figure*}    

\clearpage
\setcounter{figure}{5}

\begin{figure*}[p]    
\begin{center} \includegraphics[width=0.7\textwidth] {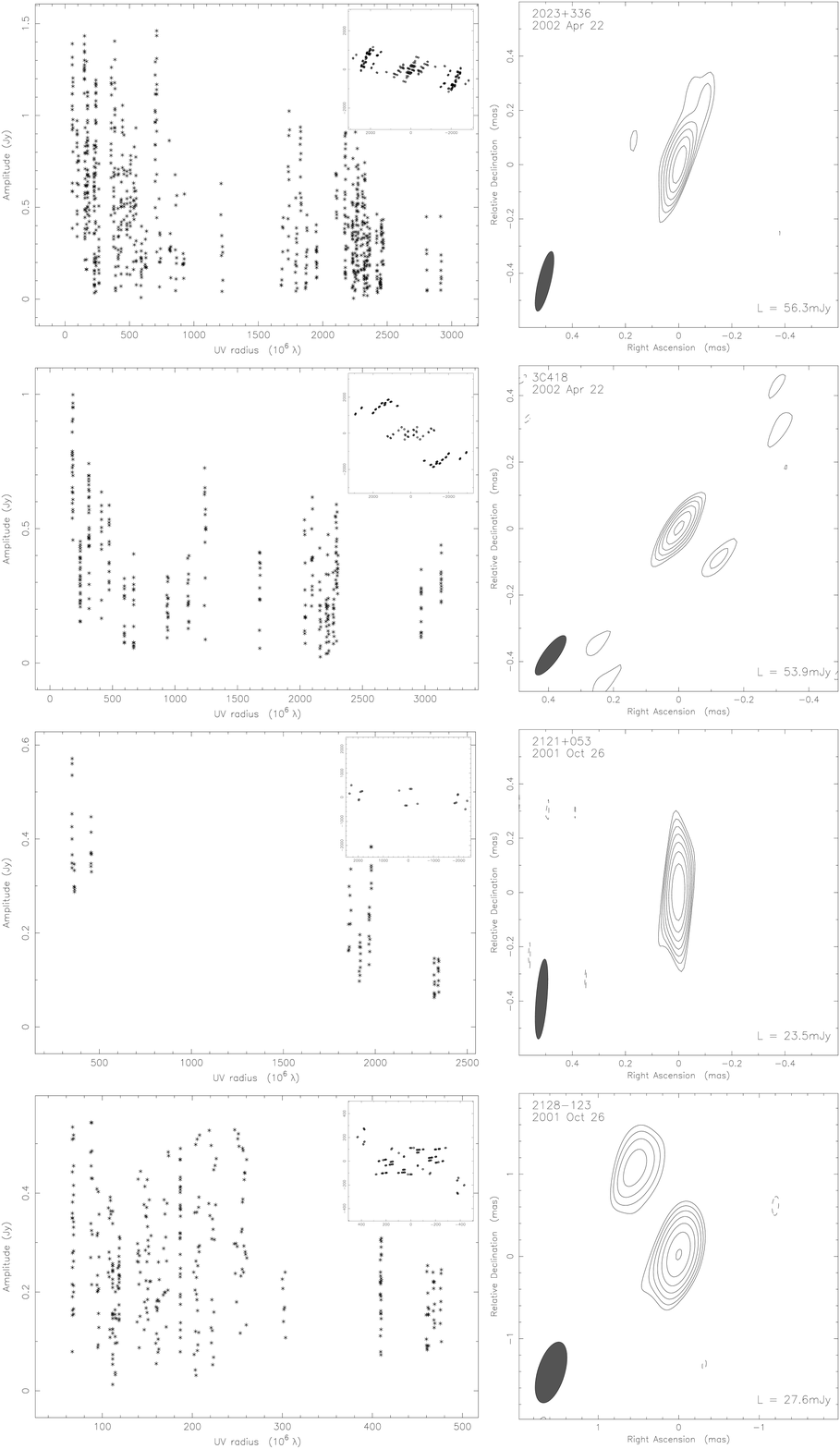}
\caption{{\it continued.}}
\end{center}
\end{figure*}    

\clearpage
\setcounter{figure}{5}

\begin{figure*}[p]    
\begin{center} \includegraphics[width=0.7\textwidth] {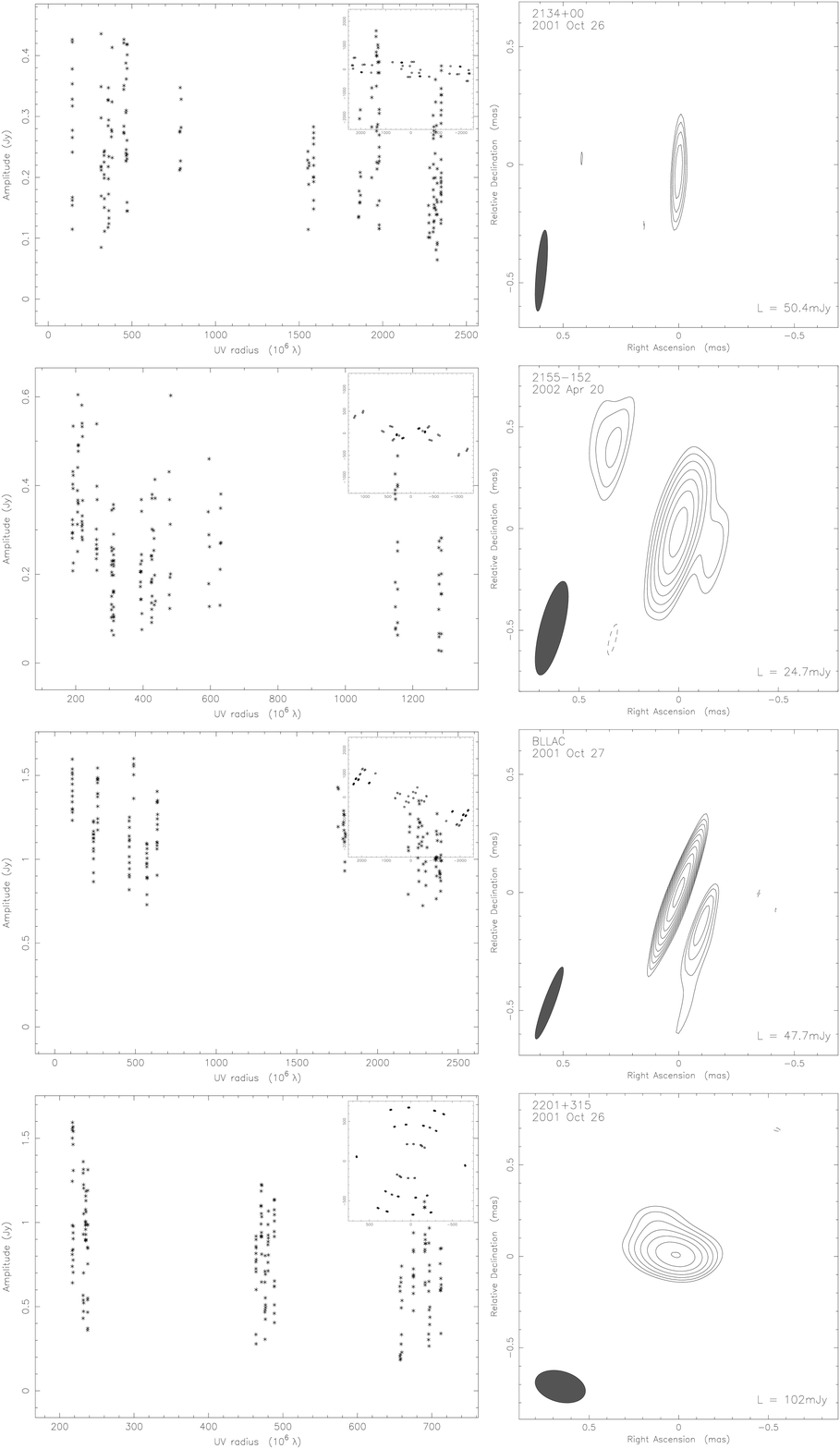}
\caption{{\it continued.}}
\end{center}
\end{figure*}    

\clearpage
\setcounter{figure}{5}

\begin{figure*}[p]    
\begin{center} \includegraphics[width=0.7\textwidth] {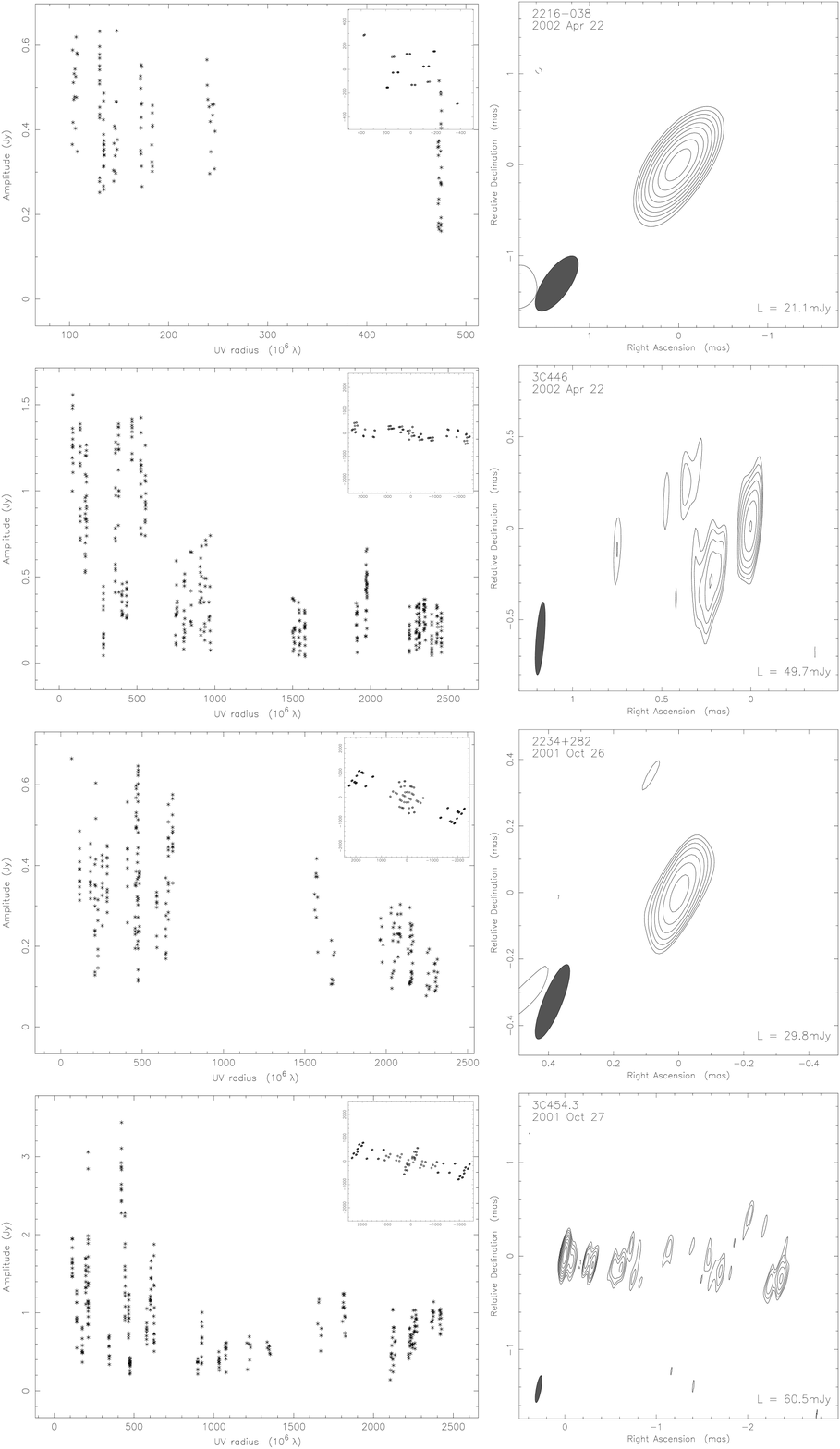}
\caption{{\it continued.}}
\end{center}
\end{figure*}    

\clearpage
\setcounter{figure}{5}

\begin{figure*}[t]    
\begin{center} \includegraphics[width=0.7\textwidth] {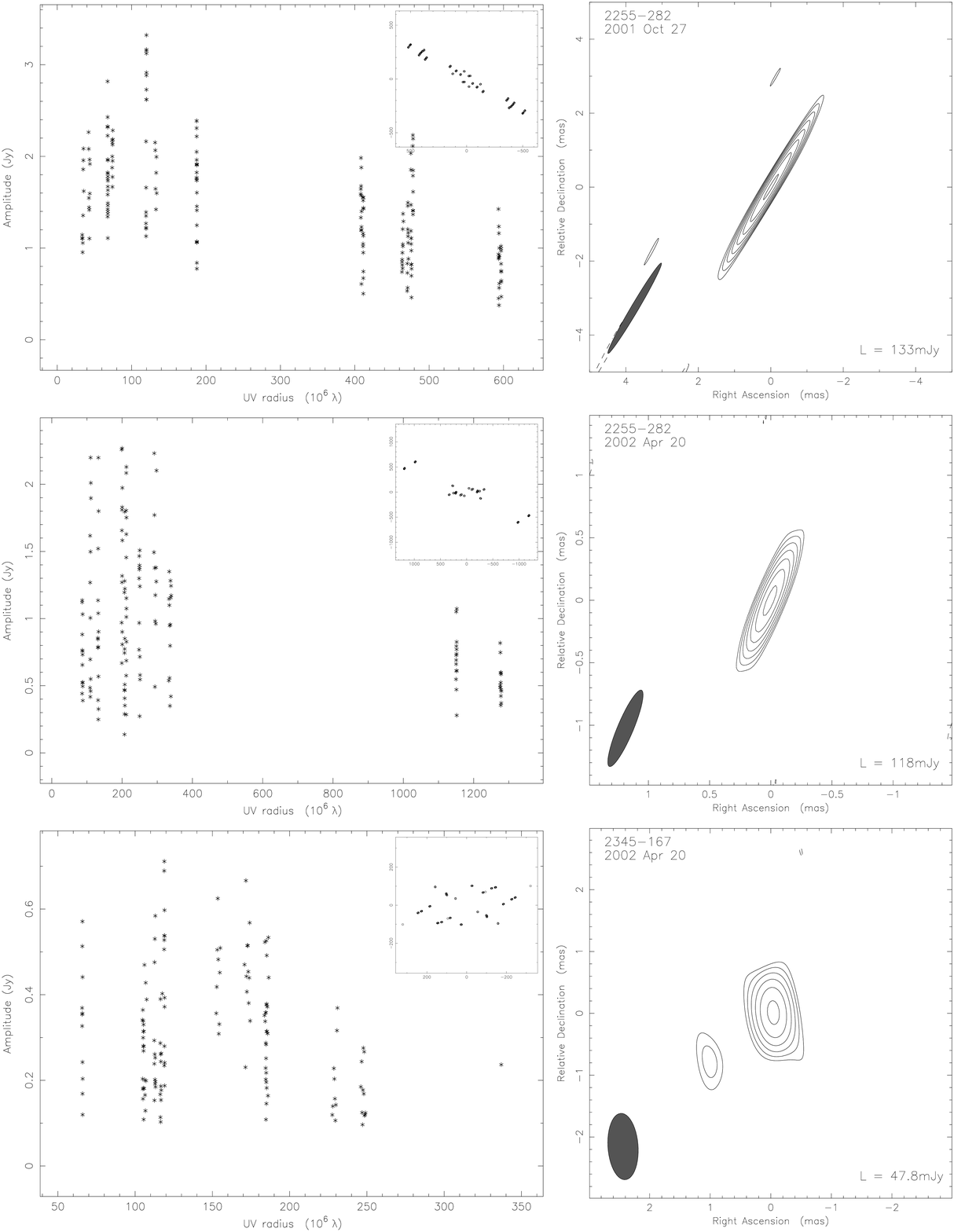}
\caption{{\it continued.}}
\end{center}
\end{figure*}    
\clearpage
\begin{figure*}    
\begin{center} 
\includegraphics[width=1\textwidth]{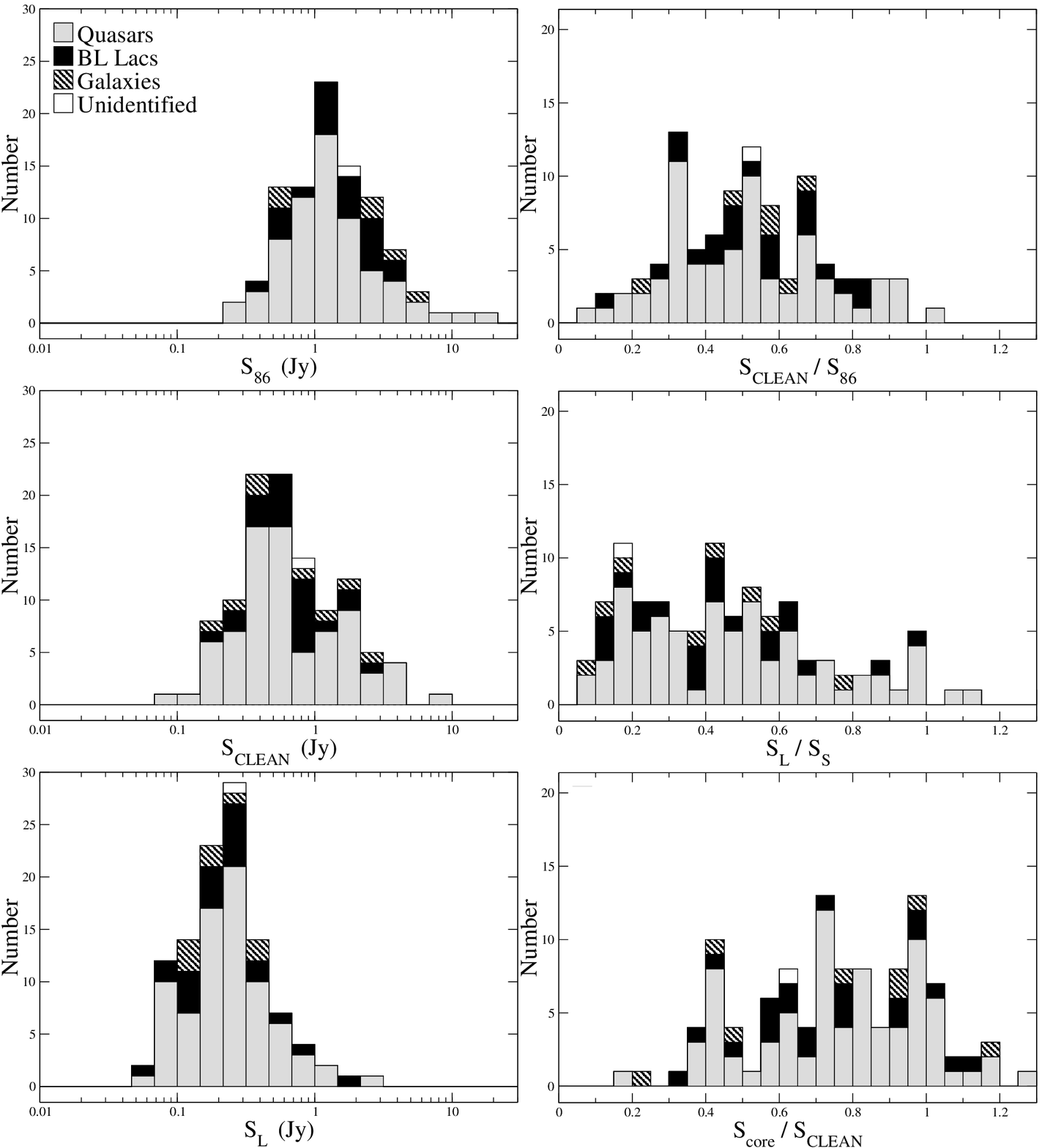}
\caption{Distributions of the total flux density, $S_{86}$ (top left), 
the CLEAN flux density, $S_{\rm CLEAN}$ (middle left),
the correlated flux density on the longest baseline, $S_{\rm L}$ (bottom left), 
compactness indices on milliarcsecond scales $S_{\rm CLEAN}/S_{86}$ (top right) 
and sub-milliarcsecond scales $S_{\rm L}/S_{\rm S}$ (middle right), 
and the core dominance $S_{\rm core}/S_{\rm CLEAN}$ of the imaged sources. 
In the top left panel, sources with the total flux density not available are excluded.
\label{fig:com1}
}
\end{center}
\end{figure*}    
\clearpage
\begin{figure}    
\begin{center} 
\includegraphics[clip, width=1\columnwidth] {radplot.all.Scln.eps}
\caption{Normalized mean amplitude of the visibility function in terms of {\it uv}-radius
for the survey sample. 
The visibility amplitude is normalized by the CLEAN flux density $S_{\rm CLEAN}$
for each source (which corresponds to points at 0\,M$\lambda$),
binned with 200\,M$\lambda$ wide bins ranging from 0 to 2600\,M$\lambda$, and averaged. 
Not all bins are sampled for all sources. 
\label{fig:com2}
}
\end{center}
\end{figure}    
\clearpage
\begin{figure}    
\begin{center} 
\includegraphics[width=0.5\columnwidth] {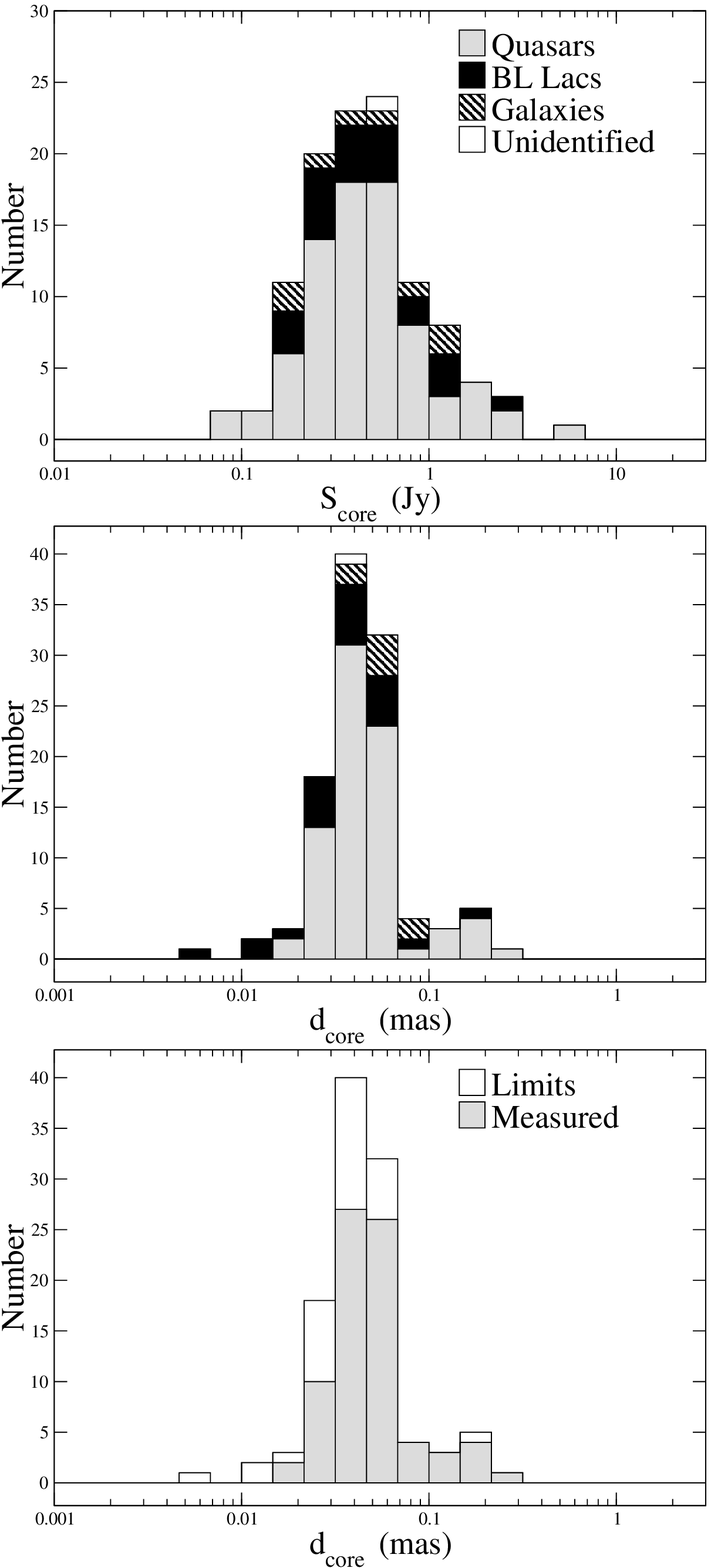}
\caption{Distributions of the flux density (top panel) and the angular
size (middle and bottom panel) of the core components for the imaged sources.
\label{fig:C4-Tb1}
}
\end{center}
\end{figure}    
\clearpage
\begin{figure}  
\begin{center} 
\includegraphics[width=1\columnwidth] {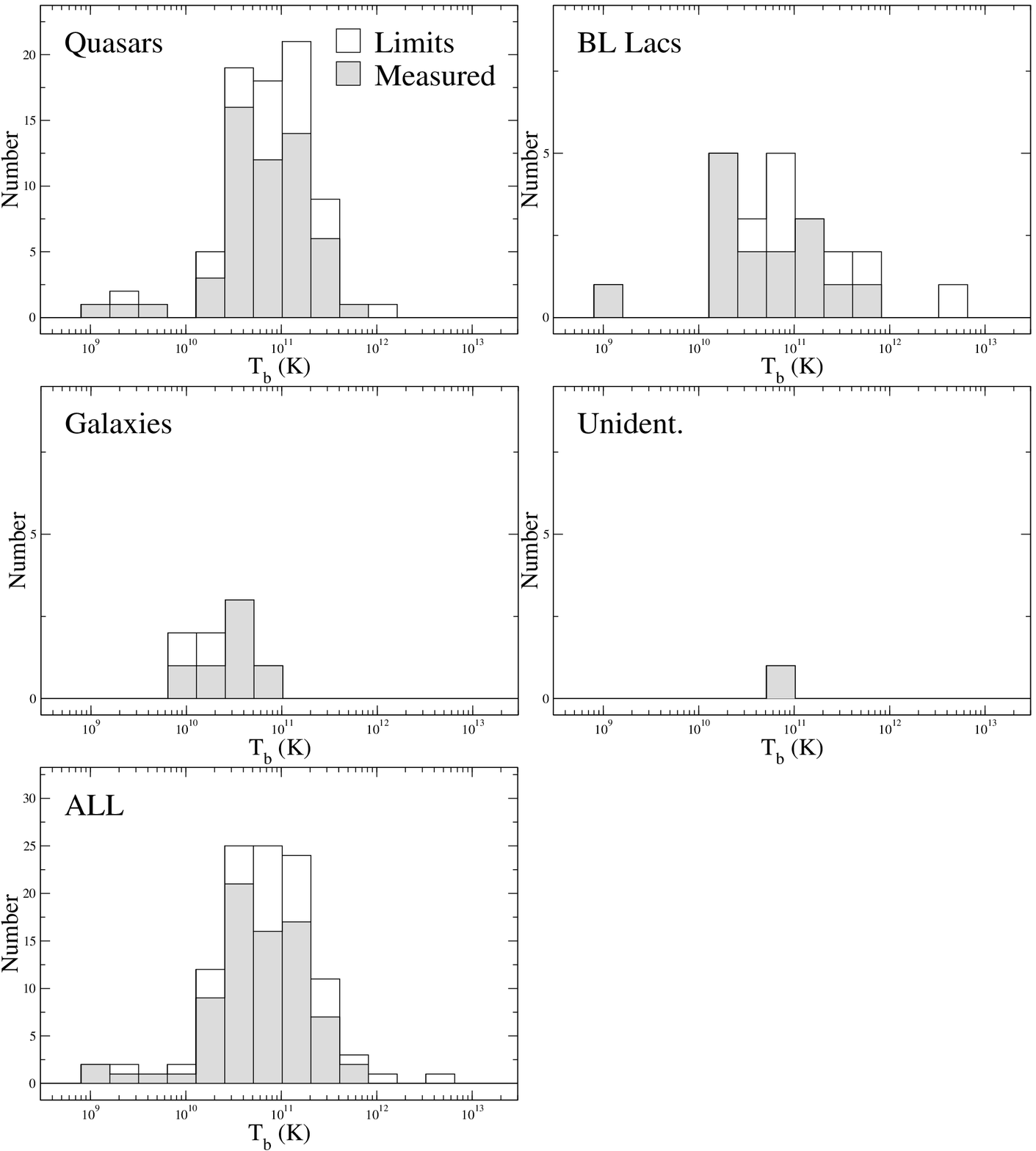}
\caption{Distributions of the measured brightness temperature of the core component in the source frame.
The brightness temperature is binned with a size of a factor of 2 in brightness temperature 
from 1$\times 10^8$\,K to the maximum value of each sample.
\label{fig:C4-Tb2}
}
\end{center}
\end{figure}    
\clearpage
\begin{figure}    
\begin{center} 
\includegraphics[width=1\columnwidth] {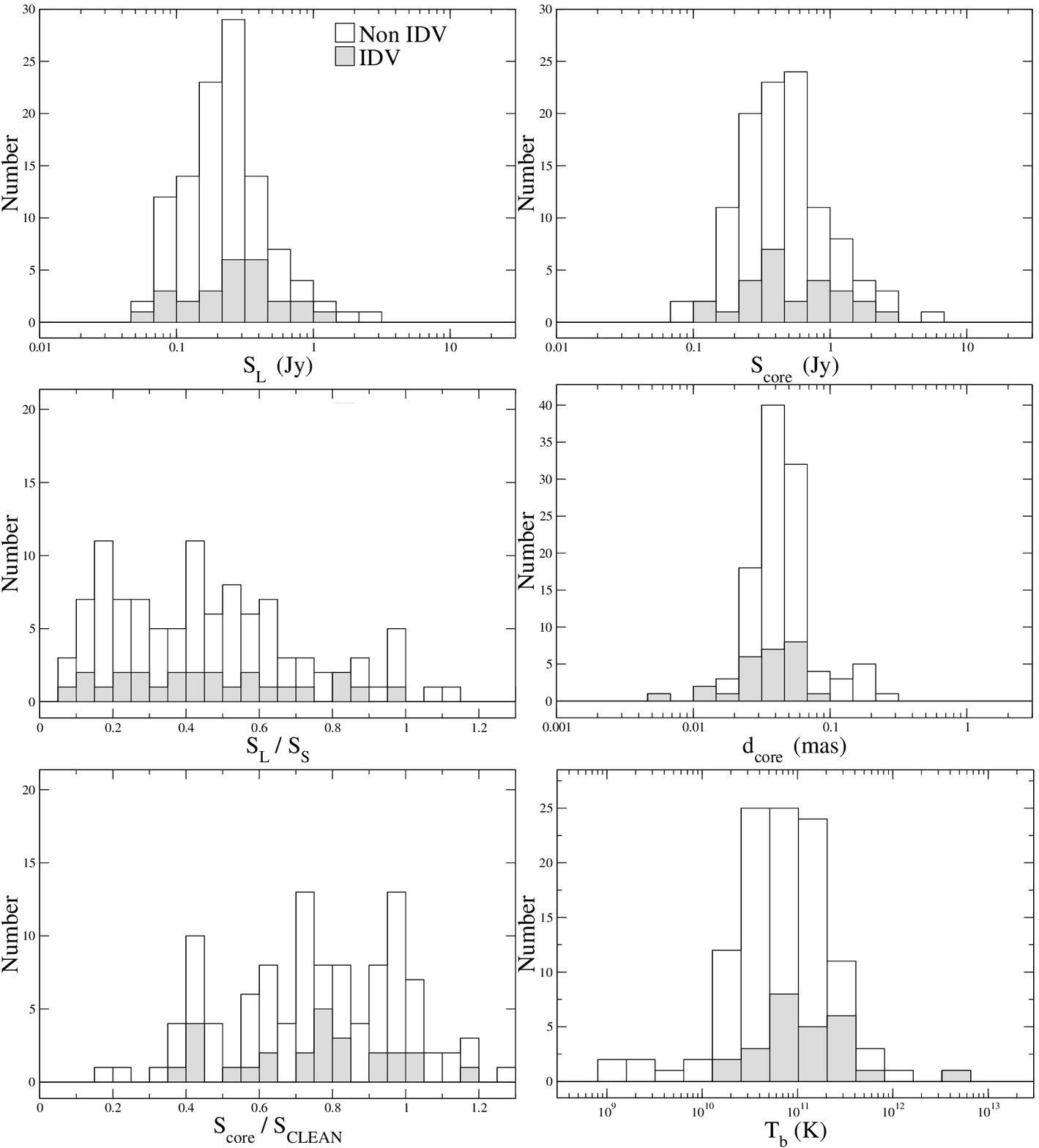}
\caption{Distributions of 
the correlated flux density at the longest baseline $S_{\rm L}$ (top left), 
the compactness index $S_{\rm L}/S_{\rm S}$ (middle left),
the core dominance $S_{\rm core}/S_{\rm CLEAN}$ (bottom left),  
the core flux density $S_{\rm core}$ (top right),
the size of core component $d_{\rm core}$ (middle right), and
the brightness temperature $T_{\rm b}$ (bottom right) 
for IDV selected and non-IDV selected sources (see text for reference).
\label{fig:IDV}
}
\end{center}
\end{figure}    
\clearpage
\begin{figure}    
\begin{center}
\includegraphics[width=1\columnwidth, clip] {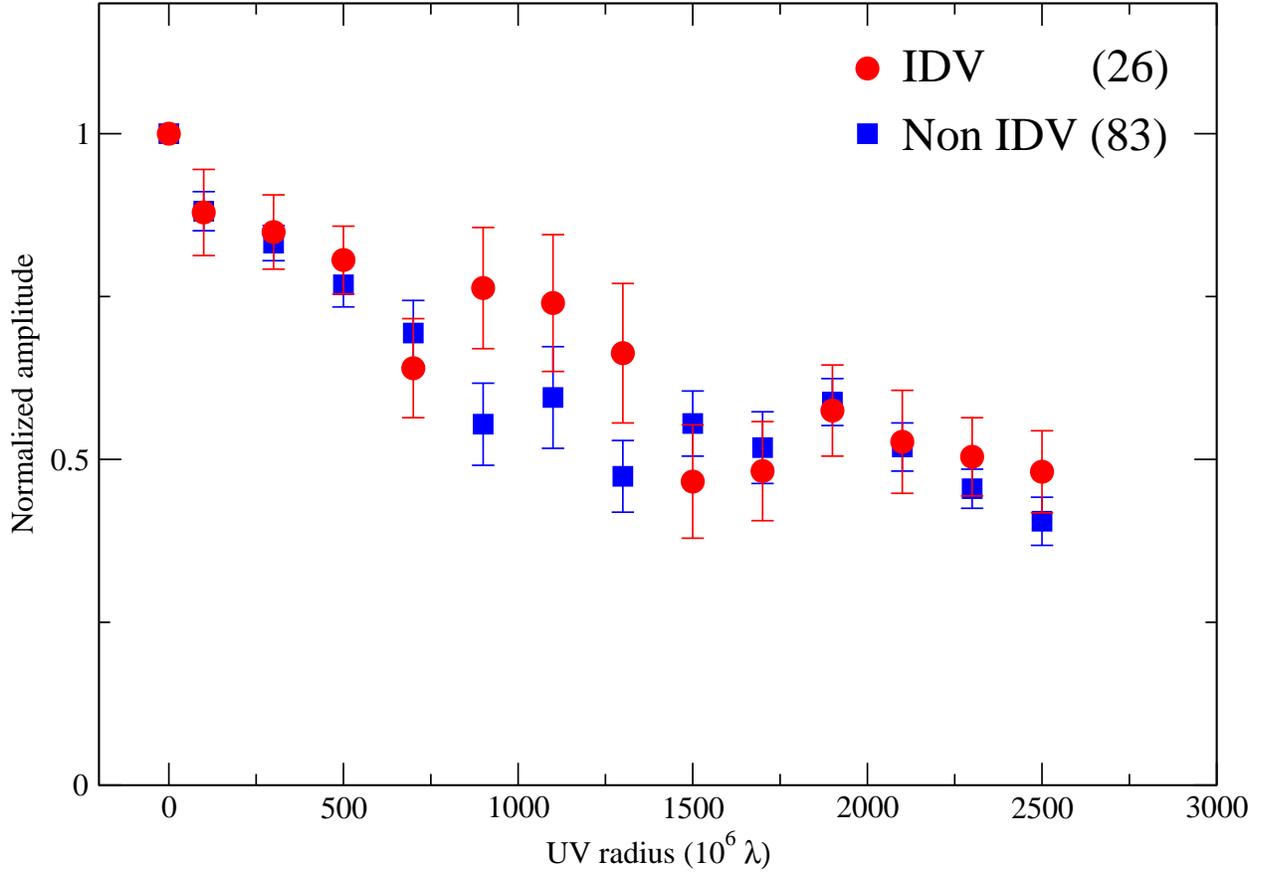}
\caption[Normalized mean amplitude of the visibility function in terms of {\it uv}-radius
         for IDV and non-IDV sources (see text for details).]
         {Normalized mean amplitude of the visibility function in terms of {\it uv}-radius
         for IDV and non-IDV sources (see text for details). 
         The visibility amplitude is normalized by the CLEAN flux density $S_{\rm CLEAN}$
         for each source (which corresponds to a point at 0\,M$\lambda$),
         binned with 200\,M$\lambda$ wide bins ranging from 0 to 2600\,M$\lambda$, and averaged. 
         Not all bins are sampled for all sources. 
\label{fig:radplotIDV}
}
\end{center}
\end{figure}    
\clearpage
\begin{deluxetable}{cccccccc}
\tabletypesize{\scriptsize}
\tablecaption{VLBI Surveys at 86\,GHz\label{table:vlbisurveys}}
\tablewidth{0pt}
\tablehead{
\colhead{}              & \colhead{}                   & 
\colhead{$\Delta S$}    & \colhead{$\Delta I_{\rm m}$} & 
\colhead{}              & \colhead{}                   & 
\colhead{}              & \colhead{}                   \\
\colhead{Survey}        & \colhead{$N_{\rm ant}$}      & 
\colhead{(Jy)}          & \colhead{(mJy/beam)}         & 
\colhead{$D_{\rm img}$} & \colhead{$N_{\rm obs}$}      & 
\colhead{$N_{\rm det}$} & \colhead{$N_{\rm img}$}      \\
\colhead{(1)}           & \colhead{(2)}                & 
\colhead{(3)}           & \colhead{(4)}                & 
\colhead{(5)}           & \colhead{(6)}                & 
\colhead{(7)}           & \colhead{(8)}                
}
\startdata
\multicolumn{1}{l}{\cite{Beasley97}}  & 3 & $\sim$0.5 & ... & ... & 45 & 12 & ... \\
\multicolumn{1}{l}{\cite{Lonsdale98}} & 2$-$5 & $\sim$0.7 & ... & ... & 79 & 14 & ... \\
\multicolumn{1}{l}{\cite{Ranta98}}    & 6$-$9 & $\sim$0.5 & $\sim$30 & 70 & 67 & 16 & 12 \\
\multicolumn{1}{l}{\cite{Lobanov00}}  & 3$-$5 & $\sim$0.4 & $\sim$20 & 100 & 28 & 26 & 17 \\
\tableline
\multicolumn{5}{l}{\bf Total number of unique objects:}
& 124 & 44 & 24 \\
\tableline
\multicolumn{8}{c}{\bf Properties of this survey}\\
& 12 & $\sim$0.2 & $\leq$10 & 50 & 127 & 121 & 109 \\
\enddata
\tablecomments{ 
{\bf Columns:} 1 - references; 
2 - number of participating antennas; 
3 - average baseline sensitivity;
4 - average image sensitivity; 
5 - typical dynamic range of images; 
6 - number of sources observed;
7 - number of objects detected; 
8 - number of objects imaged. 
}
\end{deluxetable}
\clearpage
\begin{deluxetable}{llclllcccc}
\tabletypesize{\scriptsize}
\tablewidth{0pt}
\tablecaption{Source List\label{table:sourcelist}}
\tablehead{
\colhead{Source}                & \colhead{Name}                 &
\colhead{Epoch}                 & \colhead{$\alpha_{\rm 2000}$}  &
\colhead{$\delta_{\rm 2000}$}   & \colhead{Status}               &
\colhead{$z$}                   & \colhead{Type}                 &
\colhead{$m_{\rm v}$}           & \colhead{$S_{\rm 86\,GHz}$}    \\
\colhead{(1)}                   & \colhead{(2)}                  &
\colhead{(3)}                   & \colhead{(4)}                  &
\colhead{(5)}                   & \colhead{(6)}                  &
\colhead{(7)}                   & \colhead{(8)}                  &
\colhead{(9)}                   & \colhead{(10)}                 
}
\startdata
0003$-$066  &  NRAO 5      &  C      &  00 06 13.89289  & $-$06 23 35.3356  &  $++$  & 0.347  & B  & 19.5  & 2.16  \\
0007$+$106  &  III Zw 2    &  C      &  00 10 31.00587  & $+$10 58 29.5038  &  $++$  & 0.089  & G  & 15.4  & 0.60  \\
0016$+$731  &              &  B      &  00 19 45.78642  & $+$73 27 30.0174  &  $++$  & 1.781  & Q  & 18.0  & 0.84  \\
0048$-$097  &              &  C      &  00 50 41.31739  & $-$09 29 05.2103  &  $++$  & ...    & B  & 17.4  & 0.60  \\
0106$+$013  &  4C 01.02    &  A      &  01 08 38.77104  & $+$01 35 00.3232  &  $++$  & 2.107  & Q  & 18.3  & 1.36  \\
0119$+$041  &              &  C      &  01 21 56.86169  & $+$04 22 24.7343  &  $++$  & 0.637  & Q  & 19.5  & 0.68  \\
0119$+$115  &              &  C      &  01 21 41.59504  & $+$11 49 50.4131  &  $++$  & 0.570  & Q  & 19.5  & 0.68  \\
0133$+$476  &              &  A,B    &  01 36 58.59481  & $+$47 51 29.1001  &  $++$  & 0.859  & Q  & 18.0  & $4.25^{\dagger}$  \\
\enddata
\tablecomments{ 
Column designation: 1~-~IAU source name; 2~-~common name; 3~-~observing epochs: 
{\bf A}~-~October 2001; {\bf B}~-~April 2002; {\bf C}~-~October 2002;
4~,~5~-~source coordinates; 6~-~status: ``$-$'' - not detected; ``$+$'' - detected; 
``$+m$'' - detected and only model fitted; ``$++$'' - detected and imaged; 
7~-~redshift; 8~-~optical class: {\bf Q} - quasar; {\bf B} - BL Lac object;
{\bf G} - radio galaxy; {\bf S} - star; {\bf U} - unidentified; 9~-~optical magnitude; 
7~,~8~-~information obtained from~\citet{veron2006}; 9~-~information obtained from 
NASA/IPAC Extragalactic Database, http://nedwww.ipac.caltech.edu; 10~-~total flux density [Jy]
(obtained from pointing and calibration scan measurements made at Pico Veleta);
$\dagger$: mean value of measurements on multiple epochs.
}
\end{deluxetable}
\clearpage
\begin{deluxetable}{lccccccl}
\tabletypesize{\scriptsize}
\tablewidth{0pt}
\tablecaption{Log of Survey Observations\label{table:obs}}
\tablehead{
\colhead{}                      & \colhead{}                     &
\colhead{Bit Rate}              & \colhead{Frequency}            &
\colhead{}                      & \colhead{Bandwidth}            &
\colhead{}                      & \colhead{}                     \\
\colhead{Epoch}                 & \colhead{Code}                 &
\colhead{(Mbit/s)}              & \colhead{Channels}             &
\colhead{Sampling}              & \colhead{(MHz)}                &
\colhead{Source}                & \colhead{Telescopes}           \\
\colhead{(1)}                   & \colhead{(2)}                  &
\colhead{(3)}                   & \colhead{(4)}                  &
\colhead{(5)}                   & \colhead{(6)}                  &
\colhead{(7)}                   & \colhead{(8)}                  
}
\startdata
2001 Oct 26-29  & A & 256 & 16 & 1 & 128 & 48 & VLBA + (Eb,PV,On,Mh,HA) \\
2002 Apr 20-23    & B & 256 & 16 & 1 & 128 & 35 & VLBA + (Eb,PV,On,HA)    \\
2002 Oct 24-27  & C & 256 & 8  & 2 &  64 & 60 & VLBA + (Eb,PV,PdB,HA)   \\
\enddata
\tablecomments{ 
Column designation: 1~-~observation epoch; 2~-~code of each epoch; 3~-~total recorded bit rate in Mega--bits per second;
4~-~number of baseband channels; 5~-~sampling mode [bit]; 6~-~total observing bandwidth; 7~-~number of sources observed (12 of the 127 observed 
sources were observed during more than one session);
8~-~telescopes participating: {\bf VLBA} - Fort Davis, Hancock, North Liberty, Owens Valley, Pie Town, Mauna Kea, Los Alamos, Kitt Peak;
{\bf Eb} - Effelsberg; {\bf PV} - Pico Veleta; {\bf On} - Onsala ; {\bf Mh} - Mets\"ahovi; {\bf PdB} - Plateau de Bure ; {\bf HA} - Haystack.}
\end{deluxetable}
\clearpage
\begin{deluxetable}{llrlccrcc}
\tabletypesize{\scriptsize}
\tablewidth{0pt}
\tablecaption{Participating telescopes\label{table:antenna}}
%
\tablehead{
\colhead{}                      & \colhead{}                     &
\colhead{$D$}                   & \colhead{$G$}                  &
\colhead{$T_{\rm sys}$}         & \colhead{}                     &
\colhead{$SEFD$}                & \colhead{$\Delta^{\rm 256,30s}$} &
\colhead{Threshold}                                             \\
\colhead{Name}                  & \colhead{code}                 &
\colhead{(m)}                   & \colhead{(K/Jy)}               &
\colhead{(K)}                   & \colhead{$\eta_{\rm A}$}       &
\colhead{(K)}                   & \colhead{(mJy)}                &
\colhead{(mJy)}                                                 \\
\colhead{(1)}                   & \colhead{(2)}                  &
\colhead{(3)}                   & \colhead{(4)}                  &
\colhead{(5)}                   & \colhead{(6)}                  &
\colhead{(7)}                   & \colhead{(8)}                  &
\colhead{(9)}                     
}
\startdata
Effelsberg       & Eb   & 100  & 0.140   & 130      & 0.07   & 929     & 20 & 143  \\
Haystack         & HA   & 37   & 0.058   & 200      & 0.15   & 3448    & 39 & 273  \\
Plateau de Bure  & PdB  & $6\times15$   & 0.180   & 120      & 0.65   & 667     & 17 & 121  \\
Pico Veleta      & PV   & 30   & 0.140   & 120      & 0.55   & 857     & ...& ...  \\
Onsala           & On   & 20   & 0.053   & 250      & 0.45   & 4717    & 45 & 321  \\
Mets\"ahovi      & Mh   & 14   & 0.017   & 300      & 0.30   &17647    & 89 & 621  \\
Fort Davis       & Fd   & 25   & 0.034   & 120      & 0.17   & 3529    & 40 & 278  \\
Hancock          & Hn   & 25   & 0.035   & 120      & 0.17   & 3429    & 39 & 274  \\
\enddata
\tablecomments{ 
Column designation: 1~-~name of the participating telescope; 
2~-~abbreviation of the telescope name; 
3~-~diameter; 4~-~typical zenith gain; 5~-~system temperature; 
6~-~aperture efficiency;
7~-~typical zenith SEFD obtained from the formula, $SEFD=T_{\rm sys}/G$;
8~-~baseline sensitivity on baseline to Pico Veleta, assuming a recording rate of 256 Mbps and
a fringe-fit interval of 30 seconds;
9~-~7$\sigma$ detection threshold. 
}
\end{deluxetable}
\clearpage
\begin{deluxetable}{cccc}
\tabletypesize{\scriptsize}
\tablewidth{0pt}
\tablecaption{Average antenna gain corrections\label{table:gain}}
%
\tablehead{
\colhead{Telescope}             & \colhead{Session A}            &
\colhead{Session B}             & \colhead{Session C}           \\
\colhead{(1)}                   & \colhead{(2)}                  &
\colhead{(3)}                   & \colhead{(4)}                 
}
\startdata
Eb   &  0.998 $\pm$ 0.129         &  1.079 $\pm$ 0.183  &    1.354 $\pm$ 0.555          \\
HA   &  1.160 $\pm$ 0.179         &  1.155 $\pm$ 0.315  &    1.180 $\pm$ 0.350          \\
PdB  &  ...                       &  ...                &    0.964 $\pm$ 0.150          \\
PV   &  0.971 $\pm$ 0.157         &  0.938 $\pm$ 0.156  &    0.952 $\pm$ 0.192          \\
On   &  1.033 $\pm$ 0.110         &  0.956 $\pm$ 0.121  &    ...                        \\
Mh   &  ...                       &  ...                &    ...                        \\
Fd   &  1.215 $\pm$ 0.216         &  1.213 $\pm$ 0.298  &    1.080 $\pm$ 0.321          \\
Hn   &  ...                       &  ...                &    1.055 $\pm$ 0.235          \\
\enddata
\tablecomments{ 
Column designation: 1~-~abbreviation for the name of telescopes; 
2~-~average and rms of antenna gains for observing session A; 
3~-~average and rms of antenna gains for observing session B; 
4~-~average and rms of antenna gains for observing session C. 
}
\end{deluxetable}
\clearpage
\begin{deluxetable}{lcrrrrrrrrrrrr}
\tabletypesize{\scriptsize}
\tablewidth{0pt}
\tablecaption{Image parameters\label{table:imageparameter}}
\tablehead{
\colhead{Name}                  & \colhead{Obs}                  &
\colhead{$S_{\rm 86}$}          & \colhead{$S_{\rm S}$}          &
\colhead{$B_{\rm S}$}           & \colhead{$S_{\rm L}$}          &
\colhead{$B_{\rm L}$}           & \colhead{$B_{\rm a}$}          &
\colhead{$B_{\rm b}$}           & \colhead{$B_{\rm PA}$}         &
\colhead{$S_{\rm t}$}           & \colhead{$S_{\rm p}$}          &
\colhead{$\sigma$}              & \colhead{$\xi_{\rm r}$}        \\
\colhead{(1)}                   & \colhead{(2)}                  &
\colhead{(3)}                   & \colhead{(4)}                  &
\colhead{(5)}                   & \colhead{(6)}                  &
\colhead{(7)}                   & \colhead{(8)}                  &
\colhead{(9)}                   & \colhead{(10)}                 &
\colhead{(11)}                  & \colhead{(12)}                 &
\colhead{(13)}                  & \colhead{(14)}                 
}
\startdata
0003$-$066 & C &  2.16 & 0.71 $\pm$ 0.30 &   60 & 0.29 $\pm$ 0.16 & 2450 &   461 &    49 &  -6.2 &  613 &  209 &    17 &  0.86  \\ 
0007$+$106 & C &  0.60 & 0.59 $\pm$ 0.00 &   10 & 0.35 $\pm$ 0.06 & 2470 &   457 &    54 &  -2.7 &  372 &  181 &     9 &  1.00  \\ 
0016$+$731 & B &  0.84 & 0.48 $\pm$ 0.13 &   60 & 0.25 $\pm$ 0.08 & 3130 &    75 &    51 & -11.6 &  434 &  181 &     8 &  0.91  \\ 
0048$-$097 & C &  0.60 & 0.43 $\pm$ 0.13 &   60 & 0.11 $\pm$ 0.06 & 2440 &   466 &    42 &  -3.3 &  486 &  163 &     6 &  0.75  \\ 
0106$+$013 & A &  1.36 & 0.49 $\pm$ 0.06 &  100 & 0.24 $\pm$ 0.05 & 2390 &   900 &    45 &  -8.6 &  447 &  354 &    15 &  0.62  \\ 
0119$+$041 & C &  0.68 & 0.20 $\pm$ 0.06 &   70 & 0.20 $\pm$ 0.09 & 2390 &   602 &    79 &  -9.9 &  233 &  227 &     9 &  0.77  \\ 
0119$+$115 & C &  0.68 & 0.34 $\pm$ 0.12 &   60 & 0.16 $\pm$ 0.06 & 2390 &   400 &    44 &  -7.1 &  232 &  176 &    10 &  0.63  \\ 
0133$+$476 & A &  3.96 & 1.99 $\pm$ 0.86 &   60 & 0.09 $\pm$ 0.03 & 3140 &   170 &    51 & -16.4 & 2148 &  675 &    17 &  0.80  \\
\enddata
\tablecomments{ 
Column designation: 1~-~source name; 2~-~observing epochs: 
{\bf A}~-~October 2001; {\bf B}~-~April 2002; {\bf C}~-~October 2002; 
3~-~total flux density [Jy] (obtained from pointing and calibration 
scan measurements made at Pico Veleta); 4,6~-~correlated flux density [Jy] 
measured on baselines 5,7 [M$\lambda$]; 8-10~-~restoring beam: 
8~-~major axis [$\mu$as]; 9~-~minor axis [$\mu$as]; 
10~-~position angle of the major axis [$\cdot$];
11~-~total CLEAN flux density [mJy]; 
12~-~peak flux density [mJy/beam]; 
13~-~off-source RMS in the image [mJy/beam]; 
14~-~quality of the residual noise in the image.
}
\end{deluxetable}
\clearpage
\begin{deluxetable}{lcrrrrrr}
\tabletypesize{\scriptsize}
\tablewidth{0pt}
\tablecaption{Model fit parameters of sources\label{table:modelfit}}
\tablehead{
\colhead{Name}                  & \colhead{Obs}                  &
\colhead{$S_{\rm tot}$}         & \colhead{$S_{\rm peak}$}       &
\colhead{$d$}                   & \colhead{$r$}                  &
\colhead{$\theta$}              & \colhead{$T_{\rm b}$}          \\
\colhead{(1)}                   & \colhead{(2)}                  &
\colhead{(3)}                   & \colhead{(4)}                  &
\colhead{(5)}                   & \colhead{(6)}                  &
\colhead{(7)}                   & \colhead{(8)}                  
}
\startdata
0003$-$066 & C &      210$\pm$89 &      193$\pm$60 &  $ < $ {\it 40} &             ... &             ... & $ > $ {\it 2.9}\\
           &   &      203$\pm$78 &      138$\pm$44 &       48$\pm$15 &       580$\pm$8 &   -3.0$\pm$ 0.8 &   1.9$\pm$1.2\\
           &   &       88$\pm$42 &       87$\pm$30 &  $ < $ {\it 46} &       804$\pm$8 &    5.1$\pm$ 0.6 & $ > $ {\it 0.91}\\
0007$+$106 & C &      185$\pm$93 &      196$\pm$67 &  $ < $ {\it 53} &             ... &             ... & $ > $ {\it 1.2}\\
           &   &       48$\pm$63 &       65$\pm$51 & $ < $ {\it 155} &      461$\pm$60 & -138.3$\pm$ 7.4 & $ > $ {\it 0.036}\\
0016$+$731 & B &     312$\pm$293 &     197$\pm$157 &  $ < $ {\it 28} &             ... &             ... & $ > $ {\it 18}\\
           &   &       51$\pm$51 &       56$\pm$38 &       65$\pm$44 &      115$\pm$22 &  114.6$\pm$10.8 & 0.55$\pm$0.77\\
0048$-$097 & C &      268$\pm$73 &      179$\pm$41 &       42$\pm$10 &             ... &             ... &   2.5$\pm$1.1\\
\enddata
\tablecomments{ 
Column designation: 1~-~source name; 
2~-~observing epoch:
{\bf A} - October 2001; {\bf B} - April 2002; {\bf C} - October 2002; 
3~-~model flux density of the component [mJy]; 
4~-~peak brightness of individual component measured in the image [mJy/beam]; 
5~-~size [$\mu$as]:
italic numbers indicate upper limits; 
6~-~radius [$\mu$as];
7~-~position angle [$\cdot$]; 
8~-~measured brightness temperature [$\times 10^{10}$K]: 
italic numbers indicate lower limits.
}
\end{deluxetable}
\clearpage
\begin{deluxetable}{lccccccccc}
\tabletypesize{\scriptsize}
\tablewidth{0pt}
\tablecaption{Statistics of IDV- and non IDV-selected sources
\label{table:IDV}
}
\tablehead{
\colhead{}                      & \colhead{}                     &
\multicolumn{2}{c}{$S_{\rm L}$[Jy]}                             &&
\multicolumn{2}{c}{$S_{\rm L}/S_{\rm S}$}                       &&
\multicolumn{2}{c}{$S_{\rm core}/S_{\rm CLEAN}$}                \\
\cline{3-4}
\cline{6-7}
\cline{9-10}
\colhead{Sample}                & \colhead{Number}               &
\colhead{Mean}                  & \colhead{Median}              &&
\colhead{Mean}                  & \colhead{Median}              &&
\colhead{Mean}                  & \colhead{Median}               
}
\startdata
IDV      & 26 & 0.36$\pm$0.06 & 0.26 & &  0.53$\pm$0.08  &  0.46   &&  0.75$\pm$0.04    &  0.78  \\
non-IDV  & 83 & 0.31$\pm$0.04 & 0.21 & &  0.51$\pm$0.04  &  0.44   &&  0.76$\pm$0.03    &  0.75  \\
\tableline
\tableline
 &&
\multicolumn{2}{c}{$S_{\rm core}$[Jy]}                         &&
\multicolumn{2}{c}{$d_{\rm core}$[mas]}                        &&
\multicolumn{2}{c}{${\rm log}(T_{\rm b}[{\rm K}])$}            \\ 
\cline{3-4}
\cline{6-7}
\cline{9-10}
Sample  & Number &
Mean    & Median &&
Mean    & Median &&
Mean    & Median \\             
\tableline
IDV     & 26 &  0.70$\pm$0.11  &  0.42  &&  0.039$\pm$0.004  & 0.035  &&  11.1$\pm$0.1 & 11.1 \\
non-IDV & 83 &  0.62$\pm$0.08  &  0.46  &&  0.057$\pm$0.005  & 0.043  &&  10.8$\pm$0.1 & 10.8 \\
\enddata
\tablecomments{Each mean value is presented with its corresponding 
1-$\sigma$ error, assuming a normal distribution.}
\end{deluxetable}

\begin{thebibliography}{}
\bibitem[Agudo et al.(2007)]{Agudo150} Agudo, I., et al.\ 2007, 
Exploring the Cosmic Frontier, ESO Astrophysics Symposia European Southern 
Observatory, Volume .~ISBN 978-3-540-39755-7.~Springer, 2007, p.~179, 179 
\bibitem[Alef \& Graham(2002)]{Alef2002} Alef, W., \& Graham, 
D.~A.\ 2002, Proceedings of the 6th EVN Symposium, 31 
\bibitem[Alef \& M\"uskens(2001)]{Alef2000} Alef, W., \& M\"uskens, A.\ 2001, 
Proceedings of the 15th Workshop Meeting on 
European VLBI for Geodesy and Astrometry.~Institut d'Estudis Espacials de 
Catalunya, Consejo Superior de Investigaciones Cient{\'{\i}}ficas, 
Barcelona, Spain, September 07-08, 2001.~Edited by Dirk Behrend and Antonio 
Rius., p.46, 46 
\bibitem[Attridge(2001)]{att01} Attridge, J.~M.\ 2001, \apjl, 
553, L31 
\bibitem[Beasley et al.(1997)]{Beasley97} Beasley, A.J., Dhawan,
V., Doeleman, S., \& Phillips, R.B.\ 1997, Millimeter-VLBI Science
Workshop, ed. R. Barvainis \& R.B. Phillips, MIT-Haystack Observatory, p.~53
\bibitem[Bower et al.(1997)]{Bower97} Bower, G.~C., Backer, 
D.~C., Wright, M., Forster, J.~R., Aller, H.~D., \& Aller, M.~F.\ 1997, 
\apj, 484, 118 
\bibitem[Doeleman \& Claussen(1997)]{Doeleman96} Doeleman, S.~S., 
\& Claussen, M.\ 1997, Millimeter-VLBI Science Workshop, 37 
\bibitem[Horiuchi et al.(2004)]{hor+04} Horiuchi, S., et al.\ 
2004, \apj, 616, 110 
\bibitem[Fomalont(1999)]{Fomalont99} Fomalont, E.~B.\ 1999, 
Synthesis Imaging in Radio Astronomy II, 180, 301 
\bibitem[Kardashev(2000)]{Kar00} Kardashev, N.~S.\ 2000, 
Astronomy Reports, 44, 719 
\bibitem[Kellermann \& Pauliny-Toth(1969)]{Kellermann69} Kellermann, 
K.~I., \& Pauliny-Toth, I.~I.~K.\ 1969, \apjl, 155, L71 
\bibitem[Kellermann et al.(1998)]{Kellermann98} Kellermann, K.~I., 
Vermeulen, R.~C., Zensus, J.~A., \& Cohen, M.~H.\ 1998, \aj, 115, 1295 
\bibitem[Kellermann et al.(2003)]{Kellermann03} Kellermann, K.~I., 
Lister, M.~L., Homan, D.~C., Ros, E., Zensus, J.~A., Cohen, M.~H., Russo, 
M., \& Vermeulen, R.~C.\ 2003, High Energy Blazar Astronomy, 299, 117 
\bibitem[Kovalev et al.(2005)]{Kovalev2005} Kovalev, Y.~Y., et al.\ 
2005, \aj, 130, 2473 
\bibitem[Krichbaum et al.(1995)]{Krich95} Krichbaum, T.~P., 
Britzen, S., Standke, K.~J., Witzel, A., Schalinski, C.~J., \& Zensus, 
J.~A.\ 1995, Proceedings of the National Academy of Science, 92, 11377 
\bibitem[Krichbaum et al.(1999)]{Krich99} Krichbaum, T.~P., 
Witzel, A., \& Zensus, J.~A.\ 1999, 2nd millimeter-VLBI science workshop : 
IRAM, Granada, Spain, 27-29 May 1999 / edited by A.~Greve and 
T.P.~Krichbaum. IRAM, St.~Martin d'H\'eres, France, p.5
\bibitem[Krichbaum et al.(2006)]{Krich06} Krichbaum, T.~P., 
Graham, D.~A., Bremer, M., Alef, W., Witzel, A., Zensus, J.~A., \& Eckart, 
A.\ 2006, Journal of Physics Conference Series, 54, 328 
\bibitem[Lobanov et al.(2000)]{Lobanov00} Lobanov, A.~P., et al.\ 
2000, \aap, 364, 391 
\bibitem[Lobanov(2005)]{Lob2005} Lobanov, A.~P.\ 2005, ArXiv 
Astrophysics e-prints, arXiv:astro-ph/0503225 
\bibitem[Lobanov et al.(2006)]{Lobanov06} Lobanov, A.~P., 
Krichbaum, T.~P., Witzel, A., \& Zensus, J.~A.\ 2006, \pasj, 58, 253 
\bibitem[Lonsdale et al.(1998)]{Lonsdale98} Lonsdale, C.~J., 
Doeleman, S.~S., \& Phillips, R.~B.\ 1998, \aj, 116, 8 
\bibitem[Marscher(1995)]{mar95} Marscher, A.~P.\ 1995, 
Proceedings of the National Academy of Science, 92, 11439 
\bibitem[Moellenbrock et al.(1996)]{Moellen96} Moellenbrock, 
G.~A., et al.\ 1996, \aj, 111, 2174 
\bibitem[Pagels et al.(2004)]{Pagels} Pagels, A., et al.\ 
2004, European VLBI Network on New Developments in VLBI Science and 
Technology, 7 
\bibitem[Rantakyro et al.(1998)]{Ranta98} Rantakyro, F.~T., et 
al.\ 1998, \aaps, 131, 451 
\bibitem[Readhead(1994)]{Readhead94} Readhead, A.~C.~S.\ 1994, 
\apj, 426, 51 
\bibitem[Readhead et al.(1983)]{Readhead83} Readhead, A.~C.~S., et 
al.\ 1983, \nat, 303, 504 
\bibitem[Rogers et al.(1995)]{rog+95} Rogers, A.~E.~E., 
Phillips, R.~B., \& Lonsdale, C.~J.\ 1995, \baas, 27, 1300 
\bibitem[Shepherd et al.(1994)]{Shepherd94} Shepherd, M.~C.,
Pearson, T.~J., \& Taylor, G.~B.\ 1994, \baas, 26, 987 
\bibitem[Teraesranta et al.(1998)]{Teras98} Teraesranta, H., et 
al.\ 1998, \aaps, 132, 305 
\bibitem[Urry \& Padovani(1995)]{UP95} Urry, C.~M., \& 
Padovani, P.\ 1995, \pasp, 107, 803 
\bibitem[V{\'e}ron-Cetty \& V{\'e}ron(2006)]{veron2006} 
V{\'e}ron-Cetty, M.-P., \& V{\'e}ron, P.\ 2006, \aap, 455, 773 
\end{thebibliography}
\end{document}